\documentclass[twocolumn]{aastex61}

\shorttitle{ENHANCED NUCLEAR ACCRETION RATE IN GALAXY GROUPS AT z$\sim$0.2}
\shortauthors{Baronchelli I. et al.}

\accepted{for publication in The Astrophysical Journal - March 15, 2018}

\usepackage{graphicx}

\newcommand{\mic}{$\mu$m}
\DeclareGraphicsExtensions{.pdf,.png,.jpg}
\begin{document}

\title{The Spitzer-IRAC/MIPS Extragalactic Survey (SIMES): II ENHANCED NUCLEAR ACCRETION RATE IN GALAXY GROUPS AT z$\sim$0.2\vspace{-0.5cm}}

\correspondingauthor{I. Baronchelli}
\email{ivano@ipac.caltech.edu}

\author{I. Baronchelli}
\affiliation{IPAC, Mail Code 314-6, Caltech, 1200 E. California Blvd., Pasadena, CA 91125, USA.\vspace{-0.07cm}}
\affiliation{MN Institute for Astrophysics, University of Minnesota, 116 Church St. SE,  Minneapolis, MN 55455, USA.\vspace{-0.07cm}}
\affiliation{Dipartimento di Fisica e Astronomia, Universit${\grave{\mathrm{a}}}$ di Padova, vicolo Osservatorio, 3, 35122 Padova, Italy.\vspace{-0.07cm}}

\author{G. Rodighiero}
\affiliation{Dipartimento di Fisica e Astronomia, Universit${\grave{\mathrm{a}}}$ di Padova, vicolo Osservatorio, 3, 35122 Padova, Italy.\vspace{-0.07cm}}

\author{H. I. Teplitz}
\affiliation{IPAC, Mail Code 314-6, Caltech, 1200 E. California Blvd., Pasadena, CA 91125, USA.\vspace{-0.07cm}}

\author{C. M. Scarlata}
\affiliation{MN Institute for Astrophysics, University of Minnesota, 116 Church St. SE,  Minneapolis, MN 55455, USA.\vspace{-0.07cm}}

\author{A. Franceschini}
\affiliation{Dipartimento di Fisica e Astronomia, Universit${\grave{\mathrm{a}}}$ di Padova, vicolo Osservatorio, 3, 35122 Padova, Italy.\vspace{-0.07cm}}

\author{S. Berta}
\affiliation{University of Zagreb, Physics Department, Bijeni\v{c}ka cesta 32, 10002 Zagreb, Croatia - Visiting scientist.\vspace{-0.07cm}}

\author{L. Barrufet}
\affiliation{Department of Physical Sciences, The Open University, Milton Keynes, MK7 6AA, UK\vspace{-0.07cm}}

\author{M. Vaccari}
\affiliation{Department of Physics and Astronomy, University of the Western Cape, Robert Sobukwe Road, Bellville, 7535, South Africa\vspace{-0.07cm}}
\affiliation{INAF - Istituto di Radioastronomia, Via Piero Gobetti 101, I-40129 Bologna, Italy.\vspace{-0.07cm}}

\author{M. Bonato}
\affiliation{INAF - Istituto di Radioastronomia, Via Piero Gobetti 101, I-40129 Bologna, Italy.\vspace{-0.07cm}}

\author{L. Ciesla}
\affiliation{Laboratoire AIM Paris-Saclay CEA/IRFU/SAp, Universite Paris Diderot, Orme des Merisiers, Building 709, F-91191 Gif-sur-Yvette Cedex, France.\vspace{-0.07cm}}

\author{A. Zanella}
\affiliation{European Southern Observatory, Karl Schwarzschild Stra\ss e 2, 85748 Garching, Germany.\vspace{-0.07cm}}

\author{R. Carraro}
\affiliation{Instituto de F\'\i{}sica y Astronom\'\i{}a, Universidad de Valpara\'\i{}so, Gran Breta\~{n}a 1111, Playa Ancha, Valpara\'\i{}so, Chile\vspace{-0.07cm}}
\affiliation{Department of Astronomy, Yale University, 52 Hillhouse Avenue, New Haven, CT 06511, USA\vspace{-0.07cm}}

\author{C. Mancini}
\affiliation{Dipartimento di Fisica e Astronomia, Universit${\grave{\mathrm{a}}}$ di Padova, vicolo Osservatorio, 3, 35122 Padova, Italy.\vspace{-0.07cm}}

\author{A. Puglisi}
\affiliation{Dipartimento di Fisica e Astronomia, Universit${\grave{\mathrm{a}}}$ di Padova, vicolo Osservatorio, 3, 35122 Padova, Italy.\vspace{-0.07cm}}
\affiliation{European Southern Observatory, Karl Schwarzschild Stra\ss e 2, 85748 Garching, Germany.\vspace{-0.07cm}}

\author{M. Malkan}
\affiliation{Department of Physics and Astronomy,UCLA, Physics and Astronomy Bldg., 3-714, LA CA 90095-1547, USA\vspace{-0.07cm}}

\author{S. Mei}
\affiliation{GEPI, Observatoire de Paris, CNRS, University of Paris Diderot, Paris Sciences et Lettres (PSL), 61, Avenue de l'Observatoire F-75014, Paris, France\vspace{-0.07cm}}
\affiliation{University of Paris Denis Diderot, University of Paris Sorbonne Cité (PSC), F-75205 Paris Cedex 13, France\vspace{-0.07cm}}
\affiliation{IPAC, Mail Code 314-6, Caltech, 1200 E. California Blvd., Pasadena, CA 91125, USA.\vspace{-0.07cm}}

\author{L. Marchetti}
\affiliation{Department of Astronomy, University of Cape Town, Private Bag X3, Rondebosch, 7701, South Africa\vspace{-0.07cm}}
\affiliation{Department of Physics and Astronomy, University of the Western Cape, Robert Sobukwe Road, Bellville, 7535, South Africa\vspace{-0.07cm}}
\affiliation{Department of Physical Sciences, The Open University, Milton Keynes, MK7 6AA, UK\vspace{-0.07cm}}

\author{J. Colbert}
\affiliation{IPAC, Mail Code 314-6, Caltech, 1200 E. California Blvd., Pasadena, CA 91125, USA.\vspace{-0.07cm}}

\author{C. Sedgwick}
\affiliation{Department of Physical Sciences, The Open University, Milton Keynes, MK7 6AA, UK}

\author{S. Serjeant}
\affiliation{Department of Physical Sciences, The Open University, Milton Keynes, MK7 6AA, UK\vspace{-0.07cm}}

\author{C. Pearson}
\affiliation{RAL Space, Rutherford Appleton Laboratory, Chilton, Didcot, Oxfordshire OX11 0QX, United Kingdom\vspace{-0.07cm}}
\affiliation{Department of Physical Sciences, The Open University, Milton Keynes, MK7 6AA, UK\vspace{-0.07cm}}
\affiliation{Oxford Astrophysics, Denys Wilkinson Building, University of Oxford, Keble Rd, Oxford OX1 3RH, UK\vspace{-0.07cm}}

\author{M. Radovich}
\affiliation{INAF - Astronomical Observatory of Padua, vicolo dell’Osservatorio 5, I-35122 Padova, Italy\vspace{-0.07cm}}

\author{A. Grado}
\affiliation{INAF - Osservatorio Astronomico di Capodimonte Via Moiariello 16, 80131 Napoli, Italy\vspace{-0.07cm}}

\author{L. Limatola}
\affiliation{INAF - Osservatorio Astronomico di Capodimonte Via Moiariello 16, 80131 Napoli, Italy\vspace{-0.07cm}}

\author{G. Covone}
\affiliation{INAF - Osservatorio Astronomico di Capodimonte Via Moiariello 16, 80131 Napoli, Italy\vspace{-0.07cm}}
\affiliation{Dipartimento di Fisica, University of Naples Federico II, INFN, Via Cinthia, I-80126 Napoli, Italy\vspace{-0.1cm}}

\vspace{-1.2cm}

\begin{abstract}

For a sample of star forming galaxies in the redshift interval 0.15$<$z$<$0.3, we study how both the relative strength of the AGN infra-red emission, compared to that due to the star formation (SF), and the numerical fraction of AGNs, change as a function of the total stellar mass of the hosting galaxy group (M$^{*}_{\mathrm{group}}$), between $10^{10.25}$ and $10^{11.9}$M$_{\odot}$.
Using a multi-component SED fitting analysis, we separate the contribution of stars, AGN torus and star formation to the total emission at different wavelengths. This technique is applied to a new multi-wavelength data-set in the SIMES field (23 not redundant photometric bands), spanning the wavelength range from the UV (\emph{GALEX}) to the far-IR (\emph{Herschel}) and including crucial \emph{AKARI} and \emph{WISE} mid-IR observations (4.5 \mic$<\lambda<$24 \mic), where the BH thermal emission is stronger.  This new photometric catalog, that includes our best photo-z estimates, is released through the NASA/IPAC Infrared Science Archive (IRSA). Groups are identified through a \emph{friends of friends} algorithm ($\sim$62\% purity, $\sim$51\% completeness). 
We identified a total of 45 galaxies requiring an AGN emission component, 35 of which in groups and 10 in the field.
We find BHAR$\propto ($M$^{*}_{\mathrm{group}})^{1.21\pm0.27}$ and (BHAR/SFR)$\propto ($M$^{*}_{\mathrm{group}})^{1.04\pm0.24}$ while, in the same range of M$^{*}_{\mathrm{group}}$, we do not observe any sensible change in the numerical fraction of AGNs. Our results indicate that the nuclear activity (i.e. the BHAR and the BHAR/SFR ratio) is enhanced when galaxies are located in more massive and richer groups.

\end{abstract}

\keywords{
galaxies: evolution – galaxies: groups: general – galaxies: star formation –  galaxies: supermassive black holes – infrared: galaxies – submillimeter: galaxies \vspace{-1.0cm} }

\section{Introduction}
\label{introduction}

In the last few years, the study of the nature of Active Galactic Nuclei (AGNs) and their host galaxies has been driven by the discovery of various scaling relations between their physical properties and the way they vary during cosmic time \citep[see][for a review]{2013ARA&A..51..511K}.

AGNs and starbursts are found to coexist at all redshifts \citep{2003MNRAS.343..585F, 2005Natur.434..738A}, showing a similar evolution in terms of global Star Formation Rate Density \citep[SFRD,][]{1996ApJ...460L...1L,2006ApJ...651..142H,2013A&A...554A..70B} and Black Hole Accretion Rate Density \citep[BHARD,][]{1999MNRAS.310L...5F,2007ApJ...669...67H,2008MNRAS.388.1011M,2014MNRAS.439.2736D}, with a rise through z$\sim$2 and a consecutive fall after that epoch.
The mass of Super Massive Black Holes (SMBH) and that of the bulge of their host galaxies show a tight correlation \citep{2003ApJ...589L..21M,2012MNRAS.419.2264V,2014ApJ...780...70L}, while the BHAR is related to the stellar mass M* \citep{2012ApJ...753L..30M,2015ApJ...800L..10R} through a relations similar to the so called \emph{main sequence} of the star forming galaxies existing between SFR and M* \citep{2004MNRAS.351.1151B,2007A&A...468...33E,2007ApJ...670..156D,2007ApJ...660L..43N,2014MNRAS.443...19R,2017MNRAS.465.3390A}.
Finally, \cite{2016MNRAS.458.4297G} and \cite{2015ApJ...800L..10R} found the BHAR to increase at larger SFR and specific SFR (sSFR=SFR/M*).

The correlations found between the properties of AGNs and their host galaxies suggest that a secular co-evolution must have been in place during cosmic time. The possible self-regulation of the AGN-host galaxy systems, possibly driven by AGN-feedback mechanisms, has been often invoked as an explanation for these correlations \cite[e.g][]{2012ApJ...745..178F,1996MNRAS.283.1388M,2004mas..conf..379G,2005MNRAS.361..776S,2012Natur.485..213P} even if a complete explanation of the phenomenon remains unclear, at least at lower stellar masses \citep{2017arXiv170204507F}. Recently, \cite{2017arXiv170501132A} found stochastic processes to be primarily responsible for the fueling of AGN and for their variability.

Complicating the picture here summarized, the environment in which galaxies are located plays an important role in their evolution. In the local universe, star formation seems to happen more likely in less dense environments, such as small groups or in the field, rather than in more massive clusters,
 but the SFR in cluster galaxies increases strongly with redshift \citep[e.g.][]{1984ApJ...285..426B,2006ApJ...642..188P,2008ApJ...685L.113S}. Also AGNs are influenced by environment. At low redshifts, the fraction of galaxies hosting X-ray identified AGNs is lower among cluster galaxies than in the field \citep{1978MNRAS.183..633G,1985ApJ...288..481D,1993AJ....106..831H,1999ApJS..122...51D,2004MNRAS.353..713K,2005AJ....130.1482R,2006A&A...460L..23P}. Quiescent galaxies, typically populating the inner regions of galaxy clusters, are also observed to host weaker AGNs than star forming galaxies \citep{2017arXiv170501132A}. However, the fraction of AGNs inside clusters is evolving with z more rapidly than in the field \citep{2007ApJ...664L...9E,2009ApJ...701...66M,2013ApJ...768....1M}. Also the AGN duty cycle for both quiescent and star forming galaxies evolves strongly with redshift and becomes comparable at z$\sim$2 \citep{2017arXiv170501132A}.
The mechanism thought to be responsible for the concomitant activation of the star formation and of the central black hole accretion, at least for the most X-ray luminous sources, is a large gas infall due to major mergers \citep{1988ApJ...325...74S,1991ApJ...370L..65B,2006ApJS..163....1H}. In general, the dense environment of the cluster can influence the galaxy activity through different mechanisms, such as minor mergers, harassment, ram-pressure stripping and strangulation \citep[e.g.][]{1972ApJ...176....1G,1976ApJ...204..642R,1980ApJ...237..692L,1988ApJ...325...74S,2008ApJ...672L.103K,2008MNRAS.387...79V,2008MNRAS.383..593M,2009ApJ...699.1595P}

In this framework, the study of dusty star forming galaxies and accreting SMBHs is crucial. At the peak of the cosmic star formation history, most of the ultra violet (UV) light emitted by massive young stars, and absorbed by the dusty star forming regions, is re-emitted at longer wavelengths, with a peak at $\sim$100 $\mu$m \citep[e.g.][]{2012ApJ...759..139K}. This is the far infra-red (far-IR) regime recently explored by space telescopes such as \emph{Spitzer} (24 $\mu$m-160$\mu$m) and \emph{Herschel} ($\sim$60-500 $\mu$m). The activity of the central black hole can also be heavily obscured. As for the star formation,  the dust-enshrouded BH accretion is observable at mid-IR wavelengths. In this spectral region, the dusty torus surrounding the accretion disk of a SMBH re-emits the highly energetic flux from the central engine absorbed at short wavelengths.
The \emph{AKARI} and \emph{WISE} observatories allowed to sample the spectral region interested by the peak of the AGN emission, thanks to their sensitivity at 7, 11 and 15 $\mu$m.

In the first part of this work, we present a new set of optical and mid-IR observations covering the central square degree of the Spitzer-IRAC/MIPS Extra-galactic survey (SIMES) field \citep[see][ and references therein]{2016ApJS..223....1B}. In the second part, we discuss a SED fitting technique applied to the complete data-set spanning the spectral range from the UV (\emph{GALEX}) to the far-IR (\emph{Spitzer, Hershel}) wavelengths and the results obtained.
The technique used allows us to constrain the optical emission due to stars and to disentangle the contribution of AGN torus and star formation to the total IR luminosity using a triple component fit. With this approach, we computed stellar masses (M*), star formation rates (SFR) and black hole accretion rates (BHAR) for the galaxies in our analysis sample. In our analysis, we studied the relations between the properties of the far-IR detected AGNs with their grouping properties.

Our approach is similar to that used in other works \citep{2014MNRAS.439.2736D,2016MNRAS.458.4297G}, but the strength of the data-set presented in this paper is the presence of \emph{AKARI} and deep \emph{WISE} mid-IR observations. 
The \emph{AKARI} data cover only a fraction of the square degree explored in our analysis (the green square in Figure~\ref{IRAC1_cov}; see also Table~\ref{coverage_table}). However, the \emph{WISE} observations, covering the entire sky and beeing particularly deep close to the South Ecliptic Pole (SEP), where the SIMES field is located, allow us to constrain the peak of the SMBH IR emission for almost all the sources in our analysis sample (0.15$<$z$<$0.3, see Figure~\ref{img:depths}).
Following the AGN unified model \citep{1995PASP..107..803U}, a relevant fraction of the high energy radiation produced in the inner regions of a SMBH is absorbed by the surrounding dusty torus and then re-emitted by the heated dust, in the far-IR region of the spectrum. For this reason, the IR-based selection of AGNs allows us to mitigate the incompleteness problem (20-50\%) usually affecting X-ray surveys \citep{2005ApJ...634..169D,2005A&A...444..119G,2010ApJ...717.1181P,2011ApJ...738...44A,2013A&A...555A..43G,2013ApJ...773...15W}.

The paper is organized as follows.
In Section~\ref{SEC:DATA}, we present the new set of optical (CTIO, VST) and mid-IR (\emph{AKARI}) observations covering the central region of the SIMES field and complementing the \emph{Spitzer}-IRAC, MIPS and \emph{Hershel}-SPIRE data described in \cite{2016ApJS..223....1B,2016yCat..22230001B}.
In Section~\ref{OPTICAL_PHOTO_Z_sect}, we describe the SED fitting technique used to compute photometric redshifts and the method used to improve the precision of our results on 24 \mic\ selected sources, through the use of an optical prior.
In Section~\ref{SEC:groups} we present the friend of friend algorithm used to identify group candidates in the optically covered area.
In Section~\ref{SEC:GAL_PHYS_PROP}, the triple component SED fitting technique is described together with the physical quantities obtained for each galaxy in our analysis sample.
Finally, in Section~\ref{SEC:ANALYSIS} we present the results of our analysis on the environmental dependency of the SMBH activity. 

Throughout this paper, unless differently specified, we assume AB magnitudes, \cite{2003PASP..115..763C} IMF and a standard cosmology with H$_{0}$=70 Km s$^{-1}$ Mpc$^{-1}$, $\Omega_{M}=0.3$ and $\Omega_{\Lambda}=0.7$.
In the analysis, the quantities not indicated with the underscored \emph{``group''} refer to single galaxies (e.g. M$^{*}$, BHAR, SFR and bolometric AGN fraction). Vice versa, quantities indicated with an underscored \emph{``group''} refer to the properties of the hosting group (e.g.  M$^{*}{_{\mathrm{group}}}$). The numerical fraction of AGNs identified in groups and in the field are indicated with $f_{\mathrm{group}}$ and $f_{\mathrm{field}}$ respectively.

\section{DATA}
\label{SEC:DATA}
\begin{figure}[!ht]
\centering
\includegraphics[width=8.0cm]{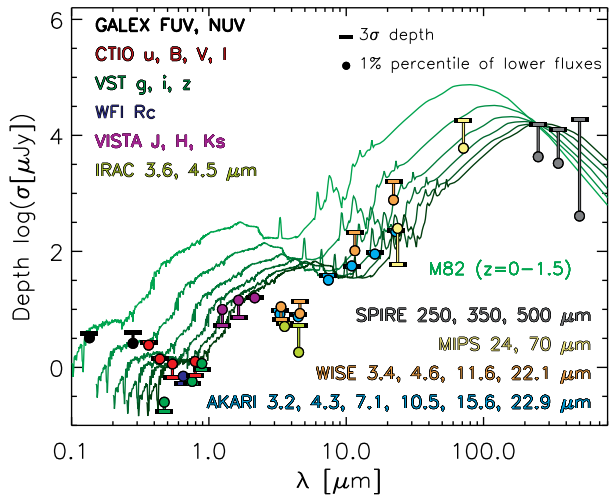}
\caption{Depth (3$\sigma$) of the multiwavelength data in the SIMES field (horizontal thick lines). The faintest fluxes in the catalog (1\% percentile) are also represented (filled circles). For comparison, the SED of the starbursting galaxy M82 \citep[green line, from ][]{2007ApJ...663...81P} is shown for 6 different redshift values in the range 0$\leq$z$\leq$1.5. The M82 SED is normalized to the 250 \mic\ - 3$\sigma$ depth.}
\label{img:depths}
\end{figure}

\begin{figure*}[!ht]
\centering
\includegraphics[width=15.5cm]{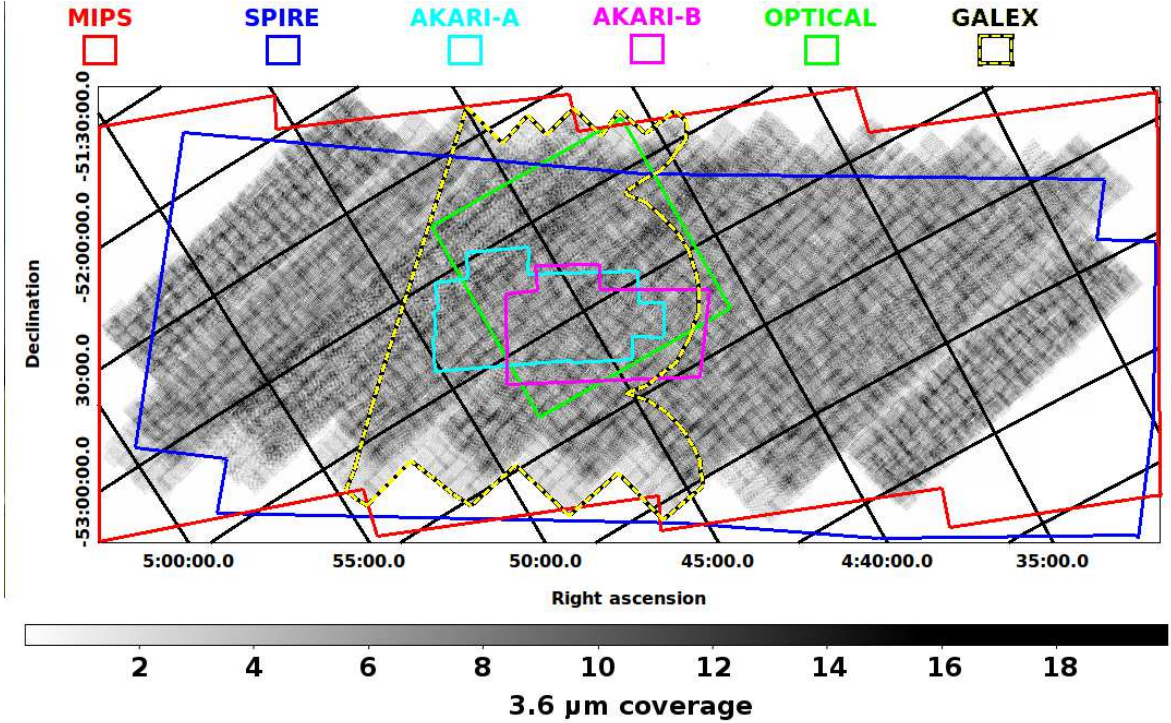}
\caption{Coverage of the SIMES field by different instruments. The background image represents the coverage at 3.6 \mic\ (\emph{Spitzer}/IRAC). The coverage at 4.5 \mic\ is similar, with a small shift ($\sim$6 arcmin) towards lower declinations. The whole field is covered in the VISTA J, H and Ks bands and by observations in the four \emph{WISE} mid-IR bands (3.3 to 22 \mic). Using different colors, we represent the coverage of \emph{Spitzer}/MIPS (24 and 70 \mic, red), \emph{Herschel}/SPIRE (250 to 500 \mic, blue), \emph{AKARI}-''A'' (3 to 11 \mic, cyan), \emph{AKARI}-''B'' (15 and 24 \mic, magenta), \emph{GALEX} (black/dashed yellow) and optical bands (green). Our analysis (Section \ref{SEC:ANALYSIS}), focuses on the region covered by both optical and SPIRE observations. We notice that even if the important \emph{AKARI} observations at 7, 11, and 15 \mic\ do not cover the entire analysis area, this is sampled at 11 \mic\ by \emph{WISE} with an high detection rate among the SPIRE and MIPS detected sources ($\sim$94\%), in the redshift explored (0.15$<$z$<$0.3).  }
\label{IRAC1_cov}
\end{figure*}

The multi--wavelength data collection used in our analysis is based on the deep \emph{Spitzer}/IRAC-3.6 \mic\ observations of the SIMES field, described in \citet[][BA16 hereafter]{2016ApJS..223....1B,2016yCat..22230001B}. Besides the measures at 3.6 \mic, the BA16 catalog includes observations at 4.5 (IRAC), 24 (MIPS), 250, 350 and 500 \mic\ (\emph{Herschel}/SPIRE), plus optical observations in the ESO/WFI R$_{c}$ band. The 3.6 \mic\ observations reach an average 3$\sigma$ depth of 5.79 $\mu$Jy at 3.6 \mic \ (see Figure~\ref{img:depths}) and cover a total area of $\sim$7 square degrees (see Figure~\ref{IRAC1_cov}). Only 3.6 \mic\ detected sources are included in this catalog.

In this work, we merged the BA16 catalog with a new set of UV, optical and IR photometric measures. The resulting catalog now includes observations in 22 photometric bands (30 bands considering observations taken by different instruments at similar wavelengths): 
Far-UV, Near-UV, u, B, g, V, R$_{\mathrm{c}}$, i, z, J, H, K$_{\mathrm{s}}$, 3.6, 4.5, 7, 11, 15, 24, 70, 250, 350, 500 \mic. This new photometric catalog is released through the NASA/IPAC Infrared Science Archive (IRSA) and it includes our best photometric redshift estimates, described in the following sections.

The complete list of bands, with the depth reached at each wavelength, is reported in Table \ref{coverage_table}. The depth is also represented, for all the bands in the catalog, in Figure~\ref{img:depths}. 
In this figure, for each filter, we report the 3$\sigma$ depths and the average flux of the faintest 1\% percentiles of the sources included in the catalog. These values can be compared with the spectral energy distributions (SED) of a starburst galaxy \citep[M82, from ][]{2007ApJ...663...81P}.
The area covered in different bands is represented in Figure~\ref{IRAC1_cov}. Our analysis refers to the fraction of the optical area (green square in Figure ~\ref{IRAC1_cov}) covered also by SPIRE observations (blue contour). We notice that this area is not fully covered by \emph{AKARI} mid-IR observations. However, the same area is fully covered by WISE observations, at $\sim$11 \mic. 

As reported in Table~\ref{coverage_table}, some of the filters (IRAC 3.6 \mic, MIPS 24 \mic\ and one among the SPIRE bands) are \emph{``required''} meaning that all the sources without a detection in one of these bands are excluded from our successive analysis. Other bands are simply \emph{``used''} when available.
Not all the filters included in the catalog are required or even used for our analysis. Having an highly precise set of measures covering the whole SIMES field at 3.6, 4.5 and 24 \mic\ (IRAC and MIPS), we did not consider the lower quality and redundant \emph{AKARI} N3, N4 (3.13 and 4.25 \mic\ respectively),  and the \emph{WISE} W1, W2 and W4 bands (3.35, 4.60, and 22.1 \mic\ respectively). However, given the importance of the mid-IR spectral region when constraining the AGN emission, we considered the \emph{AKARI} 24 \mic\ band and both the \emph{AKARI} and \emph{WISE} 11 \mic\ filters.

\begin{deluxetable*}{lcllrl}
\tabletypesize{\footnotesize}
\tablecolumns{6}
\tablewidth{0pc}
\tablecaption{SIMES Field Data included in the multiwavelength Catalog}
\tablehead{
Band & \colhead{$\lambda_{\mathrm{eff}}$\tablenotemark{[a]}} & Instrument/  & Overlap   & \colhead{1$\sigma$ depth} & \colhead{Used/required}\\
     &                     & telescope    & Area              & \colhead{ ($\mu$Jy)}  & \colhead{for the AGN}\\
     &                     &              & (deg$^{2}$)        &                                & \colhead{analysis}   }
\startdata
\label{coverage_table}
FUV                          & 154.2 nm & GALEX      & 2.34  & 1.27 $\mu$Jy & used \\
NUV                          & 227.4 nm & GALEX      & 2.34  & 1.33 $\mu$Jy & used \\
u                            & 355.5 nm & CTIO       & 0.67  & 0.89 $\mu$Jy & used \tablenotemark{[f]} \\
B                            & 433.6 nm & CTIO       & 0.67  & 0.47 $\mu$Jy & used \tablenotemark{[f]}\\
V                            & 535.5 nm & CTIO       & 0.66  & 0.22 $\mu$Jy & used \tablenotemark{[f]}\\
I                            & 793.7 nm & CTIO       & 0.66  & 0.24 $\mu$Jy & used \tablenotemark{[f]}\\
R$_{c}$ \tablenotemark{[b]} & 642.8 nm &  WFI       & 1.13  & 0.18 $\mu$Jy & used \tablenotemark{[f]}\\
g                            & 476.7 nm & VST        & 1.20  & 0.059 $\mu$Jy & used \tablenotemark{[f]}\\
i                            & 757.9 nm & VST        & 1.17  & 0.18 $\mu$Jy & used \tablenotemark{[f]}\\
z                            & 890.1 nm & VST        & 1.20  & 0.31 $\mu$Jy & used \tablenotemark{[f]}\\
J \tablenotemark{[c]}        & 1.246 $\mu$m & VISTA  & 7.74  & 1.76 $\mu$Jy & used \\
H \tablenotemark{[c]}        & 1.631 $\mu$m & VISTA  & 7.74  & 2.44 $\mu$Jy & used \\
Ks \tablenotemark{[c]}       & 2.134 $\mu$m & VISTA  & 6.01  & 5.31 $\mu$Jy & used \\
I1 \tablenotemark{[b]}       & 3.508 $\mu$m & IRAC   & 7.74  & 1,93 $\mu$Jy & required \\
I2 \tablenotemark{[b]}       & 4.437 $\mu$m & IRAC   & 7.26  & 1,75 $\mu$Jy & used\\
N3                           & 3.130 $\mu$m & AKARI  & 0.48 (A) & 3.2 $\mu$Jy & NOT used (IRAC-3.6 \mic\ instead)\\
N4                           & 4.251 $\mu$m & AKARI  & 0.48 (A) & 2.7 $\mu$Jy & NOT used (IRAC-4.5 \mic\ instead) \\
S7                           & 6.954 $\mu$m & AKARI  & 0.46 (A) & 13 $\mu$Jy & used \\
S11                          & 10.19 $\mu$m & AKARI  & 0.45 (A) & 19 $\mu$Jy & used \\
L15                          & 15.23 $\mu$m & AKARI  & 0.42 (B) & 31 $\mu$Jy & used \\
L24                          & 22.75 $\mu$m & AKARI  & 0.42 (B) & 78 $\mu$Jy & used \\
24 \mic \tablenotemark{[b,d]}     & 23.21 $\mu$m & MIPS & 7.66  & 20 $\mu$Jy\tablenotemark{[d]} & required \\
70 \mic \tablenotemark{[b,d]}     & 68.44 $\mu$m & MIPS & 7.66  & 5.1 mJy\tablenotemark{[d]} & used \\
W1                           & 3.353 $\mu$m & WISE      & 7.74  & 2.24 $\mu$Jy & NOT used (IRAC-3.6 \mic\ instead) \\
W2                           & 4.603 $\mu$m & WISE      & 7.74  & 4.58 $\mu$Jy & NOT used (IRAC-4.5 \mic\ instead) \\
W3                           & 11.56 $\mu$m & WISE      & 7.74  & 70.1 $\mu$Jy & used \\
W4                           & 22.09 $\mu$m & WISE      & 7.74  & 530 $\mu$Jy & NOT used (MIPS-24 \mic\ instead)\\
PSW (250 \mic) \tablenotemark{[b,e]}\ & 242.8 $\mu$m & SPIRE & 6.52 & 5.2 mJy & required \tablenotemark{[g]} \\
PMW (350 \mic) \tablenotemark{[b,e]}\ & 340.9 $\mu$m & SPIRE & 6.52 & 4.2 mJy & required \tablenotemark{[g]} \\
PLW (500 \mic) \tablenotemark{[b,e]}\ & 482.3 $\mu$m & SPIRE & 6.52 & 6.2 mJy & required \tablenotemark{[g]} \\
\enddata
\tablenotetext{a}{ From http://svo2.cab.inta-csic.es/svo/theory/fps3/}
\tablenotetext{b}{ For additional information on how these data are extracted or included in the catalog, see \cite{2016ApJS..223....1B}}
\tablenotetext{c}{ Original data from \cite{2013Msngr.154...35M}; Vista Hemisphere survey data release 2 expected depths.}
\tablenotetext{d}{ Original data from \cite{2011MNRAS.411..373C}. The depths reported in \cite{2011MNRAS.411..373C}, 260 $\mu$Jy and 2.4 mJy at 24 and 70 \mic\ respectively, correspond to the values where the 50\% completeness is reached. Instead, the depths reported in this table, correspond to the extrapolation, at S/N=1, of the Flux versus S/N ratio. }
\tablenotetext{e}{ Original data From \cite{2012MNRAS.424.1614O}.}
\tablenotetext{f}{ The detection in at least two of these optical bands is required, for a source, in order to be considered in the analysis.}
\tablenotetext{g}{ Only one detection at 250, 350 or 500 \mic\ is required, for a source, in order to be considered in the analysis.}

\end{deluxetable*}

\subsection{Mid--IR}
\subsubsection{WISE}
\label{WISE_sect}
The SIMES field is fully covered, as a part of the \emph{AllWISE} all--sky survey, in the \emph{WISE} W1, W2, W3, W4 bands at 3.35, 4.60, 11.56,  22.09 \mic\ respectively \citep{2010AJ....140.1868W}.
 The \emph{WISE} survey scanning strategy resulted in frame--set depth-of-coverage that increased with increasing ecliptic latitude. Moreover, the sensitivity naturally improves toward the ecliptic poles due to lower zodiacal background. Therefore, in terms of depth, the southern and the northern ecliptic poles (NEP and SEP) are privileged fields. The SIMES field is located close to the SEP area.

Using the publicly available \emph{AllWISE} tables describing the survey depth at different (RA, DEC) coordinates \footnote{http://wise2.ipac.caltech.edu/docs/release/allwise/expsup/\newline figures/sec2\_3a\_table2.tbl}, we estimated the average depths (1$\sigma$), in the SIMES area. They resulted 2.24, 4.58, 70.1, 530 $\mu$Jy for W1, W2, W3 and W4 respectively. 

At 11 \mic\ the expected depth is of the same order of that measured for the \emph{AKARI}-S11 band (19 $\mu$Jy). 
While we limit the \emph{AKARI}-S11 catalog to only sources with S/N above the 3$\sigma$ level, the \emph{WISE} measurements in this band are reported down to 1$\sigma$ level, with a prior 5$\sigma$ level detection in the W1 channel (3.35 \mic) required. This guarantees a similar spatial density of sources measured in these two filters, when a MIPS-24 \mic\ detection is present, as we require in our analysis. Among the 24 \mic\ sources detected in the \emph{WISE}-\emph{AKARI} common area at all redshifts, 89\% of the \emph{AKARI}-S11 detections have a measurement in the correspondent \emph{WISE} filter. Vice versa, the 74\% of the W3 measurements have a correspondent S11 detection. When limiting these data to the redshift range explored in our analysis (0.15$<$z$<$0.3), these fractions rise to 94\% and 96\% respectively.

In our catalog, the IRAC 3.6 \mic\ source associated to each \emph{WISE} detection is the closest geometrical counterpart inside a 2\farcs68 searching radius. This value corresponds to the square root of the sum of the variances of the IRAC 3.6 \mic\ (0.705'') and \emph{WISE} W1 (2.59'') PSFs.

\subsubsection{\emph{AKARI}--IRC}
\label{AKARI_dat_sect}

Between 3 and 24 \mic , the SIMES field is partially covered ($\sim$0.5 deg$^{2}$, see Table \ref{coverage_table} and Figure~\ref{IRAC1_cov}) by \emph{AKARI}--IRC observations \citep{2007PASJ...59S.401O} in the N3, N4, S7, S11, L15, and L24 filters ($\lambda_{\mathrm{ref}}$=3.13, 4.25, 6.95, 10.2, 15.2, 22.8 $\mu m$, P.I.: C. Pearson).

\begin{deluxetable}{lcccccc}
\tabletypesize{\footnotesize}
\tablecolumns{7}
\tablewidth{0pc}
\tablecaption{ Main Properties of the \emph{AKARI} Images\tablenotemark{a}}
\tablehead{
\colhead{Parameter} & \multicolumn{6}{c}{\emph{AKARI} Band}  \\
\colhead{}& \colhead{N3} & \colhead{N4} & \colhead{S7} & \colhead{S11} & \colhead{L15} & \colhead{L24}    }
\startdata
\label{AKARI_image_char_table}
FWHM [arcsec]                               & 4.23 & 4.23 & 5.15 & 5.62 & 5.64 & 6,86 \\
Pixel Scale [arcsec pixel$^{-1}$]                  & 1.46 & 1.46 & 2.34 & 2.34 & 2.45 & 2.45 \\
1$\sigma$ depth [$\mu$Jy] \tablenotemark{b} & 3.2  & 2.7  & 13   & 19   &  31  & 78   \\
\enddata
\tablenotetext{a}{P.I. C. Pearson }
\tablenotetext{b}{Same values as in Table~\ref{coverage_table} and reported here for convenince.}
\end{deluxetable}

We computed the depth reached in each mosaic by measuring the flux inside randomly distributed apertures with diameter $\gtrsim 2\times$FWHM (See Table~\ref{AKARI_image_char_table}) and fitting a Gaussian function to the symmetrized distribution of the fluxes. In Table~\ref{coverage_table} we report the 1$\sigma$ values of the fitted Gaussian functions.

We performed the source detection and extraction using the \emph{SExtractor} software \citep{1996A&AS..117..393B}. In this phase, the coverage maps corresponding to each band are used as weight images. Following BA16, we consider total fluxes estimated in elliptical apertures with semi-major axis ($a$) proportional to the Kron radius \citep[$R_K$,][]{1980ApJS...43..305K} of each object (i.e. SExtractor ``AUTO'' fluxes). We considered only sources with five connected pixels above a threshold of 1$\sigma$ of the local background. This choice guarantees that only $\sim$1\% of the sources included in the final catalog have fluxes lower than $3\sigma$ (see Figure~\ref{img:depths}).

For each \emph{AKARI} band, the IRAC-3.6 \mic\ counterpart is identified as the closest source found inside a search radius of 3\farcs73, 3\farcs73, 5\farcs20, 5\farcs66, 5\farcs68, and 6\farcs90 for the bands N3, N4, S7, S11, L15, and L24 respectively. For N3 and N4, the searching radius corresponds to the quadratic sum of the PSF's $\sigma$ of the IRAC--3.6\mic\ and \emph{AKARI} images. For the remaining \emph{AKARI} bands we found this choice too restrictive, given the overall quality of the \emph{AKARI} final images. Combining the IRAC-3.6 \mic\ and the \emph{AKARI} PSF's $\sigma$ leads to the exclusion of a important fraction of real counterparts. In these cases we found more reliable results using the \emph{AKARI} FWHM (2.355 $\sigma$).

\subsection{Optical Data}
\label{optical_cat_sect}

The central square degree of the SIMES field (green square in Figure~\ref{IRAC1_cov}) is covered by ESO/WFI--R$_{c}$ (P.I. T. Takeuchi, see BA16 for a detailed discussion) and VST--g, i, and z observations
, obtained as a part of the INAF VST GT VOICE Survey Project \citep[PIs :  Covone \& Vaccari,][]{2017arXiv170401495V}. The VST mosaics were produced and calibrated 
at the VSTCen of the Osservatorio Astronomico di Capodimonte (INAF) using the VST-Tube pipeline \citep[][SAIT]{2012MSAIS..19..362G}.

Half of this square degree ($\approx 64'\times 36'$) is covered by the Mosaic II camera\footnote{Retired in 2012 and replaced by The Dark Energy Camera DECam} of the CTIO-Blanco telescope in the u, B, V, and I filters (P.I. Pearson C.). In Figure~\ref{IRAC1_cov}, this corresponds to the fraction of optical area below DEC$\approx -53^{\circ}15'$. These CTIO data were reduced using standard IRAF routines included in the NOAO mosaic software MSCRED. All science frames were corrected by dividing them by the Super Sky Flat Field (SSFF) image obtained by averaging all the non--aligned and source subtracted science exposures in the same filter. The pixel scale and the FWHM of the final mosaics are summarized in Table~\ref{OPTIC_extract_param_table}.

A precise photometric calibration of the optical mosaics is obtained through the comparison with \citet[][BC03]{2003MNRAS.344.1000B} template SEDs fitted to the optical data, as described in more detail in section \ref{sec:flux_cal}. As for the \emph{AKARI} images, we calculated the depth of each optical image from the symmetrized distribution of the fluxes computed inside randomly distributed apertures with sizes$\gtrsim 2\times$FWHM. In Table~\ref{coverage_table} we report the 1$\sigma$ values of the Gaussian functions that better fitted the symmetrized histograms.

\begin{deluxetable*}{lcccccccc}
\tabletypesize{\footnotesize}
\tablecolumns{9}
\tablewidth{0pc}
\tablecaption{Main Properties of the Optical Images}
\tablehead{
\colhead{Parameter} & \multicolumn{7}{c}{Optical band}  \\
\colhead{}& \colhead{u\tablenotemark{a}} & \colhead{B\tablenotemark{a}} & \colhead{V\tablenotemark{a}} & \colhead{I\tablenotemark{a}} & \colhead{R$_{c}$\tablenotemark{b}} & \colhead{g\tablenotemark{c}} & \colhead{i\tablenotemark{c}} & \colhead{z\tablenotemark{c}}  }
\startdata
\label{OPTIC_extract_param_table}
FWHM [arcsec]                               & 2.18 & 1.74 & 1.35 & 1.42 & 0.98 & 0.89 & 1.34 & 0.78 \\
Pixel Scale [arcsec pixel$^{-1}$]                  & 0.27 & 0.27 & 0.27 & 0.27 & 0.24 & 0.21 & 0.21 & 0.21 \\
1$\sigma$ depth [$\mu$Jy] \tablenotemark{d} & 0.89 & 0.47 & 0.22 & 0.24 & 0.18 & 0.06 & 0.18 & 0.31 \\
\enddata
\tablenotetext{a}{CTIO--Mosaic II observations, P.I. C. Pearson  }
\tablenotetext{b}{ESO/WFI observations. P.I. T. Takeuchi, see \cite{2016ApJS..223....1B}.}
\tablenotetext{c}{VST observations, P.Is. :  Covone \& Vaccari, see \cite{2017arXiv170401495V}.}
\tablenotetext{d}{Same values as in Table~\ref{coverage_table} and reported here for convenince.}
\end{deluxetable*}

For all the bands listed above, the sources are detected, and their fluxes measured, using the \emph{SExtractor} software, setting a minimum threshold of 5 connected pixels at 1$\sigma$ level above the local background for u, B, and R$_{\mathrm{c}}$ and 0.75$\sigma$ for V, I, g, i and z. This combination guarantees that only $\sim$1\% of the sources in the final catalog have fluxes lower than the estimated $3\sigma$ levels reported in Table~\ref{coverage_table} (see also Figure~\ref{img:depths}). The source extraction is weighted considering the weight maps for the VST images and the RMS maps for the CTIO mosaics.

The optical detections are included in the final catalog through two separate steps. First, we searched for possible counterparts in different optical bands, inside a 1\farcs0 searching radius. In this particular case, we used the ``\emph{tmatchn}`` function of the  \emph{stilts} software\footnote{For more information, see \\ http://www.star.bris.ac.uk/$\sim$mbt/stilts/sun256/sun256.html}. 
In a matched group, each detection is linked to a detection in another band; however, for any particular pair in a group, there is no guarantee that the two detections in two different optical bands match each other, but only that it is possible to pass from one to another through a series of pair matches.
Only sources with detections in at least two different optical bands are included in the final catalog. In the second step, for each optical group, the IRAC 3.6 \mic\ counterpart is identified inside a searching radius of 0\farcs82, corresponding to the quadratic sum of the $\sigma$ of the IRAC 3.6 \mic\ PSF and the searching radius used to create the optical catalog of sources (1\farcs0).

We notice that even if only half of the central square degree is covered by all the optical bands, this does not create selection biases: the u, B, V and I bands, covering a smaller area, are shallower then the observations covering the whole square degree (see Figure~\ref{img:depths}).

\subsection{VISTA J, H, K$_{S}$}
\label{sec:VISTA_J_H_K}
The SIMES field is fully covered by VISTA observations in the J and H bands and partially (78\%) covered in the K$_{S}$ filter, as part of the VISTA Hemisphere Survey \citep[VHS, ][]{2013Msngr.154...35M}. These VISTA photometric measurements \footnote{Second data release (DR2). For a detailed description, see \\ https://www.eso.org/sci/observing/phase3/data\_releases/vhs\_dr2.pdf.} are included in our multiwavelength catalog.  
We searched the 3.6 \mic\ based catalog for the closest geometrical counterparts of the VISTA sources using a search radius of 0\farcs82, corresponding to the quadratic sum of the IRAC 3.6 \mic\ PSF's $\sigma$ and the approximated average seeing of the VISTA survey (1\farcs0)
The J, H and Ks measurements were finely calibrated using the technique described in Section~\ref{sec:flux_cal}.

\subsection{GALEX}
We included in our multi--wavelength catalog a set of measurements obtained in the near- and far--UV (2274 and 1542 \AA ) by the \emph{GALEX} space telescope in the SIMES area (DR5). As in the other cases, we associated the 3.6 $\mu m$ sources to the geometrically closest \emph{GALEX} counterpart using a 2\farcs2 searching radius, corresponding to the quadratic sum of the IRAC 3.6 \mic\ and \emph{GALEX}-NUV PSF's $\sigma$. Only a fraction of the SIMES field is covered in the UV bands and the average depth of the data (1$\sigma$) is 1.33 $\mu$Jy in the NUV band and 1.27 $\mu$Jy in the FUV channel. \emph{GALEX} observations over-impose well with the optical coverage. In Figure~\ref{IRAC1_cov} the coverage is represented using a black line.

\subsection{Refined photometric calibration}
\label{sec:flux_cal}

The photometric calibration of the bands from u to K$_{\mathrm{s}}$ is obtained in two separate steps. An initial calibration is obtained using different methods for different bands, while a subsequent refined calibration is obtained for all the bands simultaneously, using an iterative SED fitting technique. 

For the VST g, i and z bands, the initial calibration is obtained using the VST-Tube pipeline \citep[][Vaccari et al. in prep.]{2012MSAIS..19..362G}, while for VISTA J, H and K$_{\mathrm{s}}$, we refer to \cite{2013Msngr.154...35M}.
We computed the initial photometric calibration of the B, V and R$_{\mathrm{c}}$ bands using the fluxes reported in the \cite{2005yCat.1297....0Z} catalog of bright sources (including sources in the SIMES field) as a reference. For the the u and I bands, we fitted the photometric data in all the already calibrated bands (through 4.5 \mic ) with stellar templates \citep{1993KurCD..13.....K}. The reference fluxes that we use for the initial calibration of these two bands are those expected from the best fitting stellar templates, after the convolution with the filters responses.

We refined these initial heterogeneous photometric calibrations through an iterative fit of galaxy SEDs \citep{2003MNRAS.344.1000B} to a limited sub-set of high quality data. In each iteration, and for each band, we used the average difference between measured and expected fluxes (i.e. fluxes computed from the fitted SED) to correct to the calibration factors. For this process, we used only a sub-sample of $\sim$150 sources, with highly reliable fits ($\chi^{2}<1.0$), with known spectroscopic redshifts \citep{2011MNRAS.416.1862S}, and detected in 4 or more bands (excluding bands with $\lambda\geq \lambda_{\mathrm{J}}$). The SED fitting procedure is performed using the \emph{hyperz-mass} software \citep{2000A&A...363..476B}, a modified version of \emph{hyperz} used to compute the best fitting SED of sources of known redshift. We stopped the iterative process when both the average $\chi^{2}$ and the calibration corrections where stable from one iteration to the following.

\section{Photometric Redshifts}
\label{OPTICAL_PHOTO_Z_sect}
For the optically covered sources (green square in Figure~\ref{IRAC1_cov}) we computed photometric redshifts using all the data available from the u band to 4.5 $\mu m$ (13 bands). The final redshift is the combination of the \emph{hyperz} output (described in Section~\ref{chiquadminproc_2}) with an optical prior (flux in the R$_{\mathrm{c}}$, i and z bands, see Section~\ref{Z_from_OPT_flx_sec}) that we considered only for 24 \mic\ detected sources. The use of the prior does not sensibly increase the average precision of the \emph{hyperz} output but, when tested on our high-quality spectroscopic sample \citep{2011MNRAS.416.1862S}, after removing 10 probable quasars showing strong MgII emission (2798 \AA), it allows us to identify all the catastrophic \emph{outliers} (i.e. $|\Delta z /(1+z_{\mathrm{spec}})|>$0.5) and to supply a better redshift estimate in the majority of these cases. We tested the same method on a wider spectroscopic sample of COSMOS data, using similar bands \cite[][Lilly et al. 2017, in preparation]{2007ApJS..172...70L}. The final precision (dispersion in the $\Delta z /(1+z_{\mathrm{spec}})$ distribution) resulted $\sigma_{\mathrm{phot}}$=0.040$\pm$0.002 for the SIMES data ($0.01<$z$_{\mathrm{spec}}<0.9$, See Figure~\ref{precisions_SEP}) and 0.048$\pm$0.002 for the deeper COSMOS data (0.01$<$z$_{\mathrm{spec}}<$2.0).

\subsection{Redshift from SED fitting}
\label{chiquadminproc_2}

Using the \emph{hyperz} software \citep{2000A&A...363..476B}, we fitted a set of BC03 templates to the total fluxes measured in the u, B, V, g, R$_{c}$, I, i, z, J, H, K$_{s}$, 3.6 \mic\ and 4.5 \mic\ bands. In the BC03 models, an exponentially declining SFR is assumed, with $\mathrm{SFR}\propto\exp(-t/\tau)$, where $\tau=$0.3, 1, 2, 3, 5, 10, 15, and 30 Gyrs. We consider a solar metallicity Z=Z$_{\odot}$ and an extinction law in the \cite{2000ApJ...533..682C} form, with A$_{V}$ ranging from 0.0 to 4.0. The precision of this technique, when used alone, is tested using the already mentioned SIMES and COSMOS spectroscopic samples. In the COSMOS field, in particular, we could perform an equivalent test using a similar set of photometric bands. We obtained a redshift precision $\sigma_{\mathrm{Fit}}$=0.041$\pm$0.003 in the SIMES field ($0.01<$z$_{\mathrm{spec}}<0.9$, see Figure~\ref{precisions_SEP}) ) and 0.046$\pm$0.002 in the COSMOS field (0.01$<$z$_{\mathrm{spec}}<$2.0). We notice that these uncertainties are computed from the Gaussian fit of the $\Delta z$ distribution, excluding the catastrophic \emph{outliers} (i.e. $|\Delta z /(1+z_{\mathrm{spec}})|>$0.5). We identified 6 (4.2\%) and 23 (6.7\%) \emph{outliers} in the SIMES and in the COSMOS spectroscopic samples respectively.

\subsection{Redshift from optical flux: a complementary technique}
\label{Z_from_OPT_flx_sec}

In order to improve the photometric redshift precision, we integrated the results of the SED fitting with the information given by the flux observed in three optical bands. The complementary technique described in this section, when combined with the $\chi^2$ minimization results, allows the identification and correction of the \emph{outlier} redshifts.

The simple idea at the basis of this method is that the more distant a galaxy is, the fainter it appears. Consequently, a broad inverse relation between optical fluxes and redshift is expected \cite[see e.g.][]{1996AJ....112..839C}. Here we demonstrate how the precision of this kind of relations is higher for far-IR selected sources. To this purpose, we studied the correlation between optical fluxes and photometric redshifts in the COSMOS field, where a large amount of data is available \citep{2010ApJ...709..644I}.

\begin{figure*}[!ht]
\centering
\includegraphics[width=17cm]{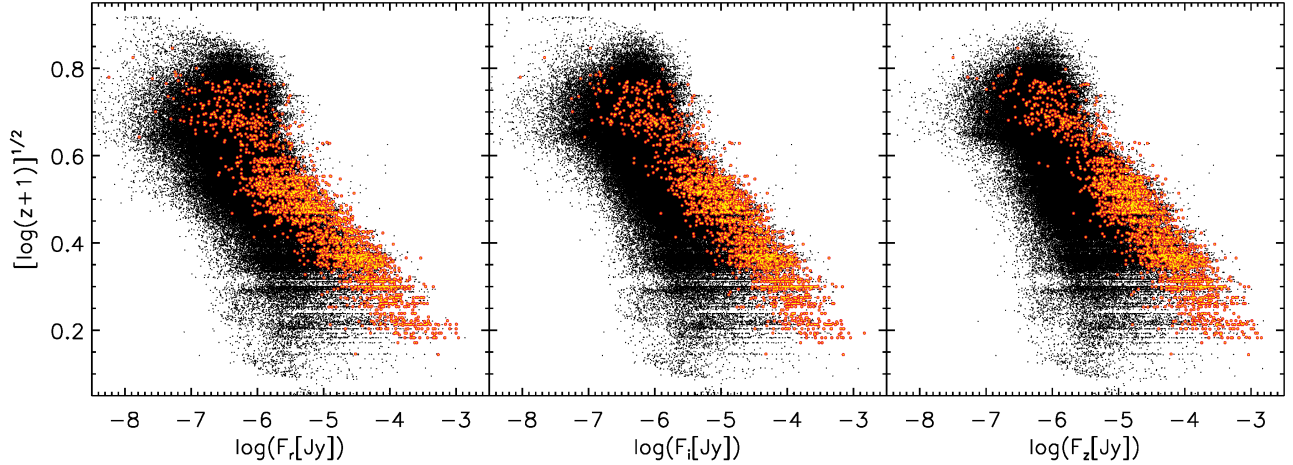}
\caption{ Observed optical flux versus redshift relation for optical fluxes in the r+, i+, z+ bands, in the COSMOS field. Sources with 24 \mic\ fluxes $>$0.3 mJy are highlighted in colour.}
\label{opt_flux_redshift_1C}
\end{figure*}

Figure~\ref{opt_flux_redshift_1C} shows the optical flux measured in three optical bands (i.e. r+, i+, z+) versus $\sqrt{\log(z+1)}$. Sources with 24 \mic\ fluxes brighter than 0.3 mJy are highlighted. The relation is evident even without a far IR selection of the sources (black dots), but if the latter is considered, the same relation became narrower \footnote{We found similar results using PACS 100 and 160 $\mu m$ selections}.

\begin{figure}[!ht]
\centering
\includegraphics[width=8.5cm]{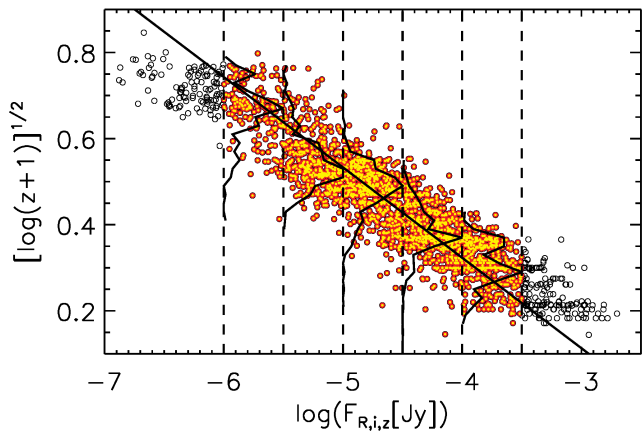}
\caption{ Combined optical flux $F_{r,i,z}$ versus redshift relation (see Equation~\ref{F_optical_EQ1}), in the COSMOS field. Only 24 \mic\ detected sources (F$_{24}>$0.3 mJy) are considered here. We excluded the brightest ($F_{r,i,z}>10^{-3.5}$Jy) and the faintest ($F_{r,i,z}<10^{-6}$Jy) sources (black circles) before computing the best linear fit. }
\label{opt_flux_redshift_1B}
\end{figure}

Starting from these observations, we computed the average ratios R$_{rz}$ and R$_{iz}$ between the fluxes measured in the z+ band, taken as reference, and those in the r+ and i+ band respectively. Then, we computed their quadratic sum as:
\begin{equation}
\label{F_optical_EQ1}
F_{r,i,z}=\sqrt{\frac{F_{z}^{2}+(R_{iz}F_{i})^{2}+(R_{rz}F_{r})^{2}}{3}}.
\end{equation}
Figure \ref{opt_flux_redshift_1B} shows the same relation between redshifts and optical fluxes already presented in Figure~\ref{opt_flux_redshift_1C}, when the three optical bands are combined together in the $F_{r,i,z}$ measure. The best linear fit to these far--IR selected data, for -6$<\log($F$_{r,i,z})<$-3.5, is:
\begin{equation}
\label{OPT_FLUX_VS_Z_EQ1}
\left[\log(z_{\mathrm{Flux}}+1)\right]^{1/2}=-0.51677-0.20997 \log(F_{r,i,z}).
\end{equation}
In the same flux range, the average dispersion measured in five bins of flux is $<\sigma_{Vz}>$=0.04, with a maximum dispersion of 0.07 in the lowest flux bin.

After calibrating the flux-redshift relation using COSMOS photometric data, as explained above, we tested this method on a sub sample of both COSMOS and SIMES spectroscopic data. For the SIMES field, we obtained F$_{r,i,z}$ from WFI-R$_{c}$, VST-z, and i band fluxes. Given the coefficients of equation~\ref{OPT_FLUX_VS_Z_EQ1}, the maximum acceptable $F_{r,i,z}$ is 3.46 mJy, since above this value the resulting redshifts would be negative. In these rare cases, we set the redshift of the sources to z$_{\mathrm{Flux}}$=0.01. 

If considered alone (i.e. not as a prior for the \emph{hyperz} output), this flux based method gives a redshift precision $\sigma_{\mathrm{Flux}}$=0.093$\pm$0.007 and 0.061$\pm$0.004 in the COSMOS and in the SIMES field respectively. These precisions, as expected, are not higher than those resulting from the SED fitting technique. However, as can be observed in Figure~\ref{Z_all_plots}, this method (z$_{\mathrm{Flux}}$) is not sensitive to the catastrophic \emph{outliers} problem affecting the SED fitting technique (z$_{\mathrm{Fit}}$). We notice that the different $\Delta z$ measured in COSMOS and in SIMES does not depend on the different redshift distribution of the two spectroscopic samples (in the SIMES field there are not spectroscopic redshifts above 1.0) and it does not substantially change considering only sources at z$_{\mathrm{spec}}<$1 in both fields.

\subsection{Combined Photometric Redshifts}
\label{Cbrcm_sect}

We combined the redshifts resulting from the SED fitting technique (Section~\ref{chiquadminproc_2}) and from the observed optical flux (Section~\ref{Z_from_OPT_flx_sec}) using a weighted mean: 
\begin{equation}
\label{WMZ1}
z_{\mathrm{phot}}=W_{\mathrm{Fit}}z_{\mathrm{Fit}}+W_{\mathrm{Flux}}z_{\mathrm{Flux}},
\end{equation} 
where the weights
are given by:
\begin{eqnarray}
\label{WMZ2}
W_{i}=\frac{1/\sigma_{i}^{2}}{1/\sigma_{Fit}^{2}+1/\sigma_{Flux}^{2}} \ \ \ \ \ \ \ \mathrm{if}\ F_{24\mu m}>0.3\mathrm{mJy} \nonumber \\
W_{\mathrm{Fit}}=1,\ \ \ \ \ W_{\mathrm{Flux}}=0 \ \ \ \ \ \ \ \mathrm{if}\ F_{24\mu m}<0.3\mathrm{mJy} 
\end{eqnarray} 

\begin{figure*}[!ht]
\centering
\includegraphics[width=15.4cm]{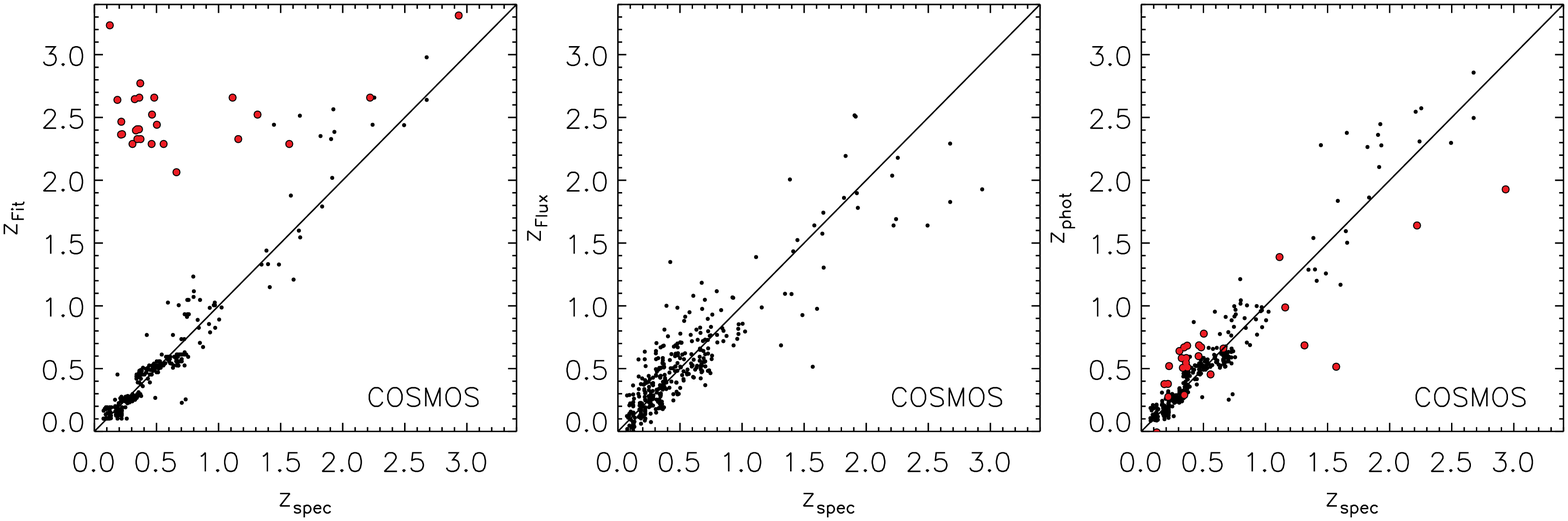}
\includegraphics[width=15.6cm]{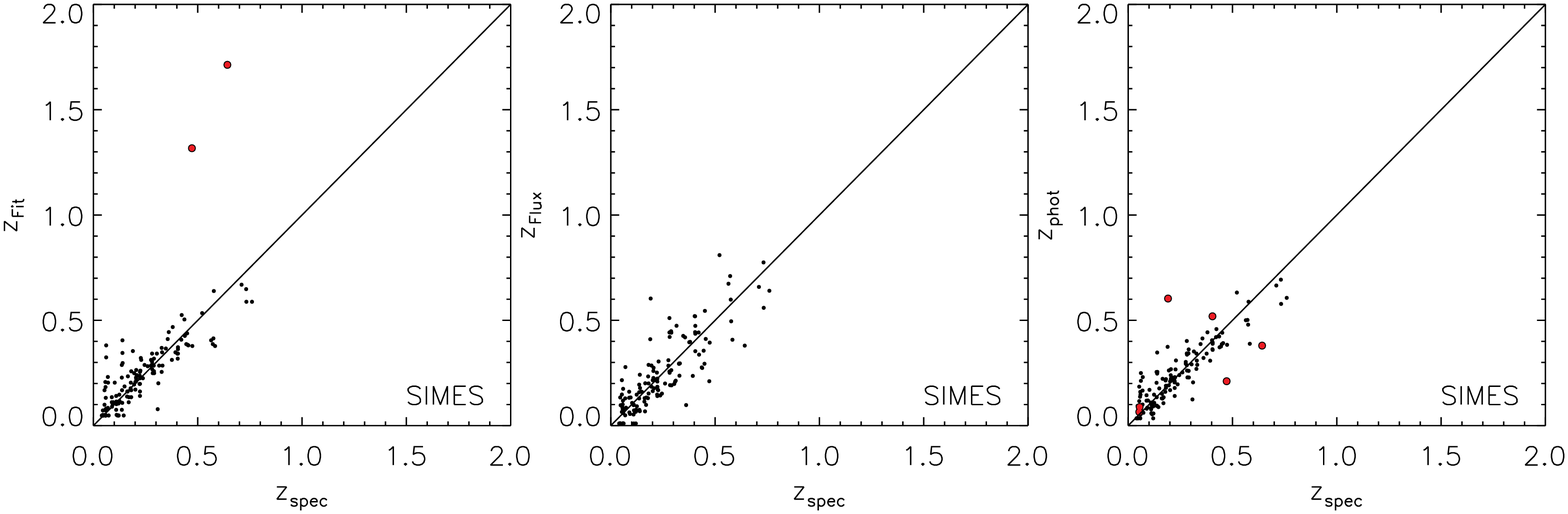}
\caption{Spectroscopic and photometric redshifts in the COSMOS field ({\bf upper panels}) and in the SIMES field ({\bf bottom panels}). The comparison is shown between spectroscopic redshifts $z_{\mathrm{spec}}$ and hyperz outputs $z_{\mathrm{Fit}}$ ({\bf left panels}), redshift obtained from optical fluxes $z_{\mathrm{Flux}}$ ({\bf central panels}), and the combined technique $z_{\mathrm{Phot}}$ ({\bf right panels}). The \emph{outliers} affecting the SED fit technique and identified through our method ($|z_{\mathrm{Fit}}-z_{\mathrm{Flux}}|>1.0$) are represented with bigger red dots before (left panels) and after the correction (right panel).}
\label{Z_all_plots}
\end{figure*}

We fixed a threshold $|z_{\mathrm{Fit}}-z_{\mathrm{Flux}}|=1.0$ above which $z_{\mathrm{phot}}=z_{\mathrm{Flux}}$. 
Using this threshold, we are able to identify a more precise solution (i.e the same precision assigned to z$_{\mathrm{Flux}}$) to the wrong \emph{outlier} solutions resulting from the SED fit strategy (see left and right panels of Figure~\ref{Z_all_plots}).

\begin{figure}[!ht]
\centering
\includegraphics[width=8.0cm]{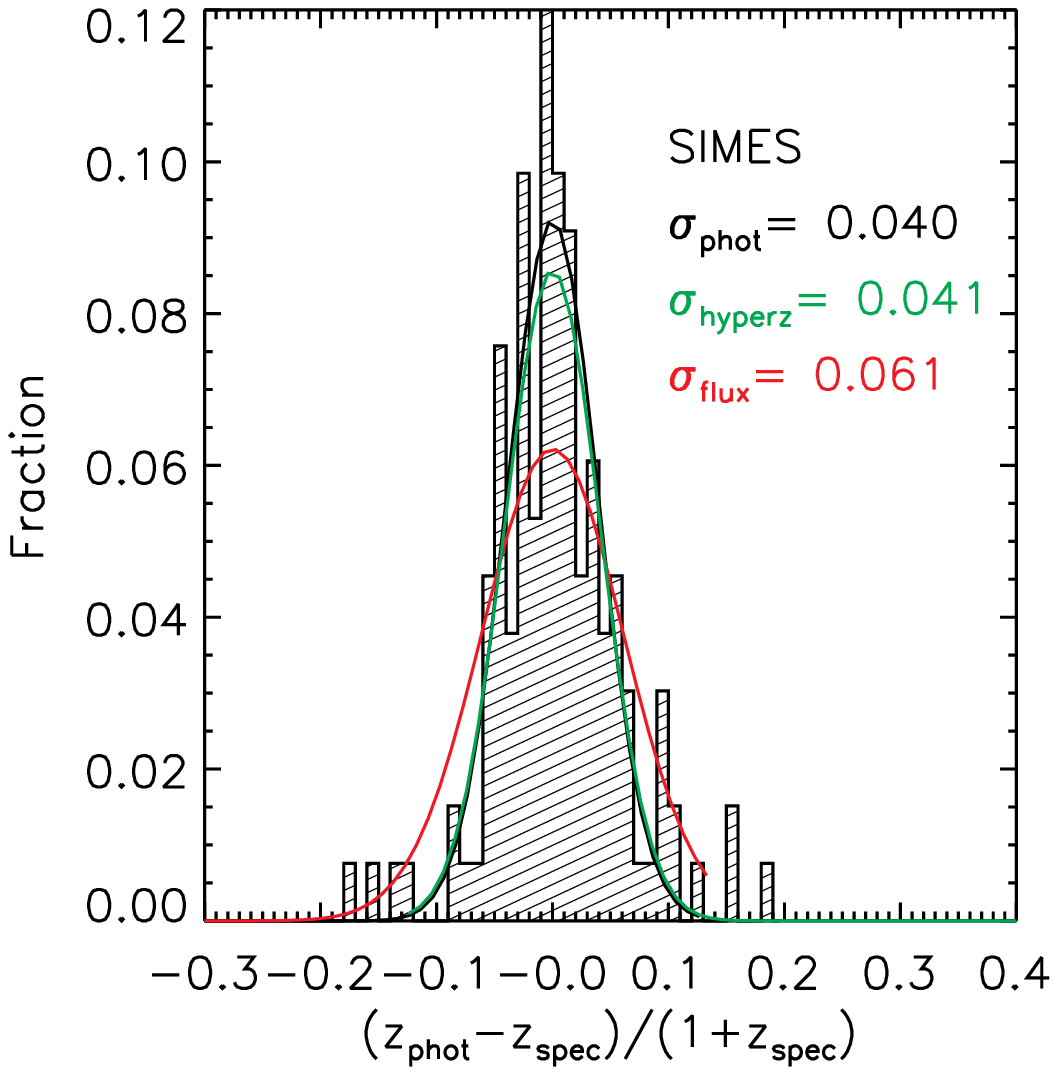}
\caption{Redshift precision of the SIMES data. The histogram and its best fitting Gaussian distribution (black line) refers to the results of the \emph{hyperz}+prior technique that we used. The precisions of the \emph{hyperz} output (green) and of the optical prior (\emph{``flux''}, red curve) considered separately are also reported for a comparison. }
\label{precisions_SEP}
\end{figure}

Excluding the \emph{outliers}, the precision of the \emph{hyperz} output does not sensibly change when the optical prior is introduced (see Figure~\ref{precisions_SEP}). However, all the 6 and 23 \emph{outliers} identified in the SIMES and COSMOS high-quality spectroscopic samples, (corresponding to the 4 and 7\% of the total) can be corrected when using the combined technique described in this section (see Figure~\ref{Z_all_plots}).

\begin{figure}[!ht]
\centering
\includegraphics[width=8.0cm]{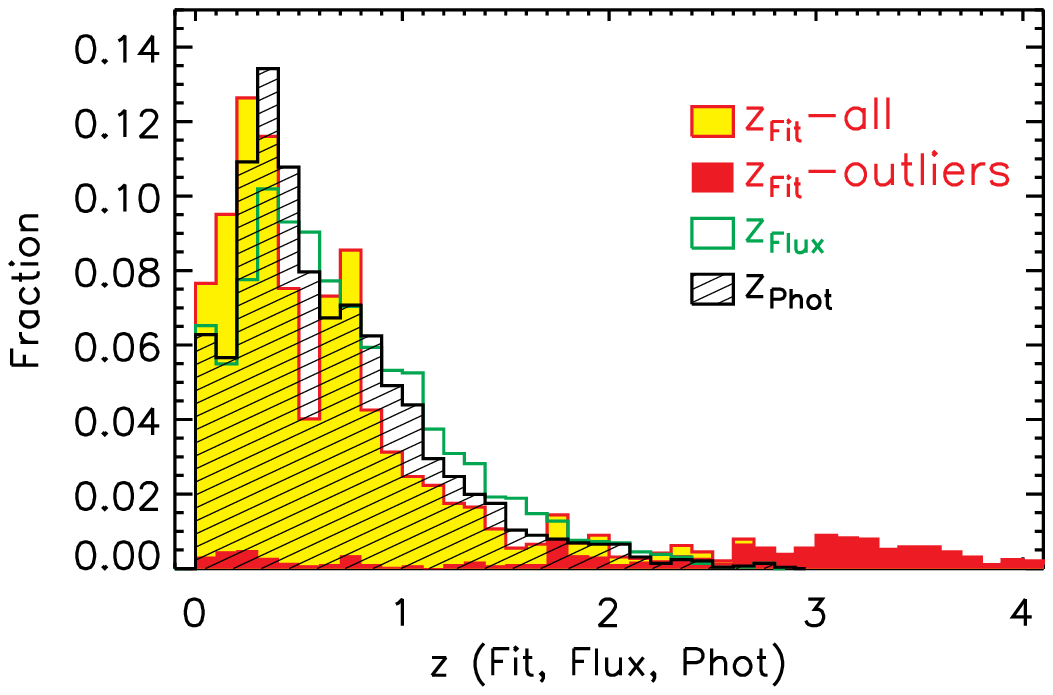}
\caption{Redshift distribution in the SIMES field for 24 \mic\ detected sources (F$_{24}>0.3$ mJy) resulting from SED fitting (red and yellow), optical flux analysis (green) and combined technique (black filled). The red filled distribution represents the identified \emph{outliers} before the correction.}
\label{z_distrib_allSEP}
\end{figure}

In Figure~\ref{z_distrib_allSEP} we show the redshift distribution ($z_{\mathrm{Fit}}$, $z_{\mathrm{Flux}}$, and $z_{\mathrm{Phot}}$) for all the sources detected at 24 \mic\ (F$_{24}>0.3$ mJy) in the SIMES field. Most of the \emph{outliers} identified using the  $|z_{\mathrm{Fit}}-z_{\mathrm{Flux}}|=1.0$ threshold are distributed above z$_{\mathrm{Fit}}\sim 1.7$, but this tail disappear in the $z_{\mathrm{Phot}}$ distribution, after applying our correction technique.

\subsection{Comparison with different softwares}
We compared the technique described in Section~\ref{Cbrcm_sect}, with a different algorithm, natively incorporating an optical prior. To this purpose, we computed the photometric redshifts for the SIMES spectroscopic sample (highest quality only) using the \emph{EAZY} software \citep{2008ApJ...686.1503B}. The SED fitting approach of \emph{EAZY} is similar to that of \emph{hyperz}, and it is based on the $\chi^{2}$ minimization. Differently from  \emph{hyperz} however, \emph{EAZY} already incorporates the possibility to include a redshift prior based on the optical magnitude of the sources.

In our first experiment, we considered the \emph{EAZY} prior based on the R magnitude of the sources.
Therefore, we limited the test sample to sources detected in the R$_{\mathrm{c}}$ band (This is not necessary using our method, where R, i or z can also be used separately).
We obtained a number of \emph{outliers} (4 sources, corresponding to 3.6\% of the sample) similar to what we find using \emph{hyperz} without our prior. All these four \emph{outliers} are identified and corrected when using our method. Excluding these sources, the z precision is consistent between \emph{EAZY} + internal prior and \emph{hyperz} without prior ($\sigma_{\mathrm{EAZY}}$=0.034$\pm$0.002, $\sigma_{\mathrm{hyperz}}$=0.038$\pm$0.003).

In a second experiment, we considered the \emph{EAZY} prior based on the K magnitude, comparing the results on a sub-set of K-detected sources. Again, the precision of the two methods is similar ($\sigma_{\mathrm{EAZY}}$=0.032$\pm$0.003 and $\sigma_{\mathrm{hyperz}}$=0.036$\pm$0.003), but \emph{EAZY} is not able to identify 5 \emph{outliers} that we are able to identify and correct using our method.

\section{Galaxy groups between z$\simeq$0.15 and 0.3}
\label{SEC:groups}
In the central optically covered square degree of the SIMES field (green square in Figure~\ref{IRAC1_cov}), $\sim$66\% of the sources in the 3.6 \mic\ based catalog of BA16 are detected in at least two bands among the U, B, V, R$_{c}$, I, g, i and z.
We used this sample of $\sim$38000 sources with a spectroscopic or photometric estimate of redshift to identify galaxy groups between z$\simeq$0.15 and 0.3.
At these redshifts, we could use a spectroscopic measure for $\sim$2\% of the sample and a photometric estimate, computed as described in Section~\ref{Cbrcm_sect}, for an additional 9\% of the sources with a far-IR measure of flux. For the remaining majority of the sources, for which a far-IR measure of flux is not available, the method described in Section~\ref{Cbrcm_sect} can not be used. In these cases, we used the photometric redshift measure obtained using the \emph{hyperz} software alone, as described in Section~\ref{chiquadminproc_2}.

Having no data available at X-rays wavelengths, where clusters are often identified \citep[e.g.][]{2004A&A...425..367B,2011ApJ...742..125G,2011A&A...534A.109P}, we find galaxy groups using an optically based friends-of-friends algorithm \citep{1982ApJ...257..423H}. This method is a simplified version of the approach described in \cite{2004MNRAS.348..866E}. We modeled the parameters of our group identification algorithm to find a compromise between the completeness of the groups selection and the precision with which the sources in these groups are identified.

In our algorithm, for each galaxy we search for possible surrounding group companions (``\emph{friends}''), inside a comoving searching radius $\ell_{\bot}=0.21(1+z)$ Mpc.
At high fluxes the SIMES sources number density is higher than that in the COSMOS catalog. To re-create the same SIMES 3.6 \mic\ flux distribution, we do not introduce simulated sources in the COSMOS data-set. Instead, when a source brighter than $F_{3.6\mu m}\simeq 53\ \mu$Jy is considered for a possible group membership, the linking radius connecting the source with another possible \emph{``friend''} is allowed to be larger. In particular, since the average projected distance between sources depends on their numerical density as $d\propto n^{-1/2}$, the linking radius used for the test in the COSMOS field is set to:
\begin{equation} 
\footnotesize
\ell_{\bot}(F_{3.6\mu m})=0.21(1+z)\min\left[\sqrt{\frac{n_{\mathrm{SIMES}}(F_{3.6\mu m})}{n_{\mathrm{COSMOS}}(F_{3.6\mu m})}},2 \right] \mathrm{Mpc}
\end{equation} 
As in \cite{2004MNRAS.348..866E}, we consider a cylindrical linking volume: for each source, the possible group companions are searched inside radial interval $\Delta z=1.5\sigma_{z}$, where $\sigma_{z}$ is the photometric uncertainty of the source considered.

We tested our algorithm using COSMOS data \citep{2010ApJ...709..644I}, comparing our results with the COSMOS galaxy and X-ray group membership catalog described in \cite{2011ApJ...742..125G}. In order to test our method on a data-set as similar as possible to the SIMES data, from the COSMOS catalog, in each 3.6 \mic\ flux bin, we eliminated randomly distributed sources, to reproduce the 3.6 \mic\ flux distribution of the shallower SIMES catalog. We also added a noise component to the COSMOS photometric redshifts, to simulate the same overall uncertainty measured in SIMES. The same test is run multiple times (100), creating each time a different dataset, with statistically similar characteristics (however, the spatial distribution of the clusters and the cluster members identified in the group membership catalog do remain the original).

In the X-ray cluster catalog of \cite{2011ApJ...742..125G}, only groups with more than 10 members are considered. Setting the same threshold in our test, 
we find an average of 450$\pm$5 member galaxies in 19.8$\pm$0.3 different groups in the redshift range $0.15<z<0.3$.
Seventy two $\pm$1\% of the groups found in the different runs are identified as such in the \cite{2011ApJ...742..125G} groups catalog and 62.1$\pm$0.4\% of the group members are reported in the reference catalog. 
The algorithm identifies 51.5\%$\pm$0.5\% of the groups and 51.2$\pm$0.6\% of the group members listed in the COSMOS X-ray.

\begin{figure}[ht]
 \centering
 \includegraphics[width=8.0cm]{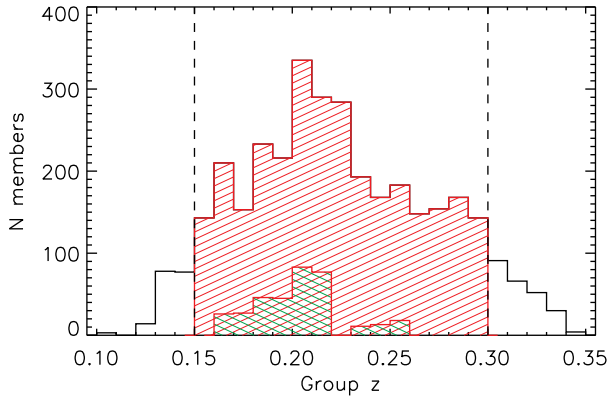}
 \caption{\label{IMG:group_z} Redshift distribution of the groups found in the SIMES field. Groups with more than 10 member galaxies are highlighted in green. In the successive analysis, only groups with an average redshift $0.15<z<0.3$ are considered (red shaded area).}
 \end{figure}

\begin{figure*}
 \centering
 \includegraphics[width=7.7cm]{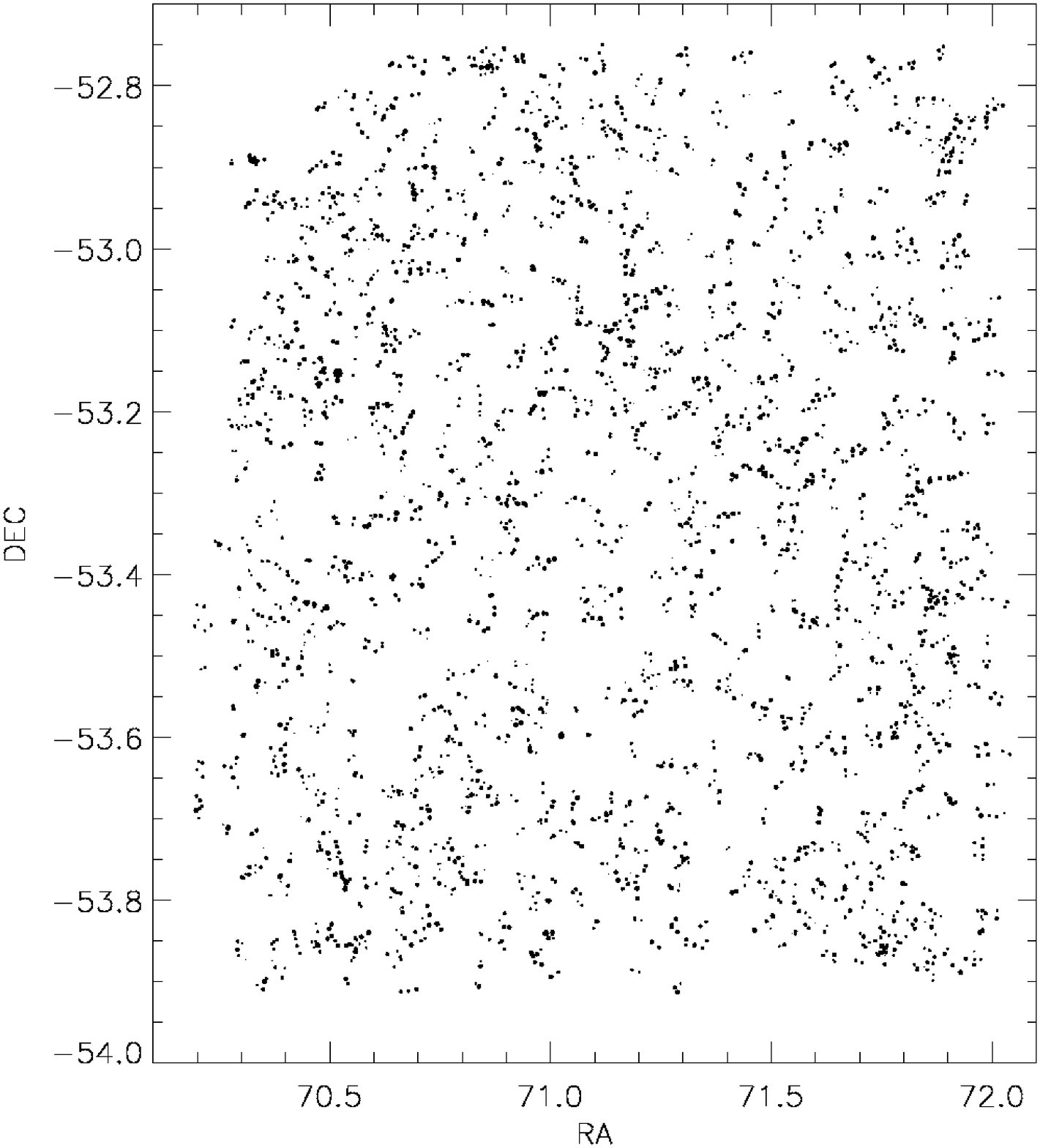}
 \includegraphics[width=8.5cm]{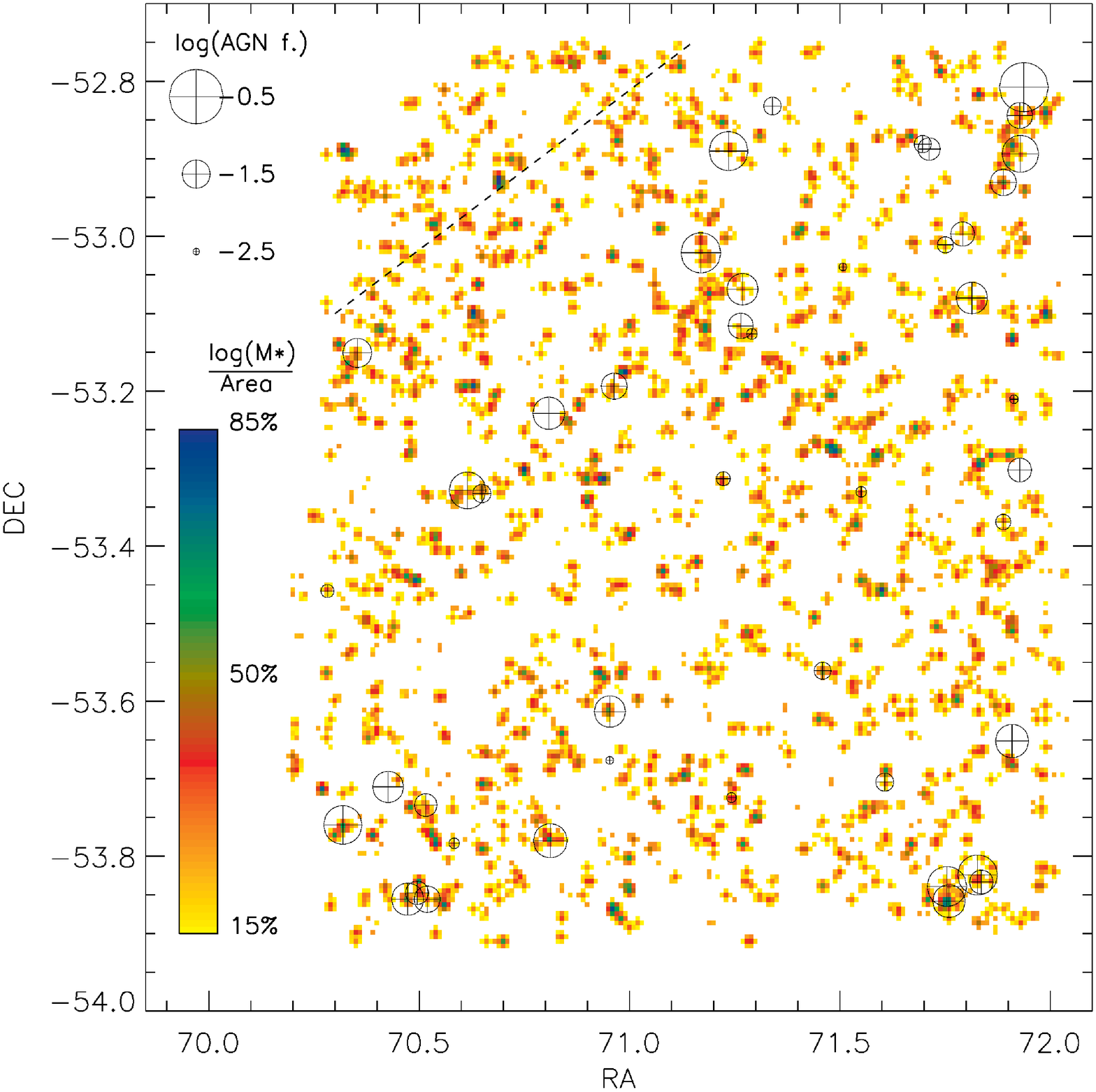}
 \caption{\label{IMG:group_distr} {\bf Left panel:} spatial distribution of the groups identified in the SIMES field between z=0.15 and z=0.3. Each point represents an actual galaxy member of a group, with bigger dots corresponding to higher stellar masses M*. {\bf Right panel:} stellar mass density of the groups in the same redshift bin. The color scale represents different fractions of the maximum value measured in the field. The AGNs identified are represented with circles proportional to the logarithmic bolometric AGN fraction. The diagonal dashed line represent the limit of the SPIRE coverage (no AGN analysis is performed above).  }
 \end{figure*}

Given an effective areal ratio $A_{\mathrm{COSMOS}}/A_{\mathrm{SIMES}}$=1.14, using the same algorithm, 395 galaxies are expected to be members of 17.4 different groups with more than 10 sources in the SIMES area.
In the SIMES field we found 22 groups with more than 10 members, for a total of 346 sources. The Poissonian uncertainty on the number of groups identified in SIMES can justify the discrepancy with the expectation from the tests run in COSMOS. Instead, the different number of group members is the consequence of the presence of two particularly rich groups in the COSMOS area, both with more than $\sim$50 sources and identified by the algorithm in each run, whereas no groups with so many members are present in the SIMES area in the same redshift interval. The presence of these two clusters justifies the difference in the number of total members identified.
Figure~\ref{IMG:group_z} shows the redshift distribution of the groups found. Groups with a minimum of 2 members are reported and the richest groups with more than 10 components are highlighted. Figure~\ref{IMG:group_distr} shows the (Ra, Dec) distribution of the same groups.

We found a total of 768 groups with at least two members. For each group, we estimate the total stellar mass M$^{*}_{\mathrm{group}}$ as the sum of the stellar masses of each single member identified. For each member galaxy, M* is computed using the \emph{hyperzmass} software \citep{2000A&A...363..476B}. For the far-IR identified sources, we could estimate a more precise value of M* using the multi-component SED-fitting approach described in Section~\ref{stellar_Mass_sect}. When possible, we used the more precise M* estimates. However, given the extinction dependent M* difference between the two methods (see description in the same section for details), all the  \emph{hyperzmass} outputs are corrected as in Equation~\ref{EQ:Mcorr}. Similarly to what already done for the computation of the photometric redshifts with \emph{hyperz} (see Section~\ref{chiquadminproc_2}), we exploited all the bands between CTIO-u and IRAC-4.5 \mic. We considered a \cite{2000ApJ...533..682C} extinction law with A$_{V}$ ranging from 0.0 to 3.0 (steps of 0.1). For the fit, we used a set of BC03 SSP models assuming SFR$\propto -t/\tau$, with $\tau=$0.3, 1, 2, 3, 5, 10, 15, and 30 Gyrs. 

\section{Physical properties of far-IR selected galaxies}
\label{SEC:GAL_PHYS_PROP}

\subsection{Sample selection}
\label{SEC:SELECTION_DESCRIPTION}
In this work, we want to study the relations between BH and SF activity and relate them to the environmental properties of the host galaxies.
To this purpose, we analyse the mid- and far-IR emission of AGNs and star forming galaxies detected at the same wavelengths.
Starting from the BA16 source catalog, based on 3.6 \mic\ detections, we selected a sample of sources detected at 24 \mic\ and, at least, in one among the 250, 350 or 500 \mic\ SPIRE bands. 

\begin{figure}
 \centering
 \includegraphics[width=7.5cm]{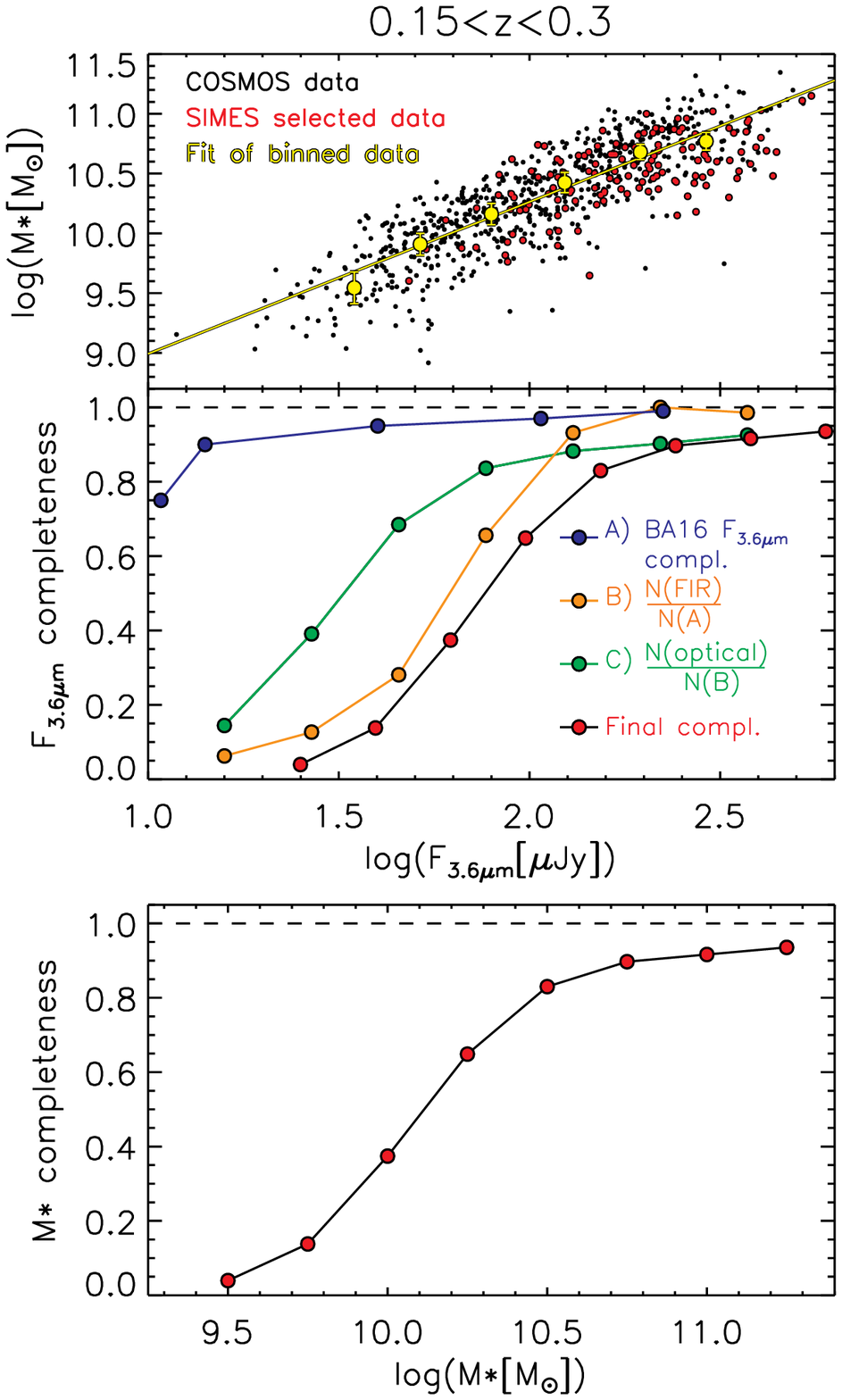}
 \caption{\label{FIG:M_compl} {\bf Upper panel:} relation between F$_{3.6\mu m}$ and stellar mass M* in the redshift interval 0.15$<$z$<$0.3, as obtained using both SIMES data, from BA16, and deeper COSMOS data from \cite{2011ApJ...739L..40R}. All the mass values are corrected as in Equation~\ref{EQ:Mcorr}. {\bf Middle panel:} the final F$_{3.6\mu m}$ completeness function of our analysis sample (black curve, red dots) is the result of the combined effects of the different selections that we applied. From the initial catalog, for which the original F$_{3.6\mu m}$ completeness curve is computed in BA16 (here represented in blue), only far IR detected sources (MIPS and SPIRE) are selected. This selection modifies the completeness of the original BA16 sample as illustrated using the yellow line. Additionally, only optically identified sources are considered in the analysis. The effects of this selection are represented with a green line. {\bf Lower panel:} mass completeness as a function of Stellar mass M*. This curve is obtained using the F$_{3.6\mu m}$ completeness function and the linear relation found between $\log$(F$_{3.6\mu m}$) and $\log$(M*).}
\end{figure}

As for the study of the environmental properties at 0.15$<$z$<$0.3, that we describe in Section~\ref{SEC:groups}, here we explore the optically covered area ($\sim$1 square degree represented with a green square in Figure~\ref{IRAC1_cov}). The optical coverage is a fundamental requirement to measure accurate photometric redshifts and stellar masses. For this reason, besides the 3.6 \mic\ and the far-IR selection, a detection is required in at least two optical bands among U, B, V, R$_{c}$, I, g, i and z. Above F$_{3.6\mu m}=100\ \mu$Jy, roughly corresponding to the M* limit that we adopt in the redshift range  0.15$<$z$<$0.3 (see Figure~\ref{FIG:M_compl}), ~87\% have 4 or more detections in these bands. 883 over a total of 1262 far-IR detected sources (70\%) do respond to this requirement ($\sim$90\% above F$_{3.6\mu m}=100\ \mu$Jy). 
The completeness curve of the original BA16 sample is modified by the far-IR and optical selections as described in Figure~\ref{FIG:M_compl}.

\subsection{SED fitting with {\sc sed3fit}}
\label{SED_FIT_SECT}

The optical-to-FIR SEDs of the galaxies in the selected sample have been analysed using the three-component SED fitting code {\sc sed3fit} by \citet[][BE13]{2013A&A...551A.100B}. 
Both photometric (see Section~\ref{OPTICAL_PHOTO_Z_sect}) and spectroscopic redshifts \citep{2011MNRAS.416.1862S}, when available (112 sources below z=1.5), where used. For the SED fitting, we used a maximum of 25 photometric bands between the far-UV and the far-IR: GALEX-FUV and NUV, CTIO-u, B, V, I, WFI-Rc, VST-g, i, z, VISTA-J, H, Ks, IRAC-3.6 $\mu$m, 4.5 $\mu$m, AKARI 7 $\mu$m, 11 $\mu$m, 15 $\mu$m, 24 $\mu$m, WISE-11 $\mu$m, MIPS-24 $\mu$m, 70 $\mu$m, SPIRE-250 $\mu$m, 350 $\mu$m, 500 $\mu$m. In the mid-IR, the 11 \mic\ band is covered by WISE and AKARI at the same time, while in the optical, we use both the CTIO-I and the VST-i observations. This reduces the number of not redundant photometric bands to 23.

Originally inspired by \citet[][DC08]{2008MNRAS.388.1595D}, {\sc sed3fit} computes the best fitting SED as the combination of stellar emission, dust heated by stellar population and an AGN (torus+disc). Stellar and dust emission are linked by the energy balance between dust absorption in the optical spectral range and the re-emission in the far-IR. 

The adopted libraries of models are: stellar emission by BC03, using a \cite{2003PASP..115..763C} IMF and adopting a \cite{2000ApJ...539..718C} extinction curve; dust emission by DC08; and torus/AGN models by \cite{2006MNRAS.366..767F,2016agnw.confE.101F} in their latest implementation by \cite{2012MNRAS.426..120F}. 

The code produces the probability distribution function (PDF) for each single model parameter and relevant derived quantity (e.g. stellar mass, SFR, dust mass, L(IR)). 
These PDFs are used to asses the degeneracy and the uncertainty of each quantity.

Each torus model in the \cite{2012MNRAS.426..120F} library is identified by six different parameters: 1) ratio between the physical inner and the outer radii (R$_{\mathrm{out}}$/R$_{\mathrm{in}}$); 2) opening angle, 3) 9.7 \mic\ optical depth in the equatorial plane ($\tau_{9.7}$); 4) radial slope of the density profile ($\beta$); 5) height slope of the density profile ($\gamma$); 6) inclination along the line of sight ($\theta$). 

Following \cite{2014MNRAS.439.2736D}, we limited the library to a subset of models covering a restricted range in parameters space (see Table~\ref{tbl:SED_FIT_param}).
The 9.7 \mic\ optical depth is limited to $\tau <$10 \citep{1992ApJ...401...99P}, and R$_{\mathrm{out}}$/R$_{\mathrm{in}}$ ratios restricted to values $\leq$100, since no evidence is found for the existence of very extended torus geometries \citep{2002ESASP.515...15W,2007A&A...474..837T,2009A&A...502...67T}. As demonstrated in \cite{2014MNRAS.439.2736D} the results obtained using the complete and the reduced libraries are consistent \citep[see also][]{2008MNRAS.386.1252H,2009MNRAS.399.1206H,2012MNRAS.423.1909P}.

\begin{deluxetable*}{lll}
\tabletypesize{\footnotesize}

\tablecolumns{3}
\tablewidth{0pc}
\tablecaption{SED Fit main input Parameters}
\tablehead{\colhead{Parameter} & \colhead{Values} & \colhead{Description}}
\startdata
\label{tbl:SED_FIT_param}
Stellar Models & BC03                       &  \\
IMF            & \cite{2003PASP..115..763C} &  \\
Extinction Law & \cite{2000ApJ...539..718C} &  \\
Dust emission  & 3 components               & PAH + hot and cold components (Star formation) \\
$R_{\mathrm{out}}/R_{\mathrm{in}}$ & 10$\div$100   & Ratio between inner and outer radii of the dusty torus (AGN) \\
$\Theta$    & 40$^{\circ}\div$140$^{\circ}$    & Dusty torus opening angle (AGN) \\
$\tau_{9.7}$ & 0.1$\div$6                    & Optical depth at 9.7\mic\ \\
$\beta$     & -1$\div$-0.5                  & Radial slope of density profile (AGN) \\
$\gamma$    & 0$\div$6                      & Height slope of density profile (AGN) \\
$\theta$    & 0$^{\circ}\div 90^{\circ}$        & Torus inclination (AGN)  \\
\enddata
\end{deluxetable*}

The unobscured bolometric AGN luminosity (L$_{\mathrm{acc}}$), representing the unabsorbed total luminosity emitted by the nuclear object, is intrinsically associated to the best fitting AGN model. L$_{\mathrm{acc}}$ accounts for the energy emitted by the central engine in the range between 10$^{-3}$-10$^{3}$ \mic\ and it is simply related to the 
black hole accretion rate (BHAR). The X-ray emission is considered negligible in the 8-1000 \mic\ to L$_{\mathrm{acc}}$ conversion (4\% of the total budget). This assumption relies on the large bolometric correction needed to convert X-ray emission to bolometric luminosity \citep[20-30,][]{2004ASSL..308..187R,2007A&A...468..603P,2007ApJ...654..731H,2009MNRAS.392.1124V,2012MNRAS.425..623L}.

\begin{figure*}
 \centering
 \includegraphics[width=18cm]{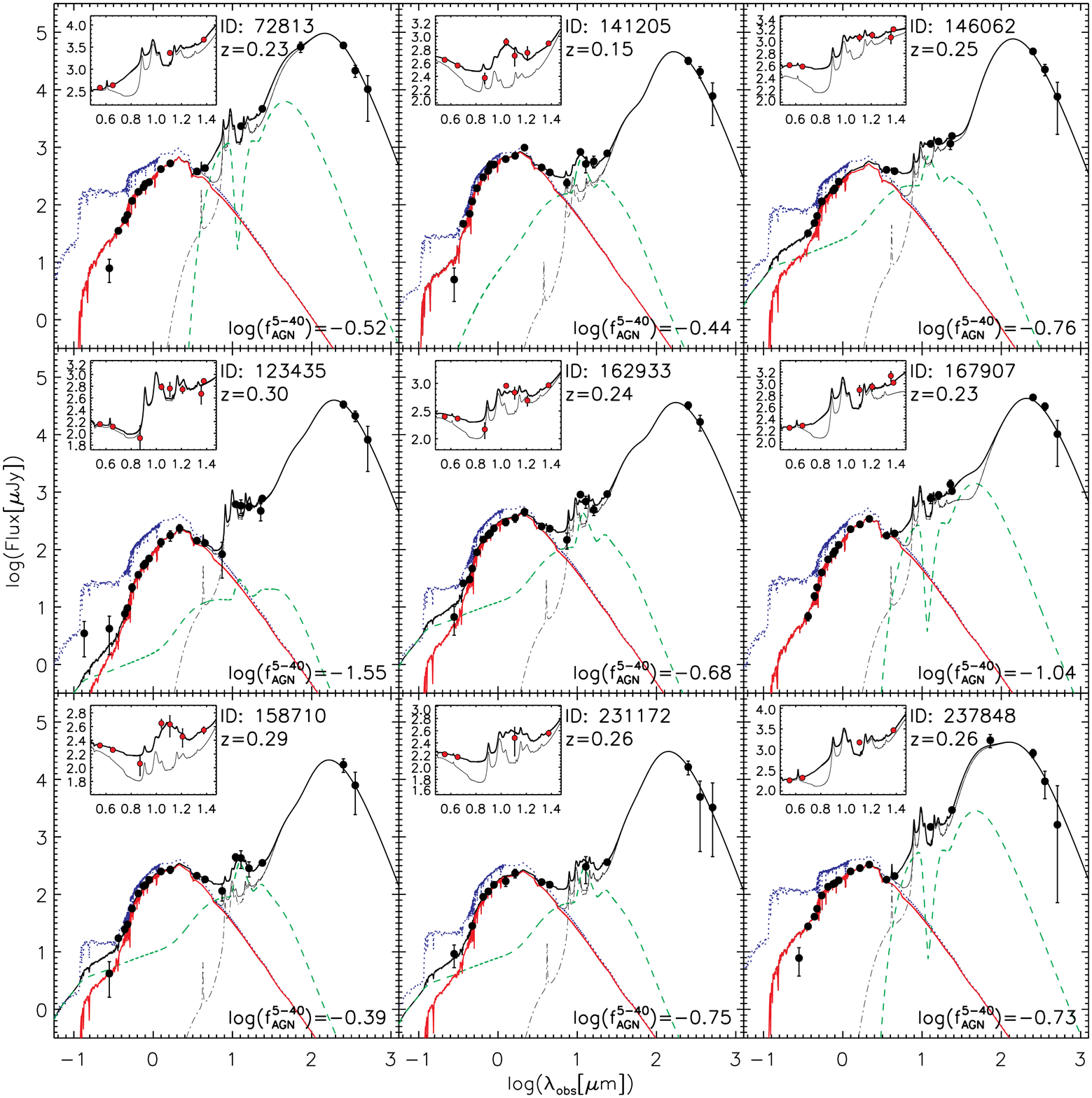}
 \caption{\label{img:SED_Examples} Examples of triple components fits obtained using the {\sc sed3fit} software \citep{2013A&A...551A.100B}. The stellar emission is represented using a solid red (obscured) and a dotted blue line (unextincted), the AGN emission with a dashed green line and the star formation region emission (SF) with a point-dashed black line. The thick solid black line is the sum of the three components (red, green and point-dashed black), while the thin black line represents the sum of stellar emission and SF only. The inserts represent the enlargement of the 3-30 \mic\ spectral region. The bottom line represents the best fitting model if the AGN effects are not considered, while in the upper thicker line, the dusty torus emission is considered. }
 \end{figure*}

After applying the SED fitting technique to the sources in our selection, we rejected ~5\% of clearly unreliable fits presenting a bad agreement between the photometric measures and the best fitting SED found.
 None of the rejected fits fall into the redshift range considered in our analysis (the majority of them are located between z=0.8 and 2.0). In Figure~\ref{img:SED_Examples}, we show 9 examples of fits obtained with the procedure described above.

\subsection{Stellar mass M*}
\label{stellar_Mass_sect}
For each of the sources, the mass in stars M* is derived from the probability distribution function (PDF) of the best fitting BC03 models. In particular, we used the median of the PDF as an estimate of M*.

As stated in Section~\ref{SED_FIT_SECT}, the SED fitting technique used accounts for a possible emission from the central super massive black hole. Depending on the best fitting AGN model, the SMBH emission can be responsible for a more or less important optical contribution. To assess the effects of the AGN on the estimated stellar masses, we compare the outputs of {\sc sed3fit} with those obtained using a SED fitting technique not involving any AGN component. For this comparative fit, we used all the available bands between u and 4.5 \mic, the hyperz--mass software \citep{2000A&A...363..476B} and BC03 templates.

\begin{figure*}
 \centering
 \includegraphics[width=18cm]{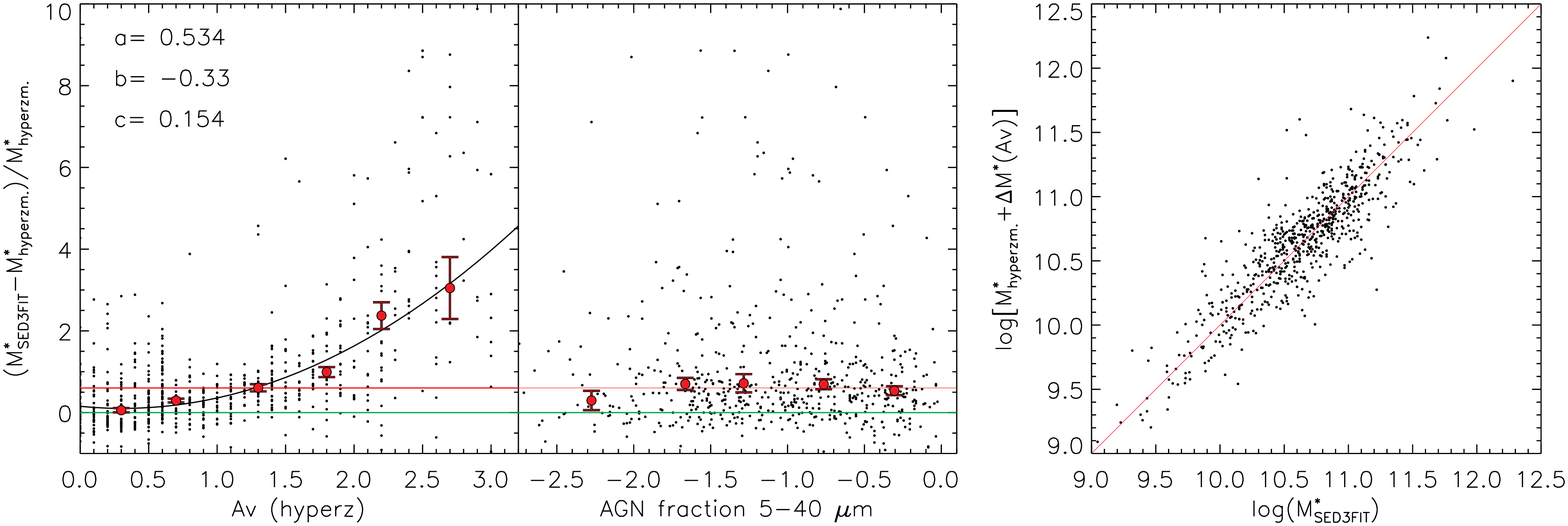}
 \caption{\label{MASS_COMPAR_1} Comparison between stellar masses (in units of M$_{\odot}$) computed considering ({\sc sed3fit}) and not considering (\emph{hyperz}) a possible AGN emission component. The green line represents the average value expected if the two methods did not produce any difference in the outputs (zero), while the red line is the average value measured. While there is no evidence for dependence on the AGN emission fraction between 5 and 40 \mic\ ({\bf central panel}), there is a clear dependence on the extinction A$_{\mathrm{V}}$ ({\bf left panel}). A similar result is found in \cite{2013ApJ...762..108L} when comparing stellar masses computed using \emph{hyperz} and \emph{MAGPHYS} \citep{2008MNRAS.388.1595D}. In the {\bf right panel}, we show the agreement between the stellar masses computed using the two methods, after correcting the \emph{hyperz-mass} results with the fit shown in the left panel. }
 \end{figure*}

The comparison between the results of the two techniques is shown in Figure~\ref{MASS_COMPAR_1}. In the same figures, the AGN fraction $f^{\lambda}_{\mathrm{AGN}}$ (i.e. fraction of emission due to the SMBH), is shown for the 5-40 \mic\ band. 

Considering all the data in our sample, we observe an average difference between the two methods:
\begin {equation}
 <\Delta M^{*}>=<\frac{M^{*}_{SED3FIT}-M^{*}_{hyperzm.}} {M^{*}_{hyperzm.}}>\sim 0.6.
\end{equation}
As can be observed in the central panel of Figure~\ref{MASS_COMPAR_1}, this difference is not likely due to the introduction of the AGN component. The difference between the outputs does not depend in a significant way on the fraction of bolometric emission attributed to the SMBH. This is expected also observing the typical AGN emission at the wavelengths where the stellar emission peaks (see Figure~\ref{img:SED_Examples}, note the the y axis is a logarithmic scale).

 Instead, as found in \cite{2013ApJ...762..108L} comparing \emph{hyperz-mass}, \emph{MAGPHYS} \citep{2008MNRAS.388.1595D} and \emph{GRASIL} \citep{1998ApJ...509..103S,2005MNRAS.364.1286V} outputs, the difference arises in particular for LIRGS and ULIRGS, when trying to compute the stellar emission hidden by the dust. This difference is more prominent for heavily obscured sources (see left panel of Figure~\ref{MASS_COMPAR_1}). Using a second degree fitting curve, we found \footnote{ For this fit, we considered equivalent uncertainties along the y axis for all the bins. This choise is made to give similar weight to differently populated bins of Av.}:
\begin {equation}
\label{EQ:Mcorr}
<\Delta M^{*}>= (0.534 A_{\mathrm{V}}^{2}-0.33 A_{\mathrm{V}} + 0.154) \ \ \ \ [M_{\odot}].
\end{equation}
Exploiting this extinction correction, we can combine stellar masses computed using \emph{MAGPHYS} or {\sc sed3fit} with \emph{hyperz-mass} results, as we do when estimating the stellar mass completeness of our sample (see Section~\ref{subsect:Mcomplet}). Given that \emph{hyperz-mass} does not consider the far-IR emission to compute the exitinction, whereas {\sc sed3fit} does, we use the results obtained with the latter as a reference.

\subsubsection{Completeness calculation}
\label{subsect:Mcomplet}
We estimated the stellar mass M* completeness in the redshift range 0.15$<$z$<$0.3 explored in our analysis.
The completeness is computed combining the 3.6 $\mu$m completeness function of the original sample, from Table 1 in \cite{2016ApJS..223....1B}, and the linear relation between log(F$_{3.6}$) and log(M*), that we measured in the same redshift range, using both SIMES and deeper COSMOS data from \cite{2011ApJ...739L..40R} (see upper panel of Figure~\ref{FIG:M_compl}). Since the \cite{2011ApJ...739L..40R} stellar masses are computed using \emph{hyperz-mass}, we corrected them according to Equation~\ref{EQ:Mcorr}, to be consistent with our results, obtained with {\sc sed3fit}.

The data used in our analysis are a sub-sample of the 3.6 \mic\ selected sample described in \cite{2016ApJS..223....1B}. This sub-sample is selected following the criteria listed in Section~\ref{SEC:SELECTION_DESCRIPTION}. In the middle panel of Figure~\ref{FIG:M_compl}, we show how this series of selections affects the completeness, as a function of the 3.6 \mic\ observed flux. First, we compute how the far-IR selection (MIPS and SPIRE) affects the completeness of the reference BA16 3.6 $\mu$m selected sample (``All''). This curve is normalized to its maximum value. Then, we compute the completeness of the final sample with respect to the 3.6 \mic\ and far-IR selected sample. The final combined 3.6 $\mu m$ completeness function is obtained multiplying these completeness curves. Using the linear relation between 3.6 \mic\ fluxes and stellar masses (upper panel of Figure~\ref{FIG:M_compl}), we obtained the M* completeness curve (lower panel of Figure~\ref{FIG:M_compl}).

\subsection{Star formation rate}
\label{SFR_sSFR_sect}
We computed the infra--red bolometric (8--1000 \mic) luminosities ($L_{\mathrm{FIR}}$) from the 50\% percentiles of the PDFs resulting from the SED fitting technique. Following \cite[KE98,][]{1998ARA&A..36..189K}, this quantity can be directly associated to the total star formation rate (SFR) of a galaxy. The underlying KE98 assumption is that, for starbursting galaxies, the contribution of stars and AGN to the total far-IR luminosity (8-1000 \mic) is negligible when compared to the far-IR luminosity originating in the dusty star forming regions. Using our SED fitting technique, we can separate the contributions to the total 8--1000 \mic\ luminosity ($L_{\mathrm{FIR}}$) due to stellar emission, AGN (if present), and star forming regions (i.e. $L_{\mathrm{FIR}}=L_{\mathrm{FIR}}^{\mathrm{stars}}+L_{\mathrm{FIR}}^{\mathrm{SF}}+L_{\mathrm{FIR}}^{\mathrm{AGN}}$). We computed the SFRs of the sources using the KE98 equation, but considering $L_{\mathrm{FIR}}^{\mathrm{SF}}$, instead of $L_{\mathrm{FIR}}$, as SFR tracer. 
The difference between $L_{\mathrm{FIR}}$ and $L_{\mathrm{FIR}}^{\mathrm{SF}}$ (and correspondingly the SFR computed) is higher than 25\% in less than 5\% of the cases (mostly extreme sources with $\log$(BHAR/SFR)$>$-1.3). 
Finally, since the original KE98 equation refers to a \cite{1955ApJ...121..161S} IMF, we applied a 0.24 dex correction factor \citep[e.g. ][]{2013A&A...557A..66B} to obtain the corresponding SFR values in a \cite{2003PASP..115..763C} IMF form. 

\begin{figure*}[ht!]
 \centering
 \includegraphics[width=15.cm]{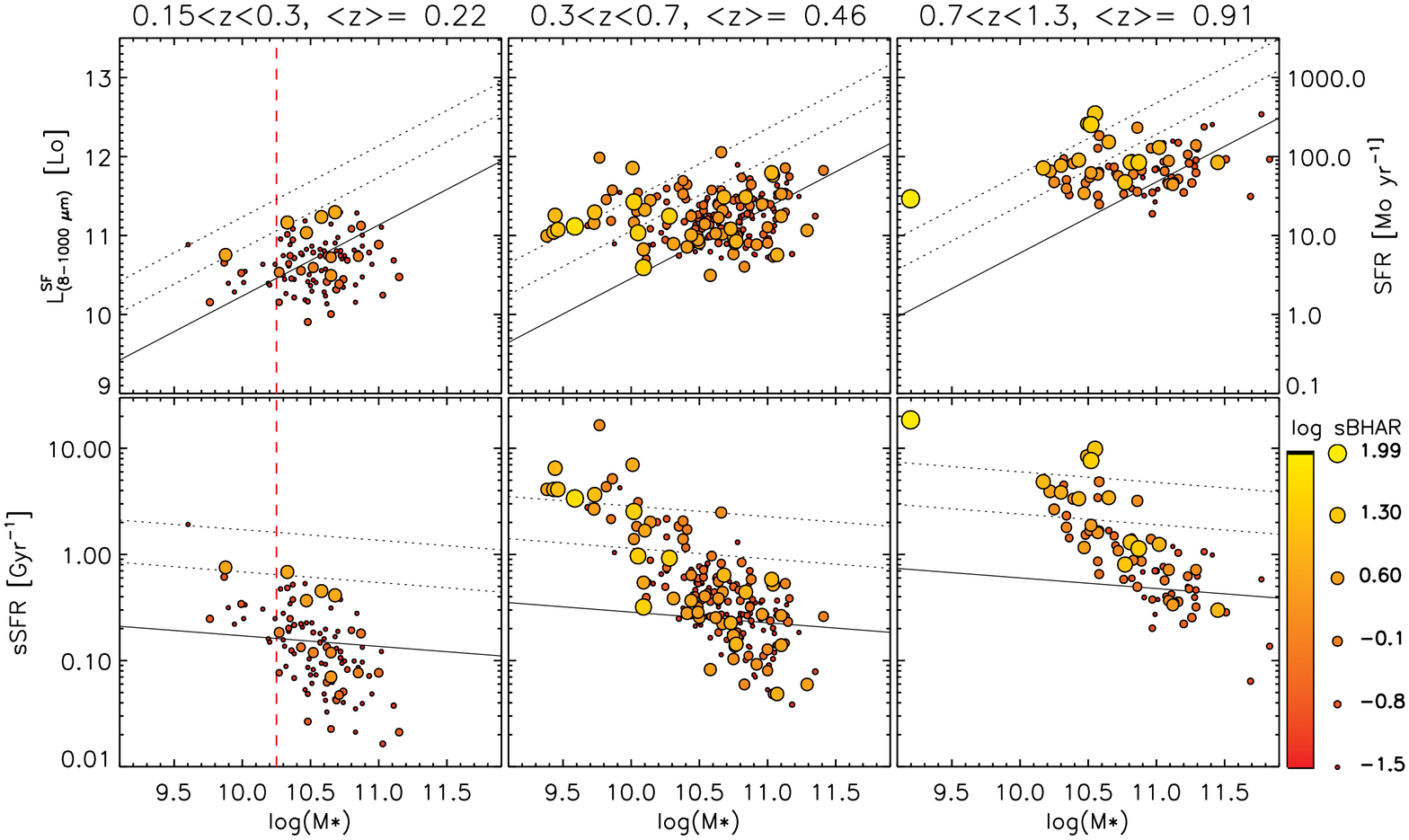}
 \caption{\label{CHECK_MS1} {\bf Upper panels:} position of the data with respect to the stellar mass - star formation rate relation (black thick line computed using Equation~\ref{EQ:SFRMS}, at the average redshift of the bin) in three redshift bins between z=0.15 and z=1.3. Bigger circles and yellow color indicate higher specific black hole accretion rate (sBHAR [Gyr$^{-1}$]). The first redshift bin (left panels) corresponds to the redshift range explored in our analysis. The mass limit used in the analysis ($10^{10.25}M_{\odot}$) is indicated with a red dashed line. The two black dotted lines indicate star formation rates 4 times and 10 times higher than the \emph{main sequence} value.
The nearly horizontal distribution of the data, especially at higher redshifts, is expected: the sample is dominated by the far-IR selection \citep[for example, see the distribution of a similar \emph{Herschel} selected sample in][]{2011ApJ...739L..40R}. In the {\bf lower panels}, the distributions are shown in the M* versus specific-SFR plots.}
 \end{figure*}

SFRs and stellar masses are expected to be related each other by the so called \emph{main sequence} of the star forming galaxies \citep{2004MNRAS.351.1151B, 2007A&A...468...33E, 2007ApJ...660L..47N,2011ApJ...739L..40R, 2014MNRAS.443...19R}. We compared the SFR of each galaxy with the expectation resulting from the \emph{main sequence} definition of \citet[][see equation 4 therein]{2007A&A...468...33E}, considering a SFR evolution as (1+z)$^{2.8}$ \citep{2012ApJ...747L..31S}. Again, we use the 0.24 dex correction factor to refer our quantities to a \cite{2003PASP..115..763C} IMF. The combination of the two equations can be expressed as:
\begin{equation}
\label{EQ:SFRMS}
\footnotesize
\mathrm{SFR}^{\mathrm{Chab}}_{\mathrm{MS}}(\mathrm{z, \mathrm{M}^{*}})= \frac{7.2}{1.7} (\mathrm{M}^{*}\times10^{-10})^{0.9} \left(\frac{1+z}{2}\right)^{2.8} \  \left[M_{\odot}\ yr^{-1}\right]
\end{equation}
In Figure~\ref{CHECK_MS1} we show the M* versus SFR for three redshift bins in the redshift range 0.15$<$z$<$1.3. The \emph{main sequence}, computed for the average redshift of each bin by using Equation~\ref{EQ:SFRMS}, is also shown for a comparison. The almost horizontal distribution of the data in the same plots is not surprising: given the far-IR selection of the sample (MIPS and SPIRE detections are required for our far-IR analysis), only the most IR-bright galaxies (i.e. the most star forming) are considered at each redshift. However, this horizontal selection does not sensibly affect galaxies above the stellar mass completeness limit computed (for the redshift range explored in our analysis, see vertical dashed line in Figure~\ref{CHECK_MS1}).

\subsection{Black hole accretion rate}
\label{BHAR_sect}
The {\sc sed3fit} software provides the AGN bolometric luminosity (L$_{\mathrm{acc}}$) as an output parameter. 
This luminosity, derived through SED fitting decomposition, is in good agreement with estimates obtained from X-rays and high excitation mid-IR lines such as [Ne V] and [O IV] \citep{2016MNRAS.458.4297G}.
Following \cite{2012ApJ...753L..30M}, and assuming an energy production efficiency $\epsilon=0.1$, we compute the BHAR of the sources in our selection as \footnote{The bolometric luminosity L$_{\mathrm{BOL}}$ in \cite{2012ApJ...753L..30M} corresponds to L$_{\mathrm{acc}}$ in this analysis.}:
\begin{equation}
\label{EQ:BHAR_Mullaney2012}
\mathrm{BHAR}=1.586\times 10^{-26} \frac{(1-\epsilon) L_{\mathrm{acc}}} {\epsilon c^{2}} \ \ \ \ \left[M_{\odot} yr^{-1}\right],
\end{equation}
where c is in units of [cm $s^{-1}$] and $L_{\mathrm{acc}}$ in units of [erg $s^{-1}$].

The specific black hole accretion rate (sBHAR) is defined as the ratio between the black hole accretion rate and the black hole mass M$_{\mathrm{BH}}$. Using a simple conversion factor \citep[e.g.][]{2012ApJ...753L..30M}, the black hole mass M$_{\mathrm{BH}}$ is obtained from the stellar mass of the host galaxy: M$_{\mathrm{BH}}$= 1.5$\times 10^{-3}$ M*.

\subsection{Gas fraction}
\label{Gas_fract_sect}
For each of the sources in our sample, we computed the gas fraction from the total dust mass M$_{\mathrm{dust}}$ (a {\sc sed3fit} output) by combining the stellar mass-metallicity-redshift (M$^{*}$-Z-z) relation of \citet[][see equation 12a therein]{2015ApJ...800...20G}:
\begin{eqnarray}
\label{EQ:M_Z_z}
\log(\mathrm{O/H})=-12+a-0.087\ [\log(M^{*}-b)^{2}],\ \ \mathrm{where}\nonumber\\
a=8.74 \ \ \mathrm{and} \ \ \ \ \ \ \ \ \ \ \ \ \ \ \ \ \ \ \ \ \ \ \ \ \ \ \ \ \ \ \ \ \ \ \ \ \ \ \ \ \ \ \ \ \ \ \ \ \ \nonumber\\
b=10.4+4.46\ \log(1+z)-1.78\ \left(\log(1+z)\right)^{2}, \ \ \ \
\end{eqnarray}
with the $\delta_{\mathrm{GDR}}$-Z relation defined in \citet{2012ApJ...760....6M}:
\begin{equation}
\label{EQ_Magdis12}
\log\delta_{\mathrm{GDR}}=10.54-0.99\ \left[12+\log(\mathrm{O/H})\right].
\end{equation}
Substituting the stellar mass M$^{*}$ in the previous equations, the gas mass can be obtained from the total dust mass M$_{\mathrm{dust}}$ using:
\begin{equation}
\label{metallicity}
M_{\mathrm{GAS}}=M_{\mathrm{dust}}\ \delta_{\mathrm{GDR}}
\end{equation}

\subsection{AGN fraction}
\label{SECT:AGN_FRACT}

Our analysis sample is made by low-redshift SPIRE detected star forming galaxies. Consequently, for all the sources in the redshift range explored (0.15$<$z$<$0.3) the presence of an AGN has only marginal effects on the computation of the total SFR. This can be visually appreciated observing the typical SEDs represented in Figure~\ref{img:SED_Examples}: the total 8 to 1000 \mic\ emission is always strongly dominated by the star formation. Consistently, we already showed in Figure~\ref{MASS_COMPAR_1} (central panel) how the presence of an AGN does not affect the computation of the stellar masses in our sample.  Similar results are found in \cite{2015A&A...576A..10C} where it is shown that including an AGN contribution in an SED fitting has marginal effects on the computation of the total stellar mass of the galaxies, unless the AGN contribution itself dominates the IR emission.

In our analysis, we compute the AGN contribution to the IR luminosity in two different spectral ranges: 8-1000 \mic\ (bolometric) and 5-40 \mic. As demonstrated in \cite{2016MNRAS.458.4297G}, even when the AGN emission contribution to the bolometric band is small ($f^{8-1000}_{\mathrm{AGN}}$$\sim$0.3\%), the AGN contribution to the 5-40 \mic\ ($f^{5-40}_{\mathrm{AGN}}$) is around one order of magnitude higher and hence detectable if photometric measures are available in the rest frame mid-IR spectral range (as in our case).

\subsubsection{Uncertainties and reliabilities}
\label{SECT:UNC_REL}

\begin{figure}[ht!]
 \centering
 \includegraphics[width=8.0cm]{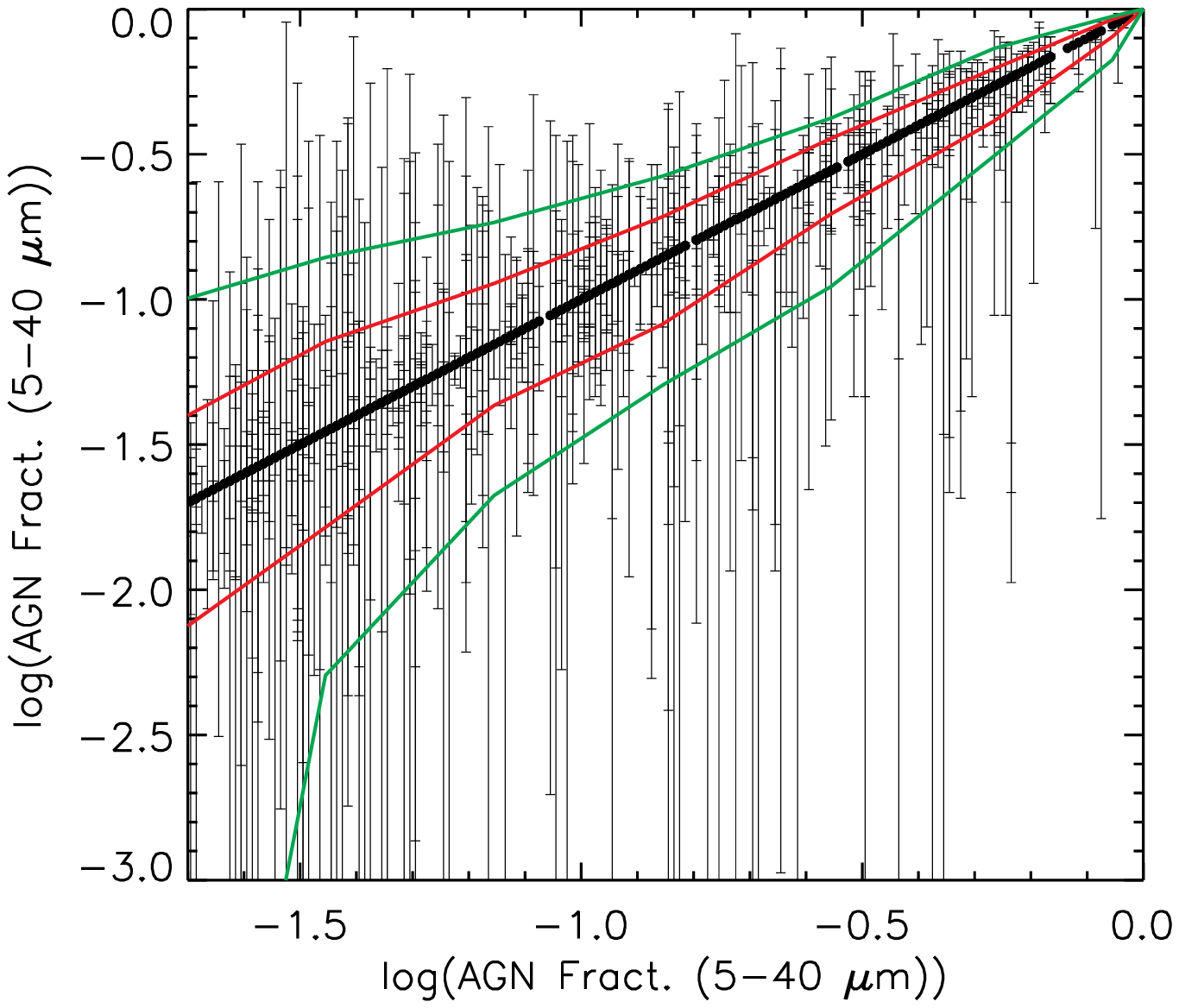}
 \includegraphics[width=8.0cm]{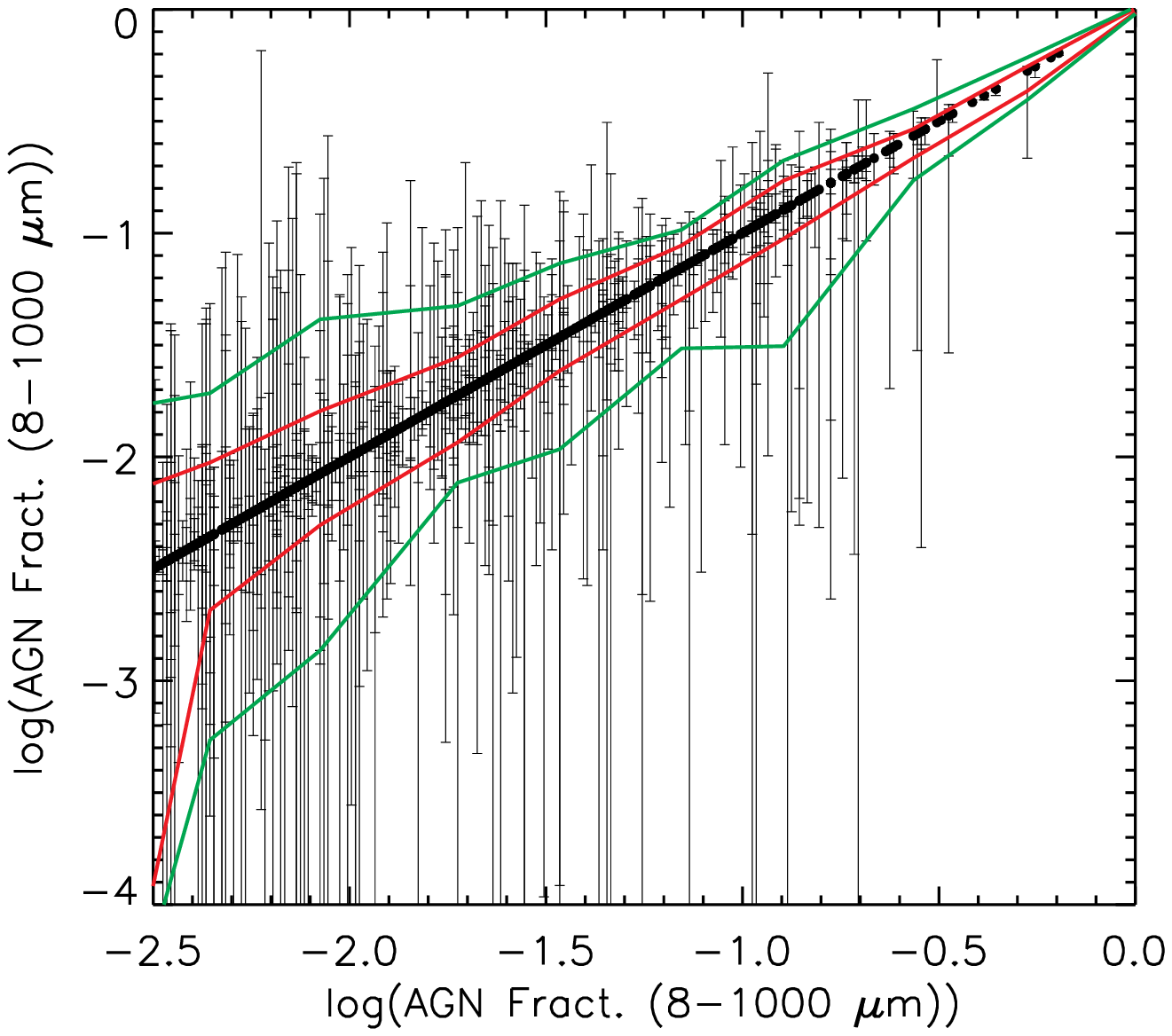}
 \caption{\label{IMG:AGNF_unc} Uncertainty (1$\sigma$) associated to the fraction of IR emission due to AGN. We consider two different bands: 5-40 \mic\ (upper panel) and 8-1000 \mic\ (bottom panel). Both axes represent the estimate of AGN emission fraction, for each galaxy. In the \emph{y} axes the upper and lower limits to these estimates are reported for each source. The red and green lines represent the median and 75\% percentiles of the upper and lower limits after binning the data in the \emph{x} axis.}
 \end{figure}

In Figure~\ref{IMG:AGNF_unc}, we show the uncertainty related to $f^{\Delta\lambda}_{\mathrm{AGN}}$ for all the sources in the parent sample with an available fit. We observe that this uncertainty can be very high on some galaxies if considered singularly, but the median uncertainty (red line in Figure~\ref{IMG:AGNF_unc}) is always smaller than $\sim$0.3 in logarithmic scale, and for the vast majority of the sources (75\%, green line in Figure~\ref{IMG:AGNF_unc}), it is never larger than 0.5. 
We stress on the fact that our analysis is based on the study of the \emph{average} AGN emission fraction of binned data, while we are not using the fitted SEDs and the corresponding SED-fitting related parameters to get information at the level of single galaxies.

Finally, we notice that the average uncertainties explode below $f^{5-40}_{\mathrm{AGN}}$$\sim$2\% or equivalently, below $f^{8-1000}_{\mathrm{AGN}}$$\sim$0.3\%. Below these limits it becomes statistically very difficult to distinguish between sources hosting low activity AGNs and sources without AGN activity at all. This is also true after averaging the behaviour of many sources.
For this reason, we set this lower limit as a threshold: AGNs are considered in our analysis only when $f^{5-40}_{\mathrm{AGN}}>$2\%.

The previous one is not the only selection criterion applied: following \cite{2017ApJ...838..127I}, we performed an additional F-test to select only the sources for which the AGN component did statistically improve the SED fit, after considering the different number of degrees of freedom. To this purpose, we run the \emph{magphys} software \citet[][\emph{magphys} does not natively include an AGN component]{2008MNRAS.388.1595D} using a set of parameters similar to those used for our analysis, AGN component excluded. Given the number of photometric bands used in the fit ($N_{j}$), the $\chi^{2}$ values obtained in the \emph{magphys} ($\chi_{1}^{2}$) and in the {\sc sed3fit} ($\chi_{2}^{2}$) runs, and the different number of components used in the two cases ($N_{1}$=2 and $N_{2}$=3), we computed $\mathcal{F}$ as:
\begin {equation}
\label{eqn:F_TEST_1}
\mathcal{F}(\chi^{2}_{N_{1}},\chi^{2}_{N_{2}})=\frac{(\chi^{2}_{N_{1}}-\chi^{2}_{N_{2}})/(d_{2}-d_{1})}{\chi^{2}_{N_{2}}/d_{2}}.
\end{equation}
In Equation~\ref{eqn:F_TEST_1}, $d_{1}=N_{j}-3 N_{1}$ and $d_{2}=N_{j}-3N_{2}$  represent the degrees of freedom corresponding to the number of components in the two runs. As in \cite{2017ApJ...838..127I}, we rejected the additional AGN component when $p(\mathcal{F},d_{1},d_{2})<0.5$. The main results of our analysis are still valid without rejecting these fits, but in that case the p-value of the relations that we find is an order of magnitude higher. Using the selection methods described, we identified an AGN component for 49$\pm7$\% of our far-IR selected sample in the redshift range $0.15<z<0.3$. 

 Concluding, in our AGN sample we included only the sources satisfying the following three criteria at the same time:
\begin{itemize}
\item{a) the SED fitting software that we use ({\sc sed3fit}) finds an AGN emission component that improves the overall fit, following the $\chi ^{2}$ minimization method;}
\item{b) the AGN contribution to the mid-IR emission, between 5 and 40 \mic\ ($f^{5-40}_{\mathrm{AGN}}$), is higher than ~2\%;}
\item{c) the AGN contribution is ``statistically required'' following the F-test described above.}  
\end{itemize}

\subsubsection{ Comparison with diagnostic methods}

\label{SEC:Diagnostic}

In this section, we compare the results of our AGN selection to two different diagnostic methods, based on mid- and far-IR photometric bands, that we designed following an approach similar to that proposed in \cite{2014MNRAS.444L..95F}. We found the combinations here proposed to be the most suitable for separating AGNs dominated systems from different types of galaxies. In our analysis, we study the galactic nucleus (SMBHs) contribution to the total mid- and far-IR emission of a sample of star forming galaxies. In this sense, we considered as AGNs only the sources satisfying the three criteria specified in Section~\ref{SECT:UNC_REL}. It is important to notice that with such a definition of ``AGN'' we are not indicating ``\emph{AGN dominated systems}'', but ``\emph{galaxies with a detected emission from the nucleus}''. This difference is important when comparing our AGN sample to diagnostic methods such as the ones proposed here. Infact, these diagnostic methods are meant to detect AGN dominated systems (i.e. AGNs showing a dominant contribution to the overall IR emission of the galaxies).

\begin{figure*}[ht!]
 \centering
 \includegraphics[width=5.35cm]{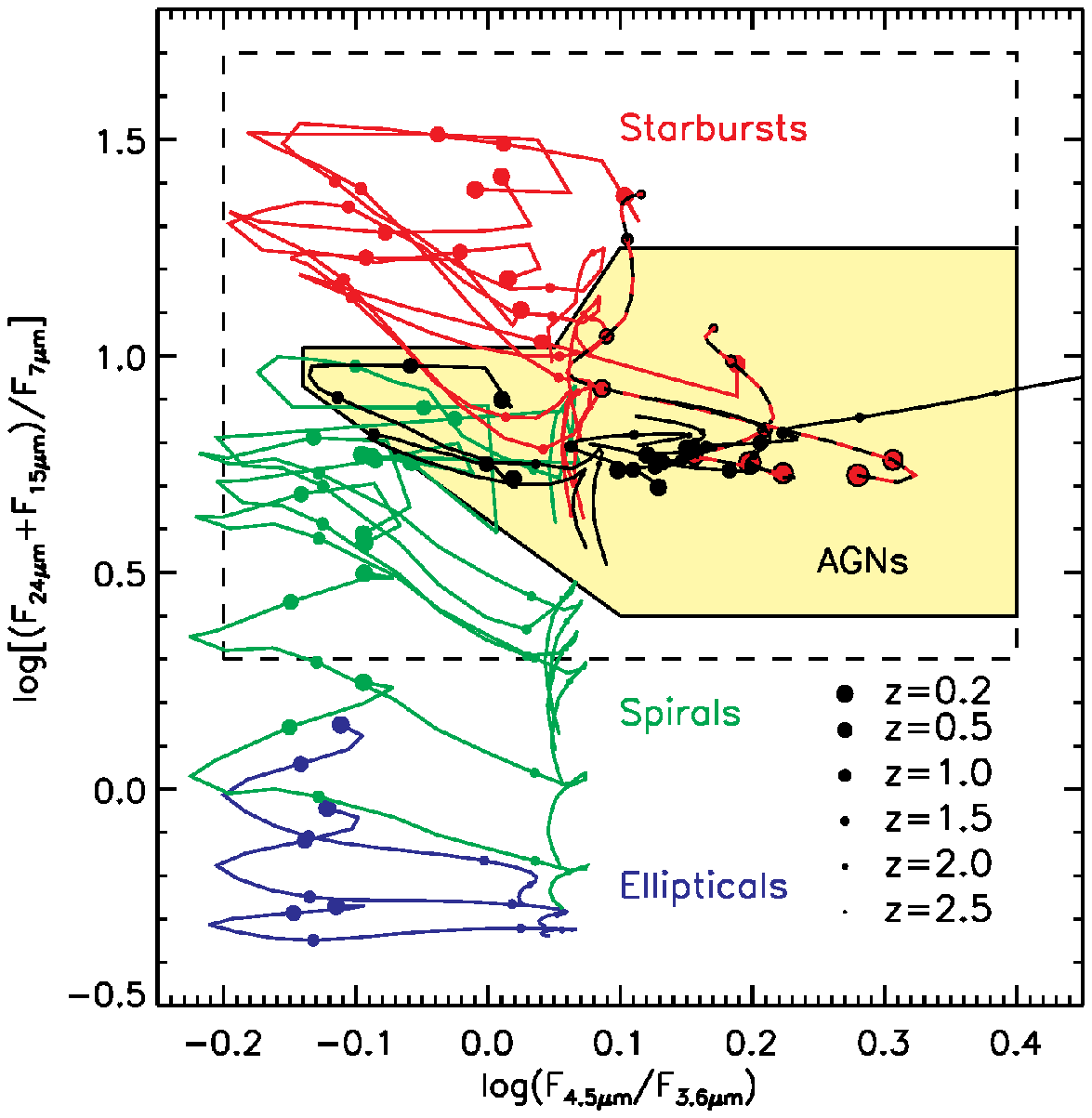}
 \includegraphics[width=6.2cm]{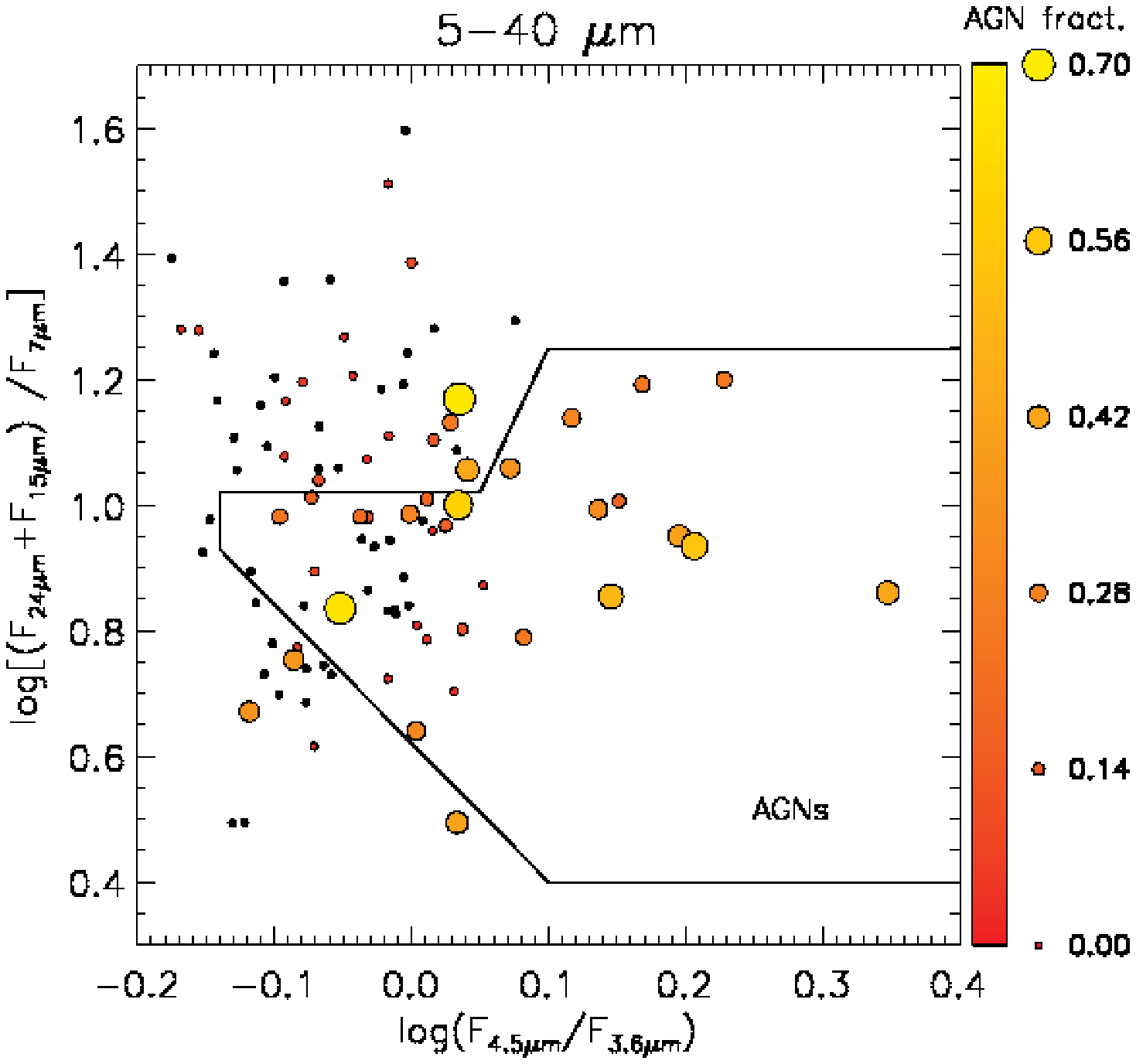}
 \includegraphics[width=6.2cm]{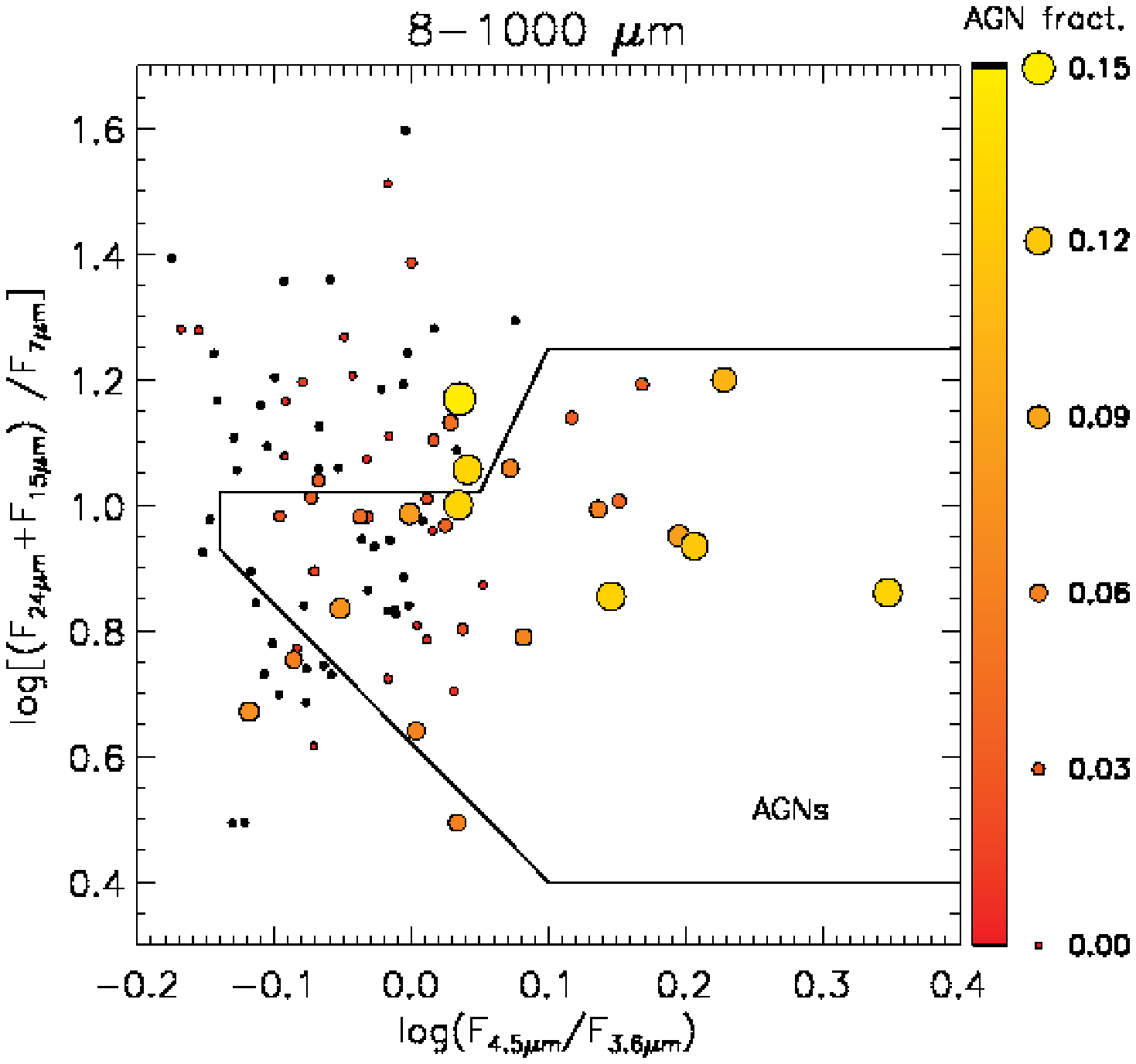}
 \caption{\label{IMG:DIAGN_1} Position of our data (no redshift selection) with respect to the first of the two AGN diagnostic methods that we propose. In the {\bf left panel} we show the tracks traced by the \cite{2007ApJ...663...81P} template SEDs in the first of the two diagnostic methods proposed, by shifting the SEDs to different redshifts. Different types of galaxies are represented using different colors: ellipticals in blue, spirals in green, starbursts in red and AGNs in black. Mixed types (AGN/starburts) are represented with dashed red/black lines. The proper AGN area of the diagnostic plot (yellow shaded area) is drawn trying to include the AGN templates avoiding the contamination from different templates as much as possible. In the {\bf central and right panels} we show the position of the sources in our sample (no redshift selection). The dimension and color of the data points represents, for each source, the AGN fractional contribution to the 5-40 \mic\  and to the bolometric (8-1000 \mic) emission. We find that 58\% of the AGNs of our sample are located inside the proper area of this diagnostic plot, with 29\% contamination. }
 \end{figure*}

In the first of the two diagnostic plots that we propose (Figure~\ref{IMG:DIAGN_1}), we combine the fluxes observed in the MIPS-24 \mic, AKARI-15  and 7 \mic\ bands, with those measured at 3.6 and 4.5 \mic\ (IRAC). On the x-axis we consider the IRAC-4.5 \mic\ to 3.6 \mic\ flux ratio \citep[as already done in e.g.][for similar purposes]{2012ApJ...748..142D}, while in the y-axis we use the mid-IR ratio $(F_{15\mu m}+F_{24\mu m})/F_{7\mu m}$ \citep[a similar quantity is proposed in the diagnostic plots presented in][]{2014MNRAS.444L..95F}. In the left plot of Figure~\ref{IMG:DIAGN_1}, we show the tracks corresponding to all the templates reported in the SWIRE library \citep{2007ApJ...663...81P} as observed at different redshifts. The templates are divided among elliplticals, spirals, AGNs and starbursts. The same library includes also two templates referring to mixed systems starburts/AGN. 
We define the ``AGN area'' of the diagnostic plot by observing the position of the corresponding \citep{2007ApJ...663...81P} templates. The area is delimited by the polygon defined by the following coordinates:
\begin{eqnarray}
x=[-0.14,-0.14,0.1 ,0.4 ,0.4,0.1,0.05] \nonumber\\
y=[1.02, 0.93,0.4,0.4,1.25   ,1.25,1.02]
\end{eqnarray} 
The amount of sources with an available measure in all these bands at the same time is mostly limited by the overlap between the areas covered by AKARI at different wavelengths. Their position in the diagnostic plot is shown in the central and right panels of Figure~\ref{IMG:DIAGN_1}.
We found that 58\% of the sources identified as AGNs at all redshifts are included in the proper AGN area of this diagnostic plot, with 29\% contamination from the other sources of our far-IR selected sample.

\begin{figure*}[ht!]
 \centering
 \includegraphics[width=5.35cm]{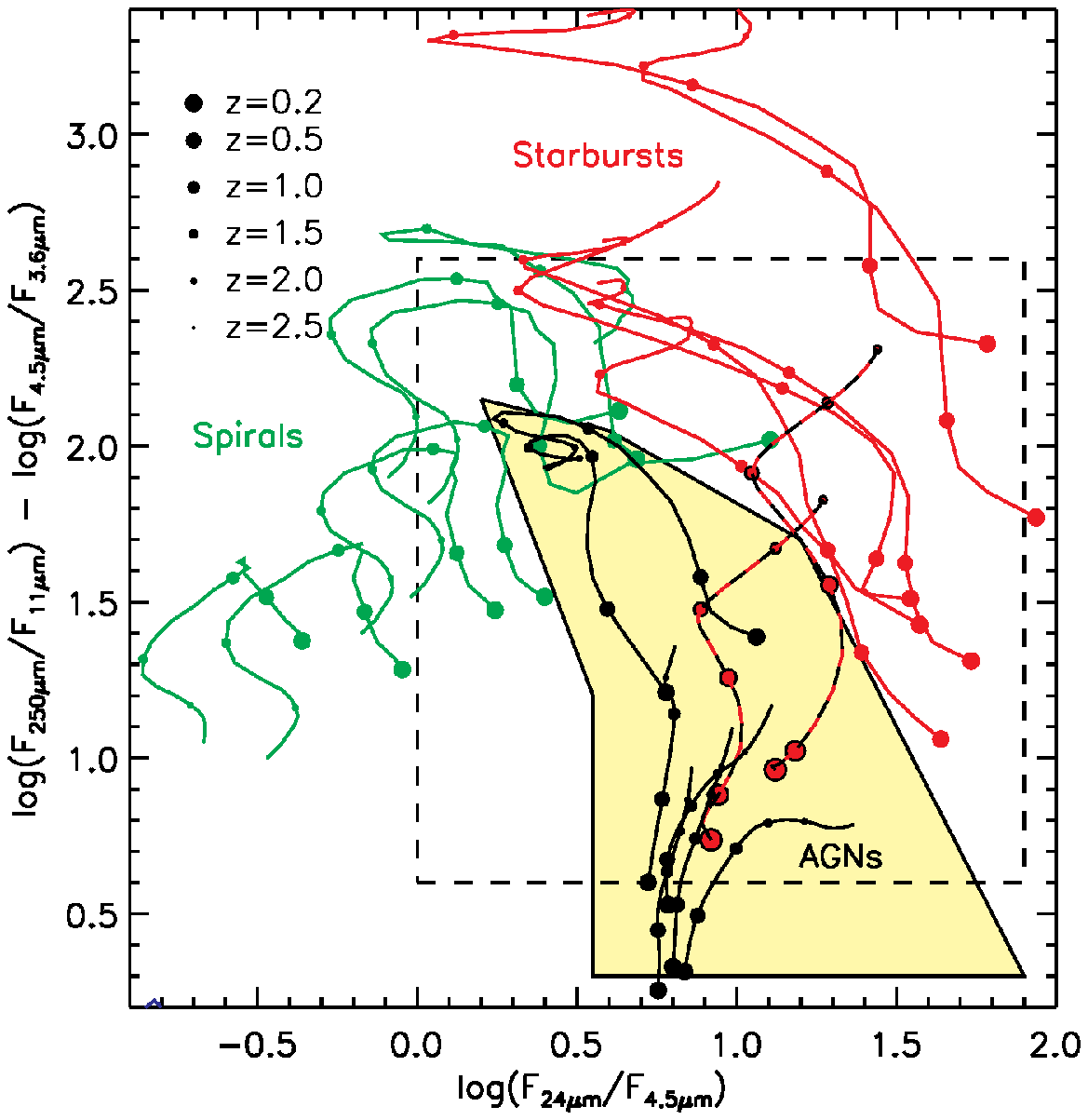}
 \includegraphics[width=6.2cm]{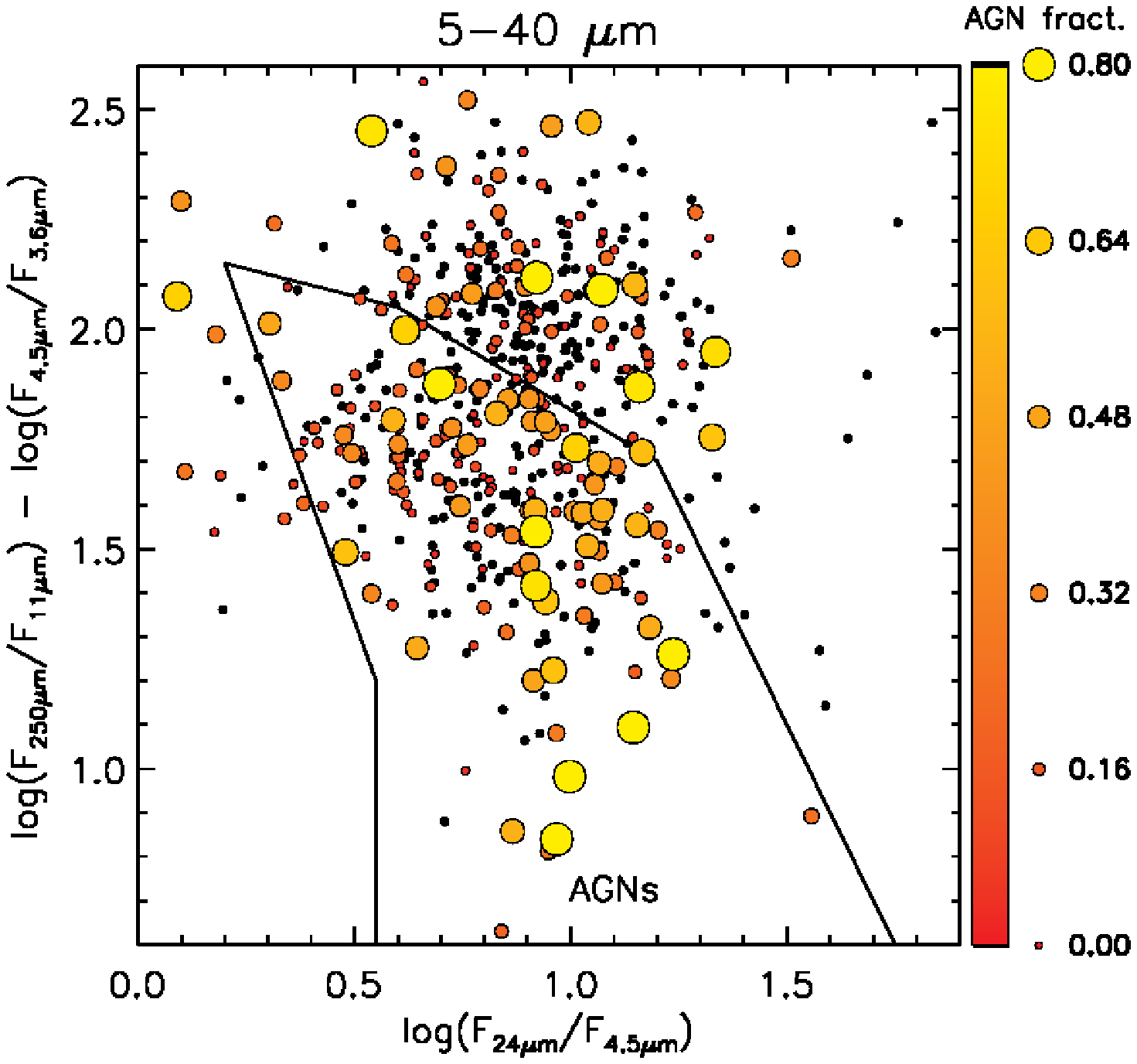}
 \includegraphics[width=6.2cm]{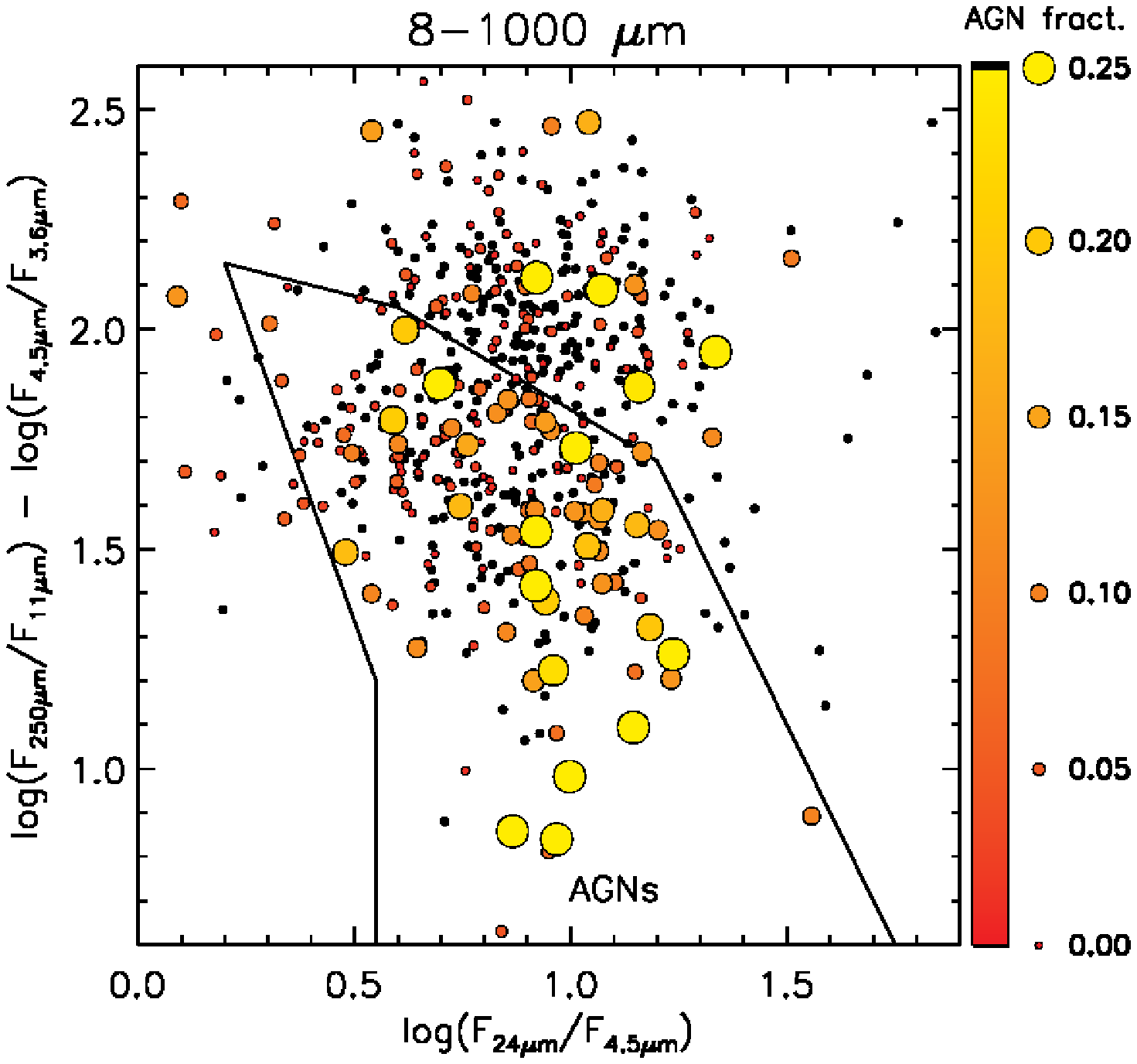}
 \caption{\label{IMG:DIAGN_2} Position of our data (no redshift selection) with respect to the second of the two AGN diagnostic methods that we propose; colors and symbols as in Figure~\ref{IMG:DIAGN_1}. In this case 63\% of the AGNs of our sample are located inside the proper area of this diagnostic plot, with 43\% contamination. }
 \end{figure*}

In the second diagnostic plot that we propose (Figure~\ref{IMG:DIAGN_2}), we combine the IRAC-4.5 \mic\ to 3.6 \mic\ flux ratio with the SPIRE-250 \mic\ to WISE-11 \mic\ ratio in a unique indicator. Then, the y axis is represented by $(F_{250\mu m}/F_{11\mu m})\times(F_{3.6\mu m}/F_{4.5\mu m})$. In this quantity, at least for galaxies at z$\leq$1.5, $(F_{250\mu m}/F_{11\mu m})$ represents an estimate of the relative AGN emission contribution with respect to the total far-IR emission (i.e. SFR). At the same time, the $(F_{4.5\mu m}/F_{3.6\mu m})$ ratio is a measure of the relative contribution of the AGN emission with respect to the total stellar emission (this is valid, at least, for the low redshifts explored in our analysis).
The x-axis of the diagnostic plot, $(F_{24\mu m}/F_{4.5\mu m})$, is mostly a measure of specific SFR (SFR/M*), but given the mid-IR nature of the two bands involved, it is also influenced by the AGN contribution. In this second case, the AGN area of the diagnostic plot is delimited by the polygon defined by the following coordinates:
\begin{eqnarray}
x=[0.2,0.55,0.55,1.9,1.2 ,0.6 , 0.2] \nonumber\\
y=[2.15,1.2,0.3,0.3,1.7  ,2.05,2.15]
\end{eqnarray} 
Using this diagnostic method, 63\% of the AGN identified are located in the corresponding AGN area of the plot, with a 43\% contamination.

\section{Discussion}
\label{SEC:ANALYSIS}
\subsection{SMBH activity in small groups}
\label{SEC:SMBH_act_groups}
We studied the dependence of the AGN activity on the total stellar mass M$^{*}_{\mathrm{group}}$ of the hosting groups in the redshift interval $0.15<z<0.3$. 
For this analysis we considered \emph{main sequence} galaxies (-0.6$<\log(\mathrm{SFR}/\mathrm{SFR(MS)})<$0.6) for which the {\sc sed3fit} software identifyied an AGN contribution to the total IR luminosity $f^{5-40}_{\mathrm{AGN}}$$>$2\%. Following \cite{2017ApJ...838..127I}, we additionally performed an F-test to select only the sources for which the AGN component did statistically improve the SED fit, at the net of the different number of degrees of freedom (see Section \ref{SECT:UNC_REL}).

In order to create a mass complete sample, we exclude from our analysis AGNs in galaxies with $\log$(M*)$<10.25$ M$_{\odot}$. However, the total stellar mass of the groups, M$^{*}_{\mathrm{group}}$, is computed using all the galaxies identified as members, without exclusions.

We divided the sample in 10 ``field'' AGNs , for which the group finder algorithm did not find any companion (all these sources have a \emph{WISE}-11 \mic\ flux measure, 1 has a spectroscopic redshift), and 35 ``group'' AGNs  with at least an identified companion (33 with an 11 \mic\ flux measure, 4 with spectroscopic redshift). The AGNs in groups are further divided into three bins of M$^{*}_{\mathrm{group}}$.

\begin{figure}[ht!]
 \centering
 \includegraphics[width=7.5cm]{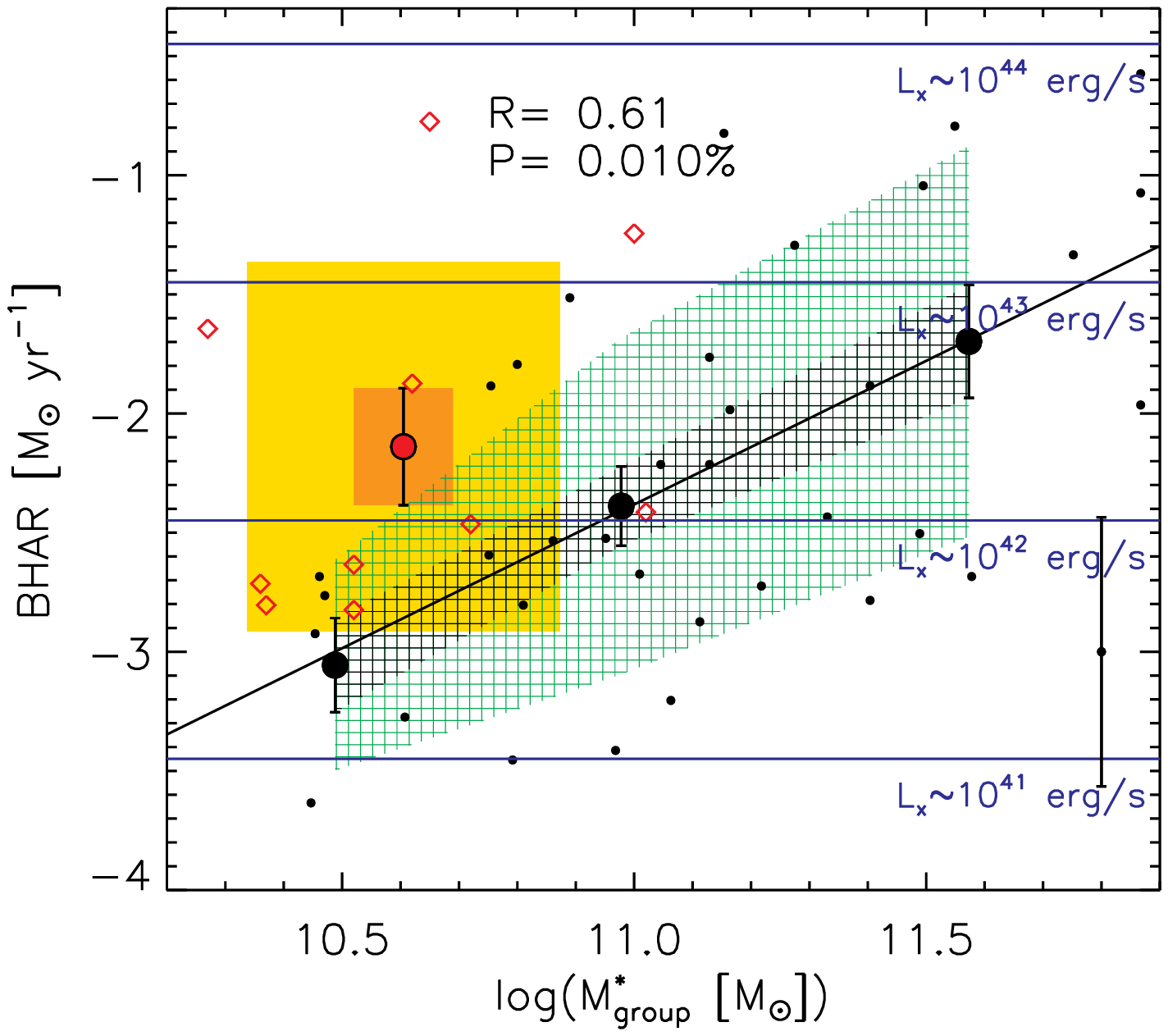}
 \caption{\label{IMG:MGROUP_BHAR} BHAR of galaxies in the field (red squares) and in groups (black circles) as a function of stellar mass. For field galaxies, the x axis represents the stellar mass M$^{*}$ of each single galaxy while, for galaxies in groups, it represents the total stellar mass of the hosting group (M$^{*}_{\mathrm{group}}$). 
We considered three bins of M$^{*}_{\mathrm{group}}$: M$^{*}_{\mathrm{group}}$[M$_{\odot}$]$<10^{10.75}$, $10^{10.75}<$M$^{*}_{\mathrm{group}}$[M$_{\odot}$]$<10^{11.25}$ and M$^{*}_{\mathrm{group}}$[M$_{\odot}$]$>10^{11.25}$, but the linear fit is computed on the underlying data points and not on the binned data (only galaxies in groups are considered for the fit). The vertical dispersion in the three bins is reported as a shaded green area, while the darker area corresponds to the 1$\sigma$ uncertainty associated to the average values of BHAR in each mass bin. Dispersion and 1$\sigma$ uncertainties are represented also for \emph{field} galaxies using different tonalities of yellow. The typical uncertainty associated to each single data-point is reported in the bottom-right corner of the plot; its value is derived from the estimated PDFs of L$_{\mathrm{acc}}$ (see Equation~\ref{EQ:BHAR_Mullaney2012}), that is an output parameter of the SED fitting). 
 The correlation coefficient R and the corresponding \emph{p}-value are reported. For an easier comparison with literature results, the BHAR values corresponding to four X-ray luminosity values are represented through horizontal blue lines. }
 \end{figure}

Figure~\ref{IMG:MGROUP_BHAR} shows the BHAR of single galaxies as a function of the total stellar mass of the hosting groups M$^{*}_{\mathrm{group}}$. We see that BHAR increases with M$^{*}_{\mathrm{group}}$, indicating that, on average, more massive groups host more accretion onto super massive BHs. We found $\log($BHAR$)\propto (1.21\pm0.27)\log($M$^{*}_{\mathrm{group}})$, with a correlation index R=0.61 and a corresponding \emph{p}-value equal to 0.010\%. 
In order to facilitate the comparison with literature studies based on X-rays data, in the same figure we indicate the BHAR values corresponding to some commonly used X-ray luminosity thresholds.
These BHAR thresholds are obtained using Equation~\ref{EQ:BHAR_Mullaney2012}, and considering a conversion factor $L_{\mathrm{acc}}=22.4 L_{\mathrm{X}}$ as in \cite{2012ApJ...753L..30M}.

\begin{figure*}[ht!]
 \centering
 \includegraphics[width=5.8cm]{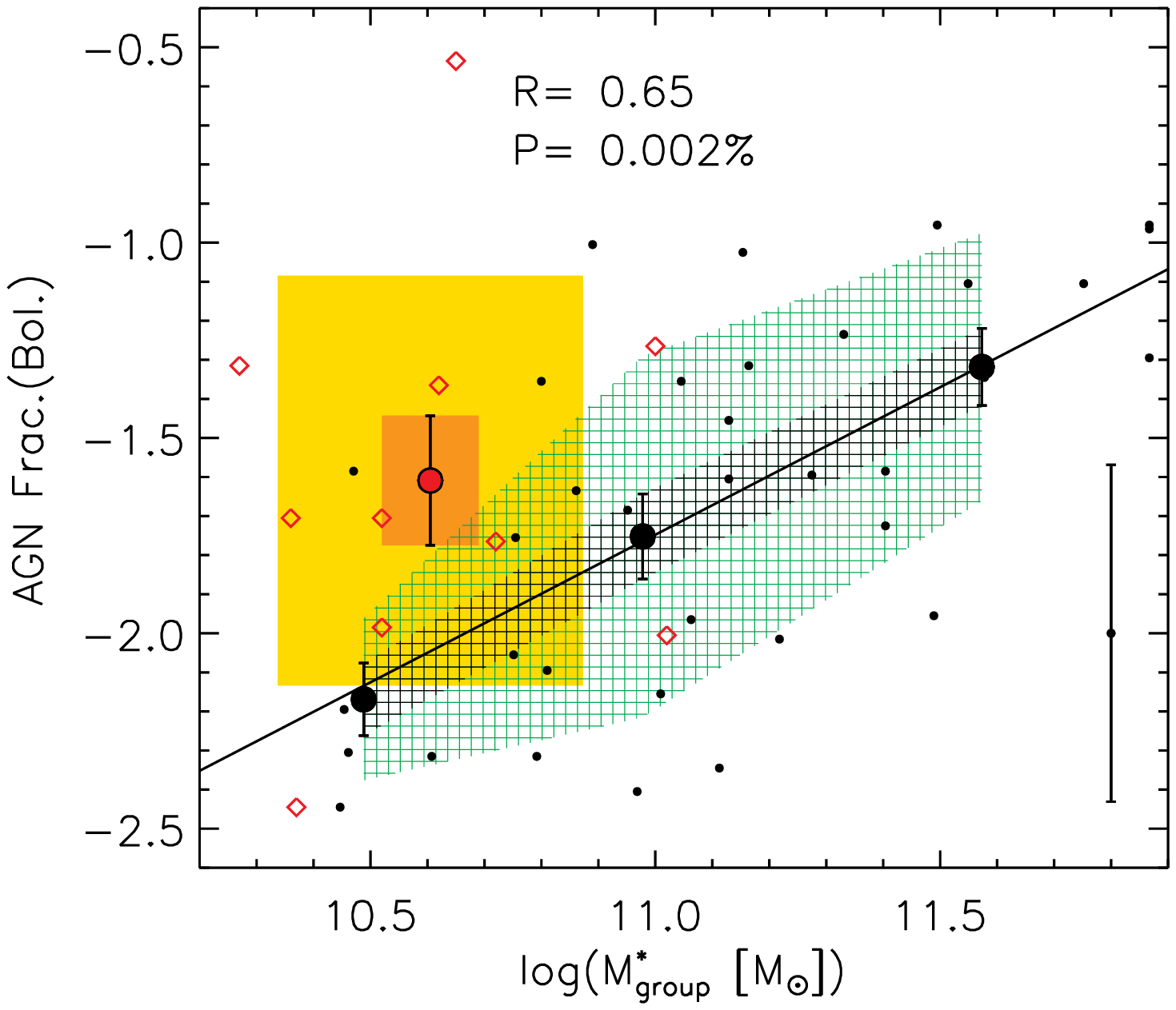}
 \includegraphics[width=5.8cm]{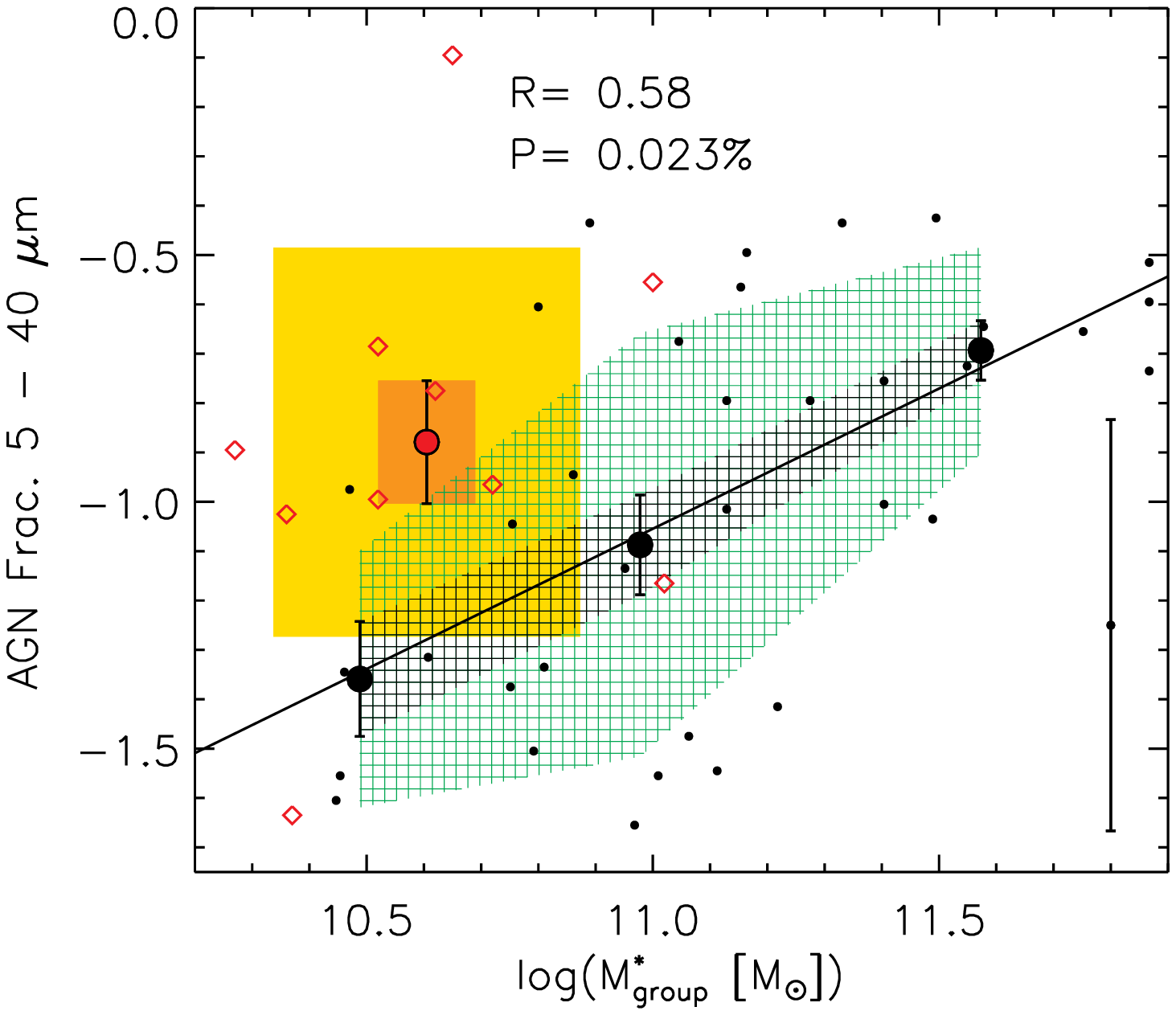}
 \includegraphics[width=5.8cm]{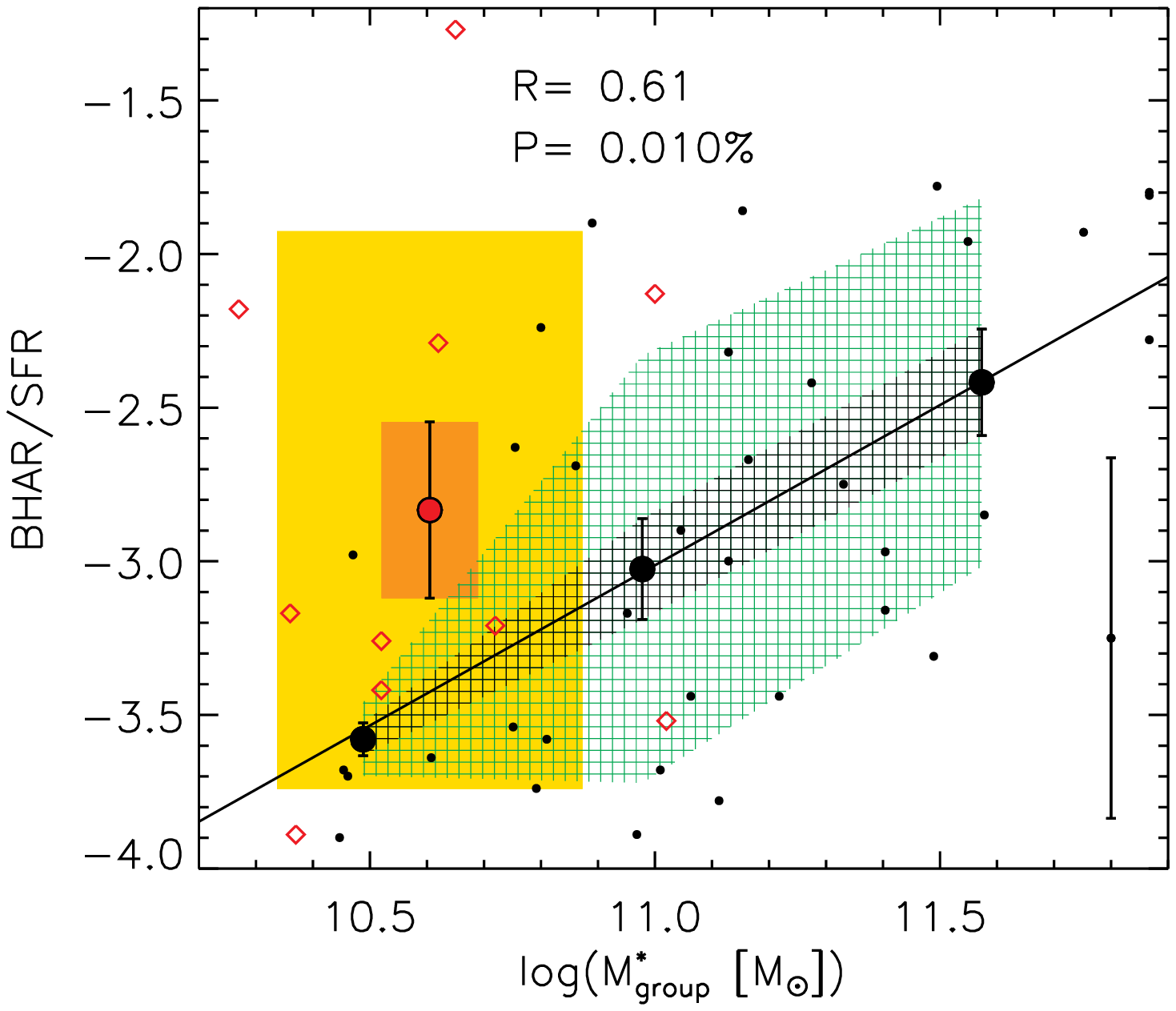}
 \caption{\label{IMG:MGROUP_AGN_act} {\bf Left and central panel:} AGN fractional contribution to the total luminosity in the bolometric (8-1000 \mic) and in the 5-40 \mic\ bands as a function of the total stellar mass of the hosting galaxy group M$^{*}_{\mathrm{group}}$, at 0.15$<$z$<$0.3. For field galaxies (red squares), the x axis represents the stellar mass M$^{*}$ of each single galaxy while, for galaxies in groups (black circles), it represents the total stellar mass of the hosting group (M$^{*}_{\mathrm{group}}$). We considered three bins of M$^{*}_{\mathrm{group}}$: M$^{*}_{\mathrm{group}}$[M$_{\odot}$]$<10^{10.75}$, $10^{10.75}<$M$^{*}_{\mathrm{group}}$[M$_{\odot}$]$<10^{11.25}$ and M$^{*}_{\mathrm{group}}$[M$_{\odot}$]$>10^{11.25}$, but the linear fit (solid line) is computed on the underlying data points and not on the binned data (only galaxies in groups are considered for the fit). The vertical dispersion of the data in the three bins of M$^{*}_{\mathrm{group}}$ is represented with a shaded green area, while the darker area corresponds to the sigma. The yellow and orange areas represent the dispersion and sigma of \emph{field} galaxies. The typical uncertainty associated to single data-points is reported in the bottom right corner of each plot. These uncertainties are derived from the estimated PDFs of each output parameter of the SED fitting that is required to compute the quantity in the y axis. The correlation coefficient R of the linear fit and the corresponding \emph{p}-value are reported. {\bf Right panel:} BHAR/SFR as a function of M$^{*}_{\mathrm{group}}$ for the same sources.}
 \end{figure*}

In Figure~\ref{IMG:MGROUP_AGN_act} we show how various AGN properties correlate with M$^{*}_{\mathrm{group}}$.
We see a moderate direct correlation between the AGN emission fraction computed for each galaxy in the bolometric (8-1000 \mic) and in the 5-40 \mic\ bands, and the total stellar mass of the hosting group. In particular, we find $\log(f^{5-40}_{\mathrm{AGN}})\propto (0.57\pm0.14)\log($M$^{*}_{\mathrm{group}})$ and $\log(f^{8-1000}_{\mathrm{AGN}})\propto (0.76\pm0.15)\log($M$^{*}_{\mathrm{group}})$. 
The AGN emission fraction in these two wide far-IR bands, is closely related to the ratio between the thermal emission of the dusty circumnuclear torus and of the molecular clouds in star forming regions. For the BHAR/SFR, we find $\log($BHAR/SFR$)\propto (1.04\pm0.24)\log($M$^{*}_{\mathrm{group}})$. The correlation coefficient R of the linear fits to the data is higher than 0.58 in all the three cases, while the corresponding p-values are lower than 0.023\%.

On average, the BHAR, BHAR/SFR ratio, AGN fraction and their dispersions are similar in field and group galaxies, if group galaxies are not divided in bins of M$^{*}_{\mathrm{group}}$.
However, the same quantities are generally higher among field galaxies than in groups, when M$^{*}_{\mathrm{group}}$$\sim <$M$^{*}_{\mathrm{field}}>$.

\begin{figure*}
 \centering
 \includegraphics[width=5.8cm]{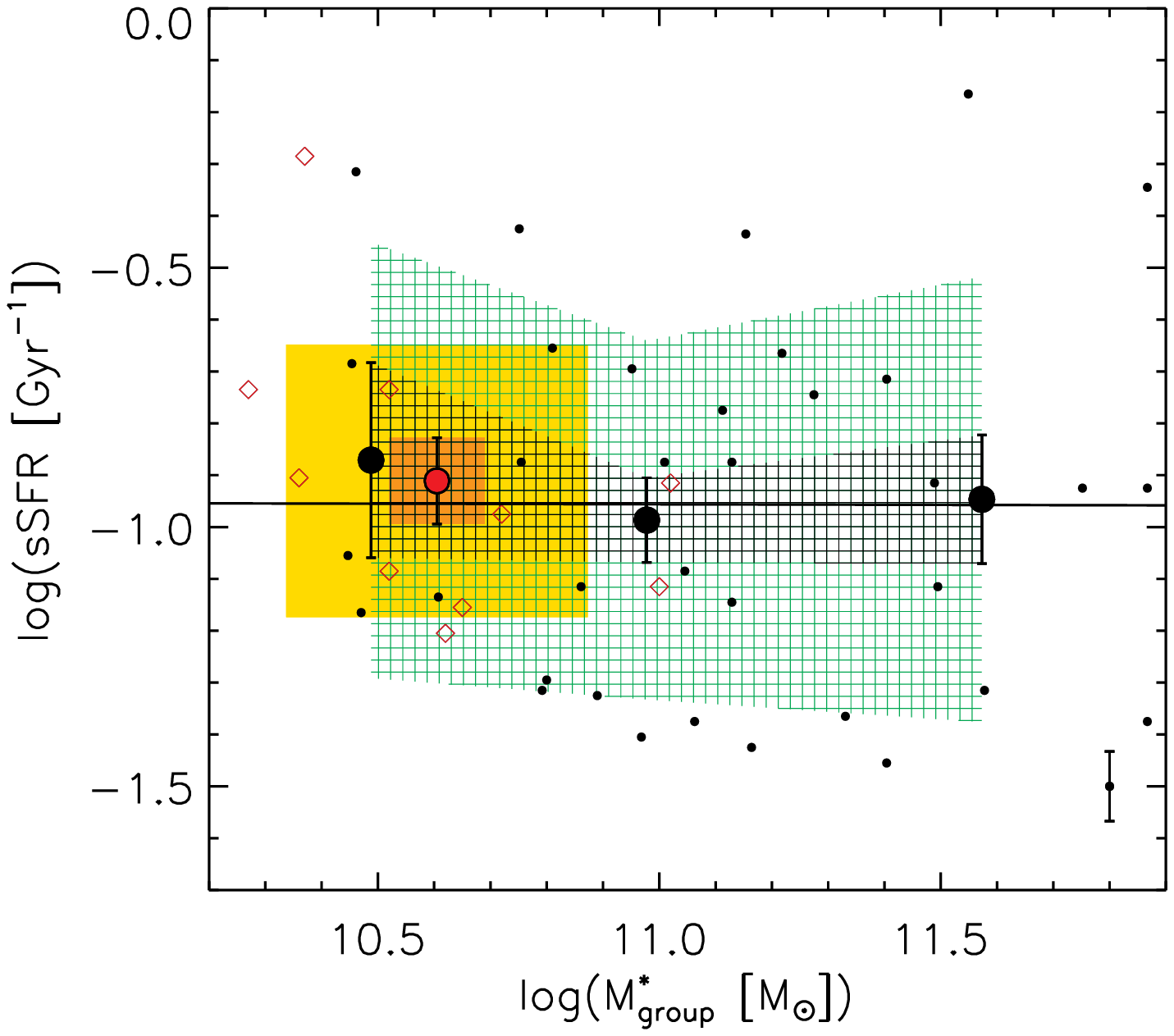}
 \includegraphics[width=5.8cm]{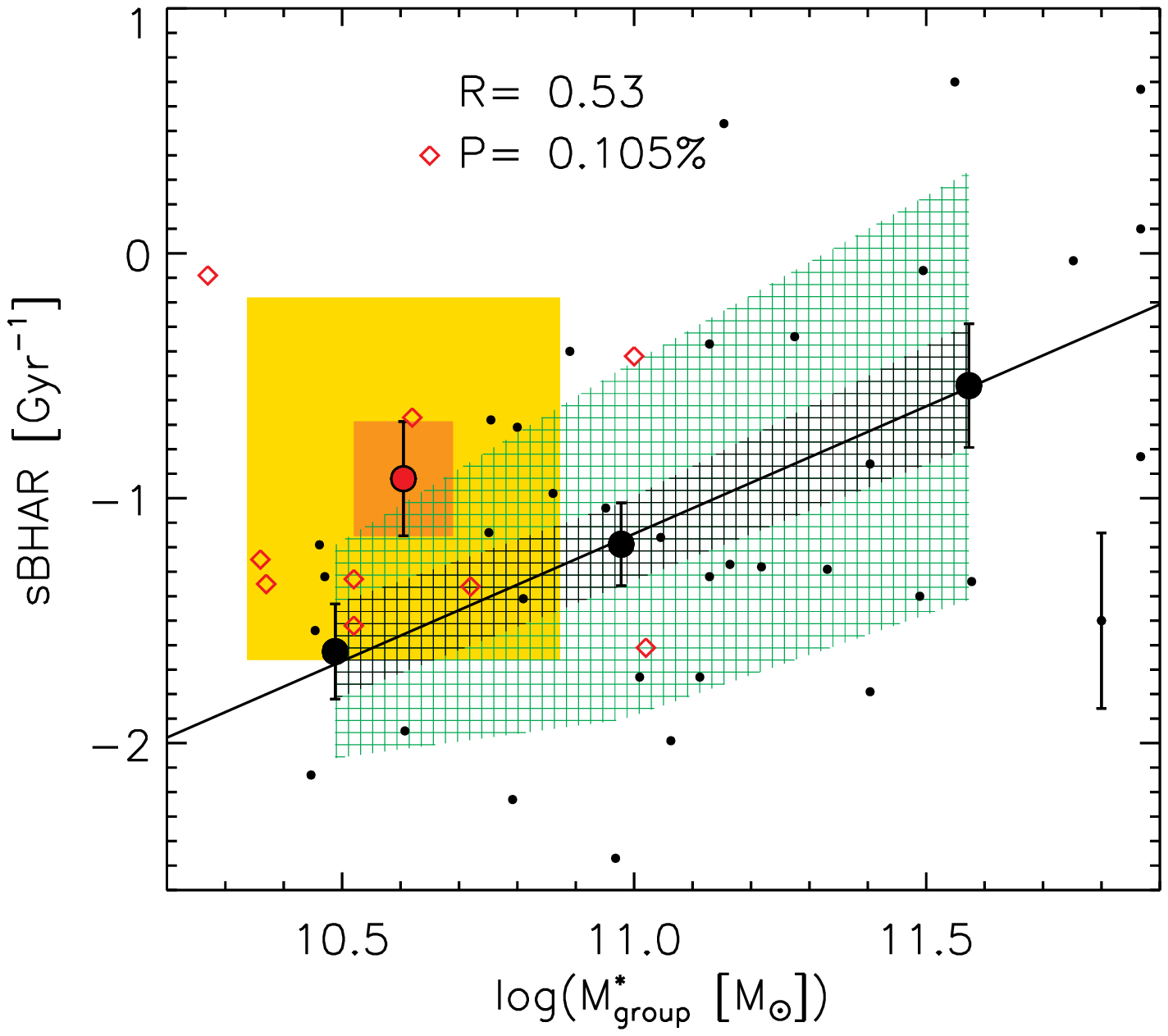}
 \includegraphics[width=5.8cm]{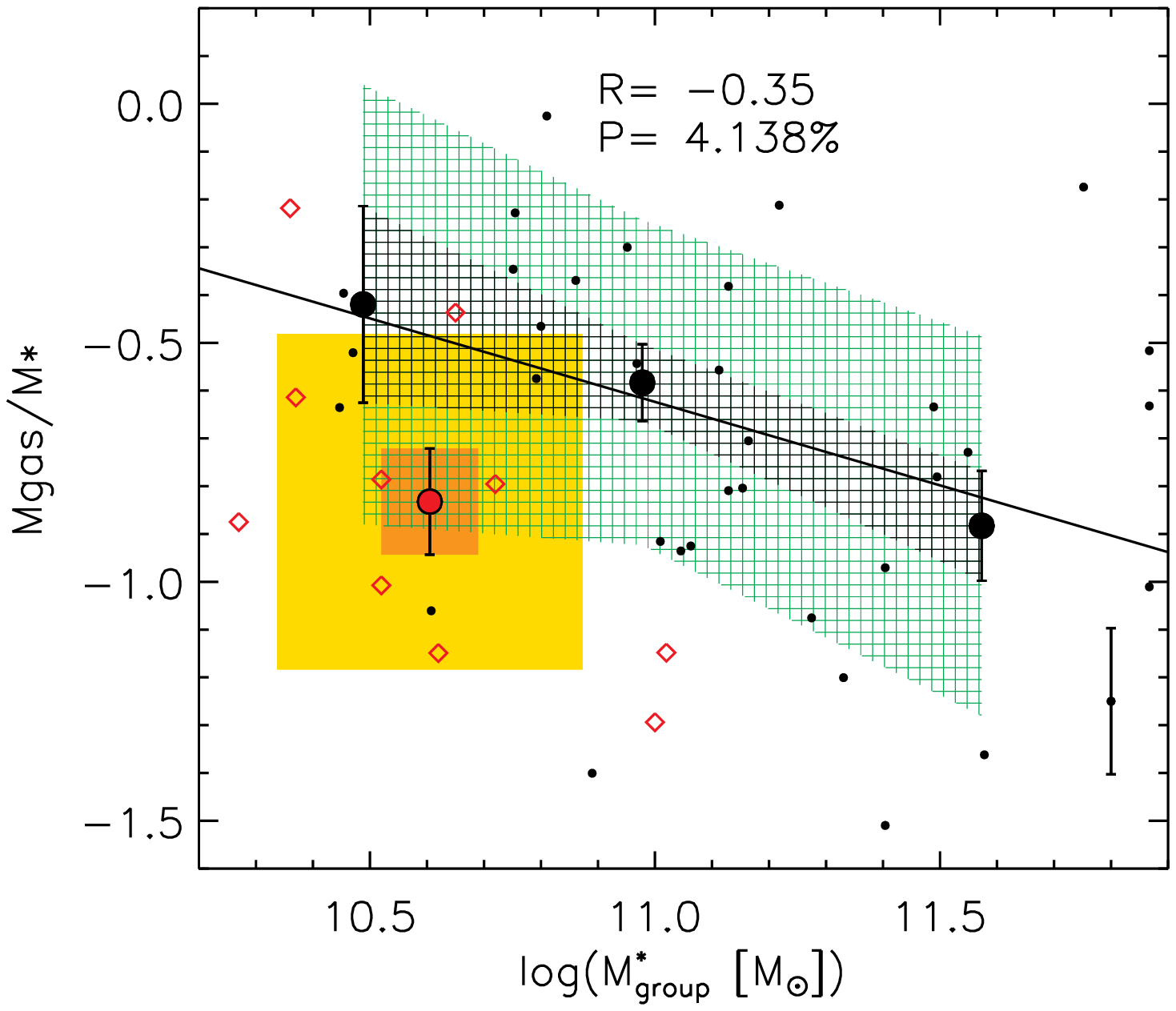}
 \caption{\label{IMG:MGROUP_SFR_BHAR_MGAS} {\bf Left panel:} sSFR as a function of M$^{*}_{\mathrm{group}}$ at 0.15$<$z$<$0.3.   No sensible differences are observed between small and bigger groups. {\bf Central panel:} sBHAR as a function of M$^{*}_{\mathrm{group}}$ for the same sources. The AGN activity per unit of group mass is higher in more massive groups. {\bf Right panel:} gas fraction as a function of M$^{*}_{\mathrm{group}}$. While a decreasing gas fraction is observed toward higher group stellar masses, this dependence is not strong enough to draw definitive conclusions. In all the plots, for field galaxies (red squares), the x axis represents the stellar mass M$^{*}$ of each single galaxy while, for galaxies in groups (black circles), it represents the total stellar mass of the hosting group (M$^{*}_{\mathrm{group}}$). We consider three bins of M$^{*}_{\mathrm{group}}$: M$^{*}_{\mathrm{group}}$[M$_{\odot}$]$<10^{10.75}$, $10^{10.75}<$M$^{*}_{\mathrm{group}}$[M$_{\odot}$]$<10^{11.25}$ and M$^{*}_{\mathrm{group}}$[M$_{\odot}$]$>10^{11.25}$, but the linear fit (solid line) is computed on the underlying data points and not on the binned data (only galaxies in groups are considered for the fit).  The typical uncertainty associated to single data-points is reported in the bottom right corner of each plot. These uncertainties are derived from the estimated PDFs of each output parameter of the SED fitting that is required to compute the quantity in the y axis. The correlation coefficient R and the correspondent \emph{p}-values are reported only when \emph{p}$<$5\%.  }
 \end{figure*}

The correlation that we identify can be explained as a sSFR decreasing with M$^{*}_{\mathrm{group}}$ and/or to an increasing sBHAR. From the first two panels of Figure~\ref{IMG:MGROUP_SFR_BHAR_MGAS}, the second of the two possibilities seems to be the most likely. While the sSFR does not seem to depend on the total stellar mass of the group, the specific BHAR increases at higher values of M$^{*}_{\mathrm{group}}$.

Following \cite{1978MNRAS.183..633G}, \cite{1985ApJ...288..481D}, \cite{1993AJ....106..831H}, \cite{1999ApJS..122...51D}, \cite{2004MNRAS.353..713K}, \cite{2005AJ....130.1482R} and \cite{2006A&A...460L..23P}, the numerical fraction of AGNs in the field ($f_{\mathrm{field}}$) is expected to be higher than in clusters ($f_{\mathrm{group}}$). However, as suggested by various works \citep{2000AJ....120...47C,2004AJ....128...68C,2001A&A...365L.110T,2006A&A...456..839T}, the AGN activity seems to be higher (or at least consistent) in small compact groups than in the field.
While in the papers mentioned above AGNs are identified using their X-ray emission, in this work AGNs are selected using the fraction of IR emission due to the Black Hole accretion as a discriminant.
As described in Section~\ref{SED_FIT_SECT}, this emission fraction can be estimated only when a far-IR detection (MIPS and SPIRE) is available (i.e. mostly for late type star forming galaxies). Consequently, we can only compute the numerical fraction of AGNs among far-IR detected star forming galaxies in the field ($F^{\mathrm{IR}}_{\mathrm{field}}$) and in groups ($F^{\mathrm{IR}}_{\mathrm{group}}$). For this reason, these values can not be directly compared with the results of the works mentioned above, where all the kind of galaxies hosting AGNs (early and late type) are considered.

\begin{figure}[!ht]
 \centering
 \includegraphics[width=8.0cm]{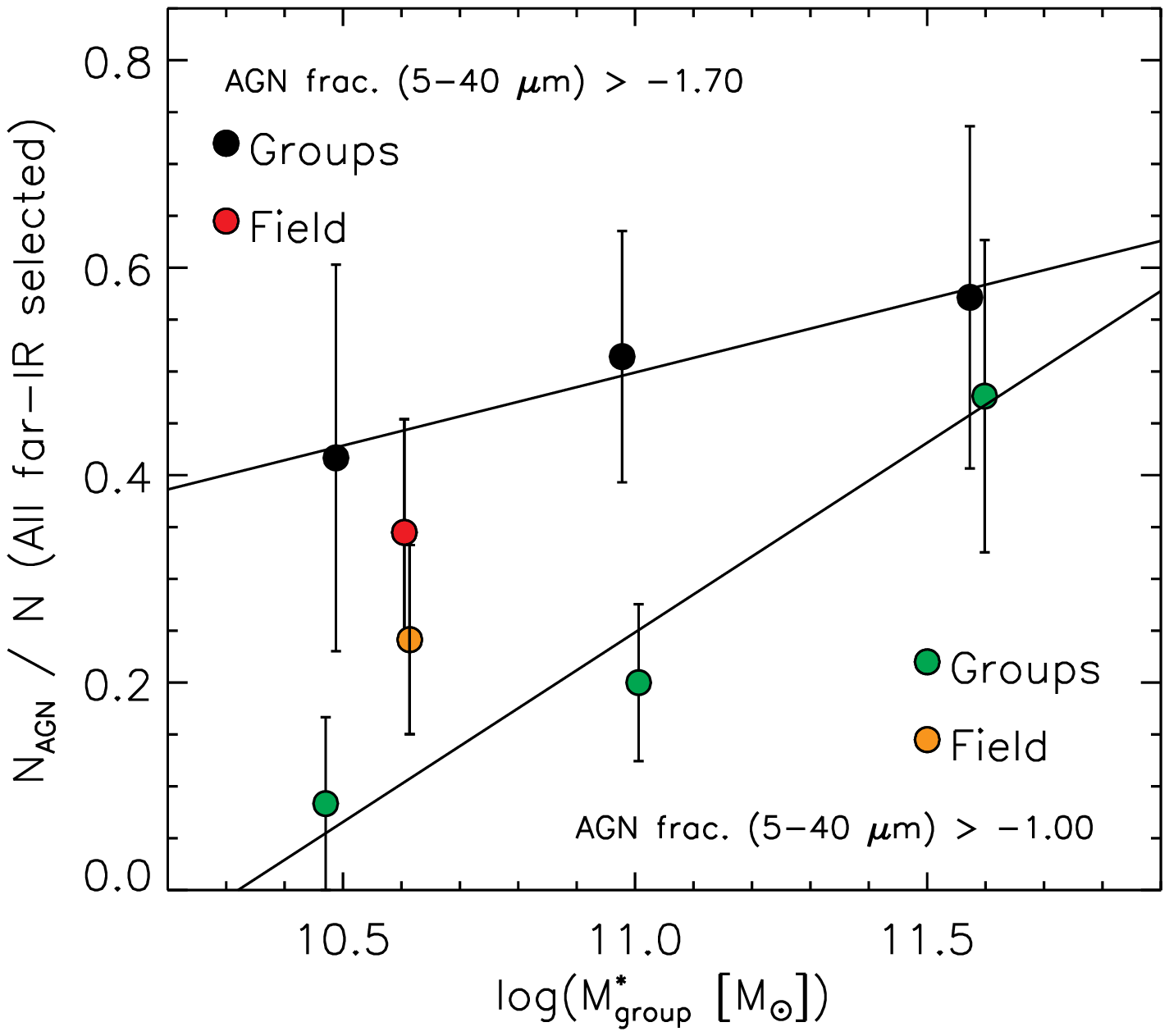}
 \caption{\label{IMG:AGN_NUM_FRACT} Numerical fraction $F^{\mathrm{IR}}$ of AGNs identified at 0.15$<$z$<$0.3, using different thresholds in $f^{5-40}_{\mathrm{AGN}}$. In our analysis, only IR-detected sources (MIPS and SPIRE) with $\log(f^{5-40}_{\mathrm{AGN}})>$-1.7 are identified as AGNs (red circle for $F^{\mathrm{IR}}_{\mathrm{field}}$, black dots for $F^{\mathrm{IR}}_{\mathrm{group}}$). The fraction of AGNs identified slightly increases with M$^{*}_{\mathrm{group}}$ but given the high uncertainties it is also consistent with a stable or even declining solution. The use of an higher threshold for the AGN identification (example: $\log(f^{5-40}_{\mathrm{AGN}})>$-1.0) brings to different results (yellow circle for $F^{\mathrm{IR}}_{\mathrm{field}}$, green dots for $F^{\mathrm{IR}}_{\mathrm{group}}$). This different behaviour is the result of a selection effect due to the fact that $f^{5-40}_{\mathrm{AGN}}$ is an increasing function of M$^{*}_{\mathrm{group}}$ (see left panel of Figure~\ref{IMG:MGROUP_AGN_act}).}
 \end{figure}

Considering as AGNs only the sources with an IR-emission fraction higher than $\log(f^{5-40}_{\mathrm{AGN}})=$ -1.7, as we do in our analysis, the numerical AGN fraction $F^{\mathrm{IR}}_{\mathrm{group}}$ shows weak or no dependence on M$^{*}_{\mathrm{group}}$ (black and red circles in Figure~\ref{IMG:AGN_NUM_FRACT}). At the same time, the fraction of AGNs in the field ($F^{\mathrm{IR}}_{\mathrm{field}}$) is smaller but still consistent with that measured in the groups. 
Given the relation observed between $\log(f^{5-40}_{\mathrm{AGN}})$ and M$^{*}_{\mathrm{group}}$ (central panel of Figure~\ref{IMG:MGROUP_AGN_act}), we expect this behaviour to be dependent on the threshold set for the AGN identification as a consequence of a selection effect. For example, using an higher threshold in $f^{5-40}_{\mathrm{AGN}}$, many of the AGNs in the lowest mass bin would not be identified as such anymore, while the new threshold would not affect by the same measure the numerical AGN fraction in the highest bin of M$^{*}_{\mathrm{group}}$. Consequently, the use of an higher threshold in $f^{5-40}_{\mathrm{AGN}}$ brings to a steeper relation between the numerical AGN fraction and M$^{*}_{\mathrm{group}}$ (Figure~\ref{IMG:AGN_NUM_FRACT}, yellow and green circles). A similar selection effect is expected if AGNs are selected on the basys of their BHAR (i.e. x-ray emission), given that also this quantity correlates with M$^{*}_{\mathrm{group}}$ in a similar way (Figure~\ref{IMG:MGROUP_BHAR}).

The central BH activity is generally thought to be ignited by nuclear inflows of gas, for example through galaxy-galaxy mergers \citep[e.g.][]{1992ARA&A..30..705B,2005ApJ...620L..79S}. Moreover, as observed in \cite{2012ApJ...758L..39T}, the AGN luminosity strongly correlates with the fraction of host galaxies undergoing a major merger (for $10^{43}<L_{\mathrm{bol}}(\mathrm{erg\ s}^{-1})<10^{46}$). Merging events must be more probable when, in a given volume, the number of galaxies that can potentially merge is higher, such as in small groups.

At the same time, however, \cite{2006A&A...460L..23P} find an anti-correlation between the fraction of AGNs in clusters and the velocity dispersion $\sigma_{v}$ of the cluster members. The merger rate is found to scale roughly as $\sigma_{v}^{-3}$ \citep{1992ApJ...401L...3M,1997ApJ...481...83M}.
Assuming that the velocity dispersion increases with the mass of the structures \citep[e.g.][]{1985ApJ...298....8H,1990A&A...237..319P}, a similar anti-correlation should be expected between the fraction of AGNs and the total observed stellar mass of the groups. 
As shown in Figure~\ref{IMG:AGN_NUM_FRACT} (filled black circles), we do not observe such a relation.
 Our results are more probably consistent with a flat or an increasing $F^{\mathrm{IR}}_{\mathrm{group}}$ at higher M$^{*}_{\mathrm{group}}$ but given the large uncertainties associated to these estimates, our results do not allow us to exclude the opposit behaviour. However, we notice that if the nuclear activity is actually driven by merger events, these must be more probable in richer (i.e. more massive) groups, at least until the the velocity dispersion start to dominate the dinamics of the structures, such as inside galaxy clusters.
In any case, we stress on the fact that the AGNs in our sample are selected only among star forming galaxies, while we are not measuring the fraction of AGNs among \emph{all} the sources (late star forming and early type).

Given these premises, the similar numerical fraction of AGNs identified at all M$^{*}_{\mathrm{group}}$ indicates that the activation of the nucleus of a star forming galaxy is not more likely in the most massive groups although, in these cases, the level of nuclear activity is increased, as shown by the BHAR/SFR ratio. As shown in the right panel of Figure~\ref{IMG:MGROUP_SFR_BHAR_MGAS}, the gas fraction of galaxies hosting an AGN tends to be similar or even lower in more massive groups. 
This observation, together with the similar sSFR measured in low and high mass groups, suggests that if a singular infall of gas is responsible for the activation of both AGN and star formation, an higher fraction of this gas must be driven to the galaxy center if the galaxy is located in a more massive group.

\subsection{Selection effects and independent confirmations}
\label{SEC:SEL_EFF}
In this section, we test the relations found in Section~\ref{SEC:SMBH_act_groups} against possible biases artificially introduced by our sample selections or by the techniques used in our analysis. We also try to find indipendent confirmations of the same relations

\subsubsection{Completeness of the groups}

\begin{figure}
 \centering
 \includegraphics[width=7.5cm]{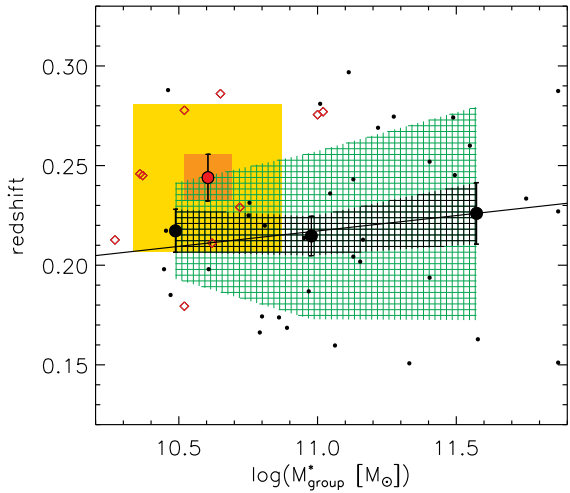}
 \caption{\label{IMG:Sel_effects} Average redshifts as a function of M$^{*}_{\mathrm{group}}$. No significant dependence is observed. The higher average redshift observed for field galaxies (in this case the x axis represents the stellar mass M$^{*}$ of the single galaxies), although not substantial, can be possibly due to the higher chance to miss group members at higher z.} 
 \end{figure}

Figure~\ref{IMG:Sel_effects} shows that the redshift of the group members does not depend on M$^{*}_{\mathrm{group}}$. This indicates that groups located in the low-z border of the redshift bin that we study (0.15$<$z$<$0.3) are not richer, or more complete, than those found in the high-redshift border of the same bin.
However, field galaxies show higher average redshifts than group members. 
This could be an effect of the decreasing completeness of the groups at higher redshifts, where the detection of low luminosity companions, connecting sources in a group through a \emph{friend of friend} algorithm, becomes more difficult. This effect should be particularly prominent for galaxies hosting AGNs, given that in the local universe their position is usually peripherical with respect to cluster centers \citep{2013MNRAS.429.1827P}. This is particularly true for the most luminous AGNs that avoid high-density regions \citep{2004MNRAS.353..713K}. For this reason, the field galaxies sample could possibly be contaminated by group members located at higher redshifts than the average, with an AGN fraction probably higher than the average.

\subsubsection{Mid- to far-IR flux ratio}
In our analisys, we are considering sources with relatively small AGN emission fractions. These values are not surprising, given that our data sample is made by low-redshift star forming sources detected above 250 \mic\ (i.e. where the emission due to star formation dominates on that from the dusty torus of the AGN). In such a sample, even a particularly IR-bright AGN would bring a relatively small contribution to the total IR emission. 

In order to ensure that the relations found are not an artificial effect introduced during the SED fitting process, we measured the ratios between the total mid-IR and the total far-IR observed fluxes for the same sources considered in our analysis. We computed the total mid-IR flux ($F_{\mathrm{mid-IR}}$) as the sum of the fluxes measured in all those bands that we expect to be more influenced by the AGN torus emission, at these low redshifts: 4.5 \mic, 7 \mic, 11 \mic\ and  15 \mic. For the 11 \mic\ band, we used both the WISE-W3 and AKARI-S11 measurements. The fluxes in all the bands are normalized using the average flux ratio between the band considered and the W3 band. In a similar way, we computed the total far-IR flux ($F_{\mathrm{far-IR}}$) as the sum of the fluxes measured at 250 \mic, 350 \mic\, and 500 \mic. In this case, the 250 \mic\ band is used for the flux normalization.

\begin{figure}
 \centering
 \includegraphics[width=7.5cm]{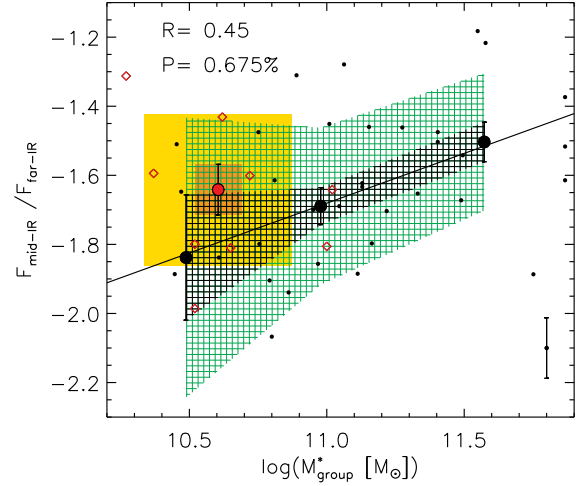}
 \caption{\label{IMG:M_MIR_FIR} Ratio between total observed mid-IR (4.5, 7, 11, and 15 \mic) and far-IR (250, 350, and 500 \mic) fluxes as a function of M$^{*}_{\mathrm{group}}$. The typical uncertainty associated to each single data-point is reported in the bottom right corner of the plot. The behaviour of the observed fluxes indipendently confirms what is found using the SED fitting technique. }
 \end{figure}

Figure~\ref{IMG:M_MIR_FIR} shows how the behaviour of the $F_{\mathrm{mid-IR}}/F_{\mathrm{far-IR}}$ ratio confirms what is found for the AGN emission fractions and for the BHAR/SFR ratio obtained through the SED fitting (Figure~\ref{IMG:MGROUP_AGN_act}). In this case, the \emph{p}-value of the relation found is higher (\emph{p}=0.7\%), but this is not unexpected, given that the SED fitting technique is meant to maximize the information obtained from single photometric bands (i.e. it should be more precise). Moreover, the observed fluxes do not take into account the redshifts of the sources, while they are considered in the SED fitting process.

\subsubsection{Influence of the M$^{*}$-BHAR relation on small groups}

As found in \cite{2012ApJ...753L..30M} and in successive works \citep[e.g.][]{ 2015ApJ...800L..10R}, the AGN activity derived from x-rays measurements is related to the stellar mass M* of each single host galaxy. 

In the small groups regime explored in our analysis, we are considering also pairs and triplets. The mass of these groups could easily be dominated by that of the AGNs identified through our selection criteria. Infact, given the M*-SFR relation (the so-called \emph{main sequence}), the far-IR selected galaxies in our sample are also the most massive ones. For this reason, the relations that we find with the total mass of the groups could possibly be only apparent and due to the underlying nature of the galaxies considered singularly.

\begin{figure}
 \centering
 \includegraphics[width=7.5cm]{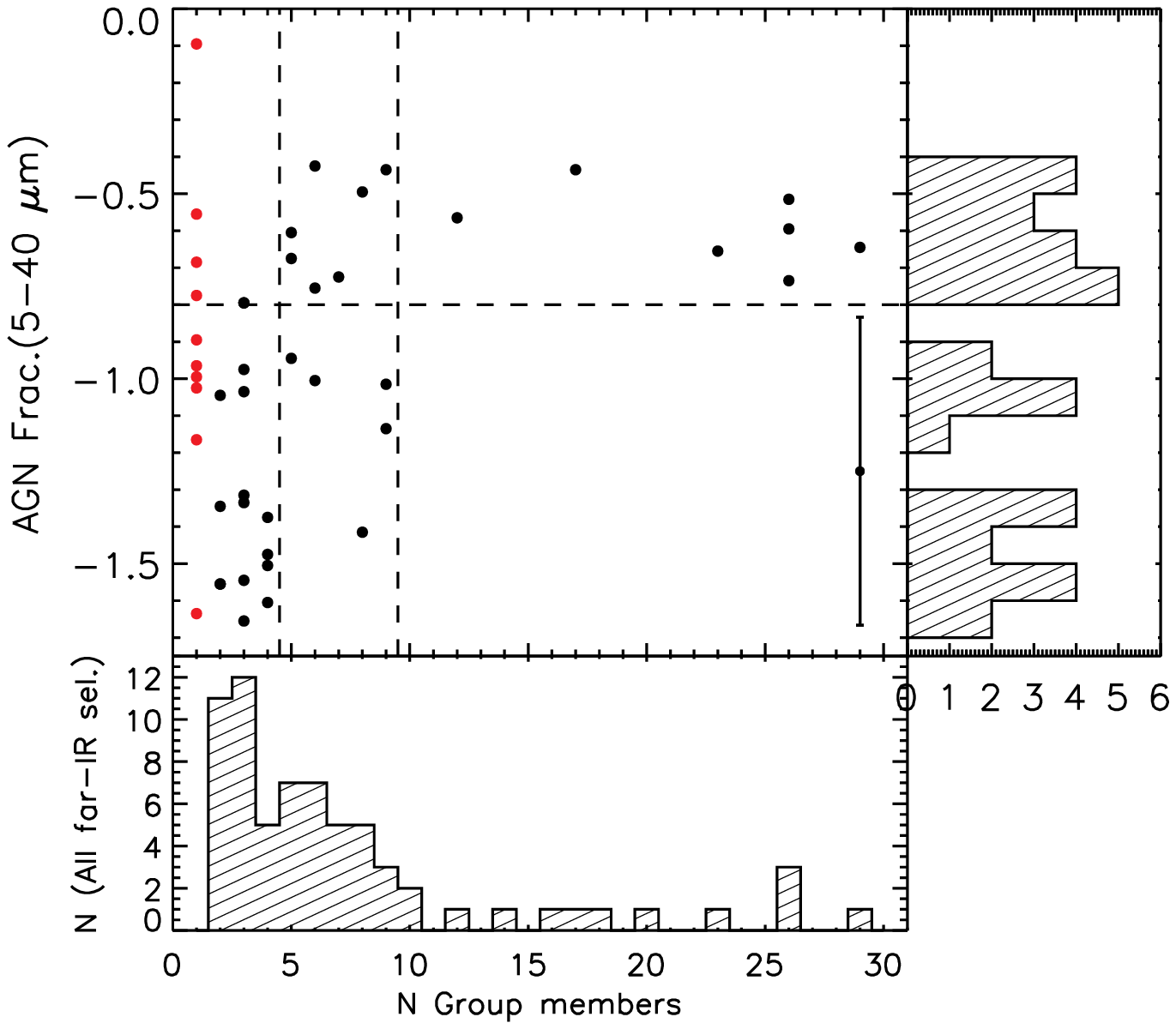}
 \includegraphics[width=7.5cm]{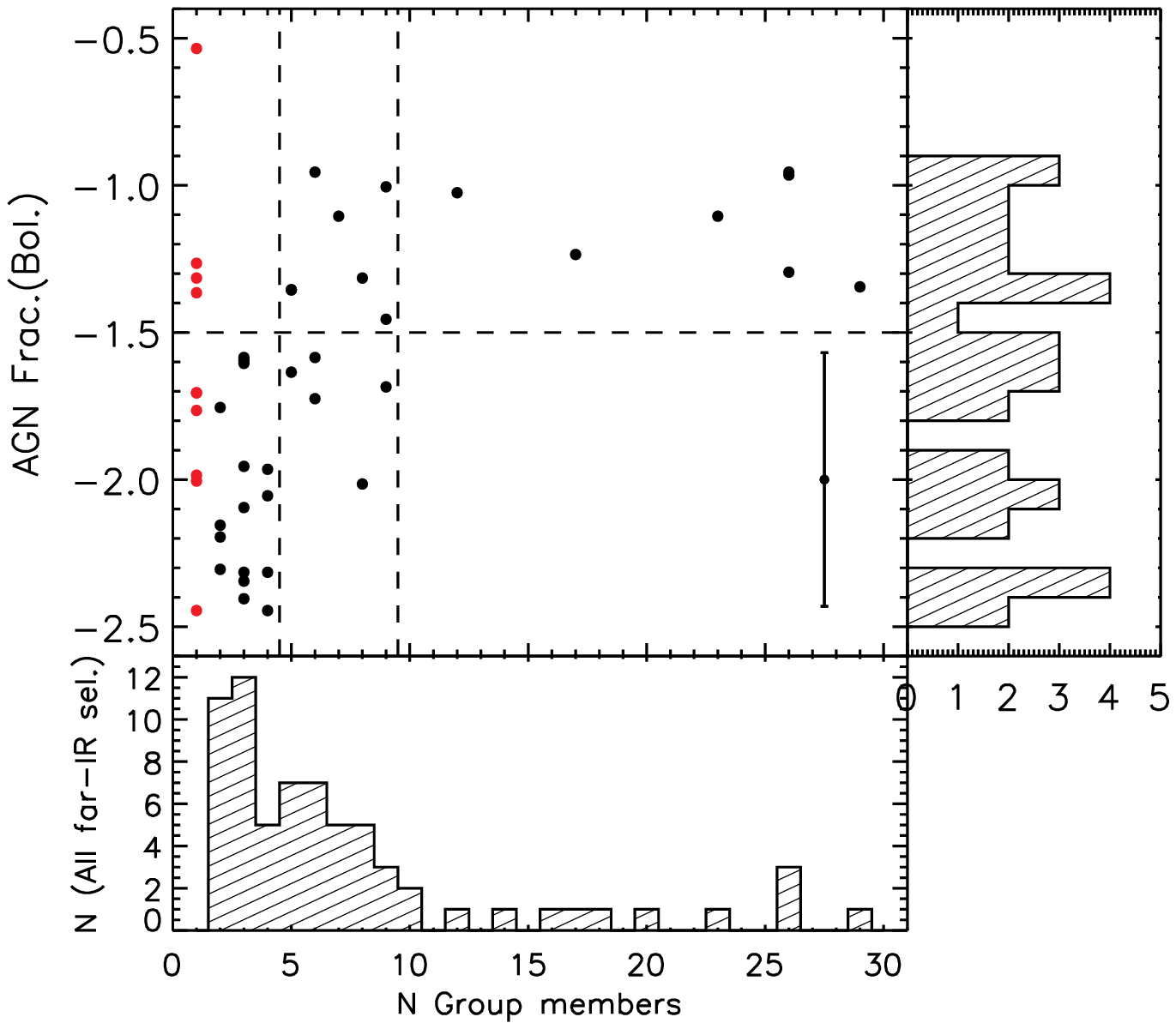}
 \caption{\label{IMG:RICH_AGNF} AGN fraction in the 5-40 \mic\ ({\bf upper panel}) and in the 8-1000 \mic\  ({\bf bottom panel}) bands as a function of the richness of the groups. Field galaxies are represented using red filled circles (they are conveniently located at N=1).  The typical uncertainty associated to each single data-point is reported in the bottom right corners of each plot. These uncertainties are derived from the estimated PDFs of each correspondent output parameter of the SED fitting. We visually delimit two clearly separable regimes of AGN activity using black dashed lines: among the members of the richest groups, the logarithmic AGN fraction (when detected) is always higher than -0.8 and -1.5 in the 5-40 \mic\ and in the bolometric bands respectively. Vice versa, very small groups (less than 5 members), never show an AGN fraction higher than that limit. While the histograms in the right show the distribution of only the sources visible in the plot, the histograms in the bottom show the distribution of all the far-IR selected galaxies (with and without AGN component). }
 \end{figure}

In order to see if the environment plays a real or only an apparent role, we studied the dependence of the AGN emission fraction on the richness of the groups. This parameter is not influenced by the M$^{*}$ of the selected AGNs.
As shown in Figure~\ref{IMG:RICH_AGNF}, the AGN emission fraction depends on the total number of group members similary to what happens with M$^{*}_{\mathrm{group}}$. This confirms that the relations that we find with M$^{*}_{\mathrm{group}}$ are not due to a selection effect.
In the same Figure, we can broadly identify three different regimes of AGN activity. In very small groups (less than 5 members) the AGN activity, when detected, is always lower than $\log(f^{5-40}_{\mathrm{AGN}})$=-0.8 (or -1.5 in the bolometric band). Vice versa, in the richest groups (more than 10 and less than 30 members), the AGN fraction is always higher than the previous limit. Between these two extremes, groups with more than 5 but less than 10 members show a mixed behaviour, but with an higher minimum of possible AGN fraction.

\section{Summary and conclusions}
\label{SECT:Conclusion}

We presented the multiwavelength catalog of sources detected in the SIMES field. This new catalog, including observations in 30 photometric bands (23 excluding the redundant bands covered by different instruments), is released through the NASA/IPAC Infrared Science Archive (IRSA). Our best photo-z estimates are included.
Using these photometric data, we studied the AGN emission fraction and the ratio between black hole accretion and star formation rates (BHAR/SFR), for a sample of star forming galaxies located in the field and in small galaxy groups. 
The redshift range that we explored goes from z=0.15 to z=0.3.

In the mass range 10.25$<\log($M$^{*}_{\mathrm{group}}[$M$_{\odot}])<$11.9, we found that the relative importance of the AGN activity in a star forming galaxy is an increasing function of the total stellar mass of the hosting group, with $\log($BHAR/SFR$)\propto (1.04\pm0.24)\log($M$^{*}_{\mathrm{group}})$.
We suggest this behaviour is due to an increasing efficiency of the BH accretion in larger groups, because the sSFR does not depend on M$^{*}_{\mathrm{group}}$, while we do observe an increased average sBHAR among more massive groups. 
The average value of the BHAR/SFR ratio for Field AGNs, and its dispersion, do not allow us to find significative differences between the behaviour of field galaxies and galaxies in groups.

Differently from the fraction \emph{of IR emission} due to the central black holes, the  \emph{numerical} fraction of AGNs (i.e. fraction of sources identified as AGNs in the far-IR selected sample, $F^{\mathrm{IR}}_{\mathrm{group}}$) shows weak or no dependence on M$^{*}_{\mathrm{group}}$, with the fraction of field AGNs lower but consistent with that measured among group members. However, we found that the slope of this relation does depend on the threshold set for the selection of the AGNs: using higher thresholds in $f^{\lambda}_{\mathrm{AGN}}$ (or in BHAR) brings to steeper relations between $F^{\mathrm{IR}}_{\mathrm{group}}$ and M$^{*}_{\mathrm{group}}$. We warn that a similar bias could affect those surveys where AGNs are selected by similar quantities, such as the x-ray emission.  

The results here summarized indicate that at these scales, an higher density environment is more effective in \emph{driving} an higher rate of nuclear accretion of star forming galaxies, than in \emph{activating} it. In particular, the nuclear accretion is faster if the galaxy is located in a more massive group.

If the AGN activity is driven by merging events (or by gas infall) the chance of an isolated galaxy to be subject to such an event must be lower than in a small group. On the other hand, the high velocity dispersion characterizing the richest clusters lower the probability of these events, as suggested by \cite{2006A&A...460L..23P}. The expected net result is an AGN activity that increases from low mass groups towards higher masses until a turning point is reached, where the velocity dispersion of these structures prevails on the effects of the increased number of possible interactions.

\acknowledgements
This work is based on: 
observations obtained with AKARI, a JAXA project with the participation of ESA; 
data products from the Wide-field Infrared Survey Explorer, which is a joint project of the University of California, Los Angeles, and the Jet Propulsion Laboratory/California Institute of Technology, funded by the National Aeronautics and Space Administration; 
data from the \emph{Spitzer Space Telescope}, which is operated by the Jet Propulsion Laboratory, California Institute of Technology under a contract with NASA; 
data from \emph{Herschel}, an ESA space observatory with science instruments provided by European-led Principal Investigator consortia and with important participation from NASA; 
observations made with the NASA Galaxy Evolution Explorer. GALEX is operated for NASA by the California Institute of Technology under NASA contract NAS5-98034; 
Further fundamental observations were obtained with the ESO-\emph{VST} and VISTA, and CTIO observatories.
We gratefully acknowledge Simon Lilly and Alvio Renzini for supplying us with the extended sample of spectroscopic redshifts measured in the COSMOS field.
M.V. acknowledges support from the European Commission Research Executive Agency (FP7-SPACE-2013-1 GA 607254), the South African Department of Science and Technology (DST/CON 0134/2014) and the Italian Ministry for Foreign Affairs and International Cooperation (PGR GA ZA14GR02). 
R.C. acknowledges financial support from CONICYT Doctorado Nacional No. 21161487. 
I.B. thanks Stefano Rubele for the useful discussions concerning the technical details of the optical data reduction, Lee Armus for his comments on the AGN fraction reliability and Gabriele Rodeghiero and Maurizio Pajola for their helpful comments about data representation.

\bibliographystyle{apj}
\bibliography{biblio2}{}

\end{document}